# Towards a Theory of Requirements Elicitation: Acceptability Condition for the Relative Validity of Requirements

Ivan J. Jureta[*,†] John Mylopoulos[‡] Stéphane Faulkner[§]

October 30, 2018

**Abstract**

A requirements engineering artifact is valid relative to the stakeholders of the system-to-be if they agree on the content of that artifact. Checking relative validity involves a discussion between the stakeholders and the requirements engineer. This paper proposes (i) a language for the representation of information exchanged in a discussion about the relative validity of an artifact; (ii) the acceptability condition, which, when it verifies in a discussion captured in the proposed language, signals that the relative validity holds for the discussed artifact and for the participants in the discussion; and (iii) reasoning procedures to automatically check the acceptability condition in a discussions captured by the proposed language.

[*]Fonds de la Recherche Scientifique – FNRS, Brussels, Belgium.
[†]Information Management, University of Namur; ivan.jureta@fundp.ac.be.
[‡]Department of Computer Science, University of Toronto; jm@cs.toronto.edu.
[§]Information Management, University of Namur; stephane.faulkner@fundp.ac.be.

# Contents

# 1 Introduction

A basic question for requirements engineering (RE) is how to find out what the stakeholders of a system-to-be "really need" [8]. In response, RE stars with the *elicitation* of requirements, the purpose of which is the initial investigation of the goals, functions, and constraints of the system-to-be, as they are stated by the stakeholders. Despite the difficulty of making a clear distinction between the various specific tasks that RE can involve [8], elicitation is acknowledged to be one of the three fundamental tasks in RE, in addition to *validation* and *modeling/specification* [14, 22].

Modeling/specification depends heavily on elicitation, for it is elicitation that provides the application domain- and project-specific information that is, potentially in a changed form, represented in RE artifacts (i.e., models and specifications). RE artifacts capture the elicited information in a format that lends itself to specific analysis, which the stakeholders themselves have difficulty to perform, such as, e.g., the verification of the internal consistency of requirements. Neither elicitation, nor modeling/specification can be successful without validation. As Goguen and Linde observe, "[t]here are very good reasons why [stakeholders] often do not, or cannot, know exactly what they need; they may want to see models, explore alternatives, and envision new possibilities" [8]. A key purpose of requirements validation is to seek feedback from the stakeholders on RE artifacts so as to inform further iterations of elicitation and/or modeling/specification. To check the validity of an RE artifact is to determine if what it says about the system-to-be is in line with what the stakeholders "really need".

**Problem.** The aim of validation is ambitious: through repeated and intertwined performance of validation together with elicitation and modeling, we would indeed hope to arrive at RE artifacts that capture *exactly* what the stakeholders *really* need. Such *absolute validity* should be distinguished from what we call *relative validity.* While the former certainly stands as an ideal to aim for, the latter is achievable in practice and is the concern of this paper.

*Relative validity is concerned with whether the stakeholders agree on the content of an* RE *artifact.* Validity is in this sense *relative to the stakeholders.* An RE artifact is therefore valid in this sense if the *stakeholders* agree that what it says about the system-to-be is acceptable to them. Stated otherwise, this form of validity will verify if all the concerns, which the stakeholders raised, are answered.

It is safe to say that we cannot know if the stakeholders agree on an artifact if we do not give them the possibility to raise their concerns. The engineer can inform them in this task by providing graphical animations of a behavior model [21], the results of checking of predefined properties on models made from parsed text [7], explicit accounts of (the inconsistencies between) different viewpoints on the system-to-be [14]. In each of these cases, the engineer will be producing information to present in a potentially summarized form to the stakeholders, and then discuss it. *Checking relative validity inevitably leads to a discussion between the stakeholders and the requirements engineer.*

**Contributions.** This paper focuses on the modeling and analysis of a *discussion* between the stakeholders and requirements engineers about the relative validity of an RE artifact. By building on contributions in design rationale research, argumentation research in artificial intelligence, and graph traversal algorithms, our aims are to: (i) provide a simple but expressive propositional model of the explicit exchange of information in what is usually called a discussion; (ii) based on the model of a discussion, propose a condition, called the *acceptability condition* on an artifact (denoted **AC**), such that if it holds, then it signals that the relative validity verifies for that artifact and for the participants in that discussion; and (iii) if a concrete discussion is recorded (as is the case when discussions are realized in forum-like applications), then check automatically if **AC** holds at some point in the discussion. To meet these aims, we propose the *A*cceptability *E*valuation framework, henceforth ACE. ACE can be seen as a simple propositional reasoning framework, that is independent of the RE method that produces the artifact, and of the application domain.

**Organization.** Basic assumptions are first discussed, and a preliminary definition of the acceptability condition is proposed (§3). The two components of ACE are then presented: the language to record the information relevant to the evaluation of acceptability (§4), and (ii) algorithms for retrieving the recorded information and evaluating its acceptability (§5). The full definition of the acceptability condition is then given in light of the complete ACE framework (§6). Notes some implementation considerations are then outlined (§7) and the related work is discussed (§8). We end with a summary of conclusions and point to future work (§9).

## 2 Baseline

What is typically called a discussion is, roughly speaking, a complex exchange of information between potentially many participants. Various properties of a discussion can be studied, such as its topic, purpose, (dis)organization, and so on. We focus on discussions about RE artifacts, the purpose of which is to reach a conclusion about the relative validity of the artifacts. We are interested in the specific traits of the structure of such discussions. These are the inference, attack, and preference relationships between pieces of information offered in a discussion. Inferences connect premises to conclusions, attacks connect somehow opposing information, and preferences compare in terms of desirability the conditions described via the various pieces of information offered in the discussion.

The range of discussions we focus on is not confined to specifc RE methods and artifacts. Leite and Freeman [14] observed that "the whole process of [RE] is a web of subprocesses, and it is very difficult to make a clear distinction between them". A subprocess in that web amounts to the application of some RE method to specific inputs, in order to produce an output, itself fed into the application of another method, and so on. To remain general, we can say that the application of any RE method, i.e., any subprocess in the complex RE web, fits the abstract input-transformation-output pattern. Namely, in a given application domain $D$, information elicited or produced by another RE method acts as the input $I_D$ to a domain-independent RE method, symbolized by the function $T$. The latter produces the domain-specific output $O_D$, i.e., $O_D = T(I_D)$. E.g., the refinement of a requirement asks for an abstract requirement and domain-specific knowledge as its inputs $I_D$, and results in a set of less abstract requirements as its output $O_D$, while the transformation $T$ establishes the relations, such as consistency, that must verify between inputs and outputs. Observe that $I_D$ and $O_D$, or any part thereof is clearly an RE artifact. Moreover, we can view the application of a method to specific inputs, i.e., $T(I_D)$ as an artifact itself: there really seems to be no strong argument not to allow the participants to discuss the engineer's choice of applying $T$ to $I_D$.

Discussion performed to the aim of checking the relative validity of the application of an RE method, i.e., $O_D = T(I_D)$, or equivalently, the relative validity of individual artifacts $O_D$, $T(I_D)$, and $I_D$, consists of offering information in favor or against these, and providing opinions about the relative desirability of the offered information. If I agree with you, I can provide additional information to support your position; if we disagree, I can offer information against your positions; if I have no further information to offer in favor of or against that which has been offered, I can say which of the already present conclusions I prefer to others. $O_D = T(I_D)$ will be acceptable if and only if no information offered against any of the components of $O_D = T(I_D)$ holds by the end of the discussion. *Acceptability signals agreement.* It is reasonable to interpret agreement as relative validity. It is by analysing a discussion that we can determine if there is agreement about the artifacts being validated, and thereby if they are valid relative to the participants in the discussion. If the parties agree that the given inputs $I_D$ transformed by the application of the method $T$ give $O_D$, then they agree that $O_D = T_D(I_D)$ holds, so that the given method application is acceptable, denoted $\mathbf{AC}(I_D, T(I_D), O_D)$.

## 3 Acceptability Condition

**Definition 3.1.** ***AC.*** *The application of the* RE *method $T$ to the input $I_D$ to produce the output $O_D$ is acceptable, denoted $\boldsymbol{AC}(I_D, T(I_D), O_D)$ if and only if:*

$$\boldsymbol{AC}(I_D) \wedge \boldsymbol{AC}(T(I_D)) \wedge \boldsymbol{AC}(O_D) \tag{1}$$

In order to apply to *any* method, ACE sees any RE artifact, or part thereof, as a proposition. In conceptualizing a *proposition*, we follow McGrath's [16] stipulation that propositions "are the sharable objects of the attitudes [(i.e., what is believed, desired, etc.)] and the primary bearers of truth and falsity". Regardless then of the syntax and semantics of the RE method deployed to produce an artifact, the artifact itself is a conjunction of propositions. Symbols $p$, $q$, and $r$, indexed when needed, denote individual propositions. $\mathit{In}(I_D)$, $\mathit{In}(O_D)$, and $\mathit{In}(T(I_D))$ denote the sets of propositions, respectively in $I_D$, $O_D$, and $T(I_D)$. We assume that all propositions in $I_D$, $T$, and $O_D$ are visible to all participant in a discussion about the relative validity of these artifacts. A participant having information in favor or against any proposition in $\mathit{In}(I_D)$, $\mathit{In}(O_D)$, and $\mathit{In}(T(I_D))$ will voice that information. We evaluate the acceptability of the individual propositions in $\mathit{In}(I_D)$, $\mathit{In}(O_D)$, and $\mathit{In}(T(I_D))$ in order to verify $\mathbf{AC}(I_D, T(I_D), O_D)$:

$$\mathbf{AC}(I_D, T(I_D), O_D) \text{ iff } \forall p \in \mathit{In}(I_D) \cup \mathit{In}(O_D) \cup \mathit{In}(T(I_D)),\ \mathbf{AC}(p) \tag{2}$$

The acceptability of a given proposition $p$ is automatically verified in ACE via an algorithm that analyzes the information given in favor or against $p$, and captured via a language defined in the following section.

# 4 Language

We first present exemplified overview of the language (§4.1), then provide a detailed definition of the language (§4.2).

## 4.1 Overview of the Language

All information relevant for the evaluation of acceptability is encoded into a directed labeled graph $G$, with the set of vertices $V(G)$ and lines $L(G)$, and the labeling functions $\lambda_V$ and $\lambda_L$ for vertices and lines, respectively. Any one proposition $p$ or a conjunction of propositions in any of $\mathit{In}(I_D)$, $\mathit{In}(O_D)$, and $\mathit{In}(T(I_D))$ is captured by exactly one vertex $v \in V(G)$. As all lines carry the same label $\forall l \in L(G),\ \lambda_L(l) = \mathtt{To}$, there is no need to write this label in graphs. There are four labels for vertices: $\forall v \in V(G), \lambda_V(v) \in \{\mathtt{i}, \mathtt{I}, \mathtt{C}, \mathtt{P}\}$. $G$ together with the labeling functions and the propositions forms the syntax of the language. Consider the following example for illustration.

Suppose that the aim is to build a system that would deliver music on-demand: a user visits a website, chooses songs from a database, and can play them in the audio player on the website. The following is an important goal:

(Ex.1) $g_1$: *Generate revenue from the audio player.*

We refine it by the conjunction of the three goals $g_2$, $g_3$, and $g_4$ below:

(Ex.2) $g_2$: *Display text ads in the audio player.*

(Ex.3) $g_3$: *Target text ads according to users' profiles.*

(Ex.4) $g_4$: *Maintain the player free to all users.*

We therefore have $I_D = g_1$ and $O_D = g_2 \wedge g_3 \wedge g_4$. The applied RE method is the standard AND-refinement of a goal [5]. We capture the application of AND-refinement in the example via the graph shown in Ex.1.

(Ex.5)

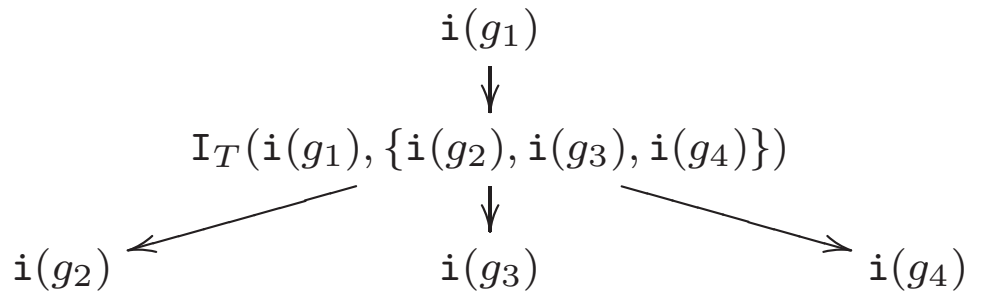

The refined goal $g_1$ and the components $g_2$, $g_3$, $g_4$ of its refinement are assigned the label $\mathtt{i}$ because of their role as the inputs and outputs to the application of an inference rule, denoted $\mathtt{I}_T(\mathtt{i}(g_1), \{\mathtt{i}(g_2), \mathtt{i}(g_3), \mathtt{i}(g_4)\})$. The label $\mathtt{i}$ is assigned to an *information* vertex, which serves as the input and/or output to the application of an *inference rule*, corresponding to the label $\mathtt{I}_T$. An inference rule vertex in $G$ represents the application of some particular rule of deductive or ampliative inference to inputs in order to obtain the given outputs. An example of deductive inference is modus ponens. Inference or reasoning is ampliative when a conclusion is inferred, which includes information absent from the premises, from which the conclusion is inferred. Examples of rules of ampliative inference are induction by enumeration, reasoning with analogies, causal reasoning. Refinement, as any method is an inference rule, so that the application of AND-refinement is represented by $\mathtt{I}_T(\mathtt{i}(g_1), \{\mathtt{i}(g_2), \mathtt{i}(g_3), \mathtt{i}(g_4)\})$ in the graph. Any transformation $T$ translates into at least one I vertex; it will equate to more vertices when we are not content with evaluating the acceptability of the application of the method as a whole (i.e., the black box approach), but are interested in evaluating in detail the acceptability of the various steps or other considerations called for in the application of the method. The meaning of each line is derived from the vertices it connects, and its direction. The line from $\mathtt{i}(g_1)$ to $\mathtt{I}_T(\mathtt{i}(g_1), \{\mathtt{i}(g_2), \mathtt{i}(g_3), \mathtt{i}(g_4)\})$ is understood as stating that the former is the input to the application of the given inference rule, $\mathtt{I}_T$.

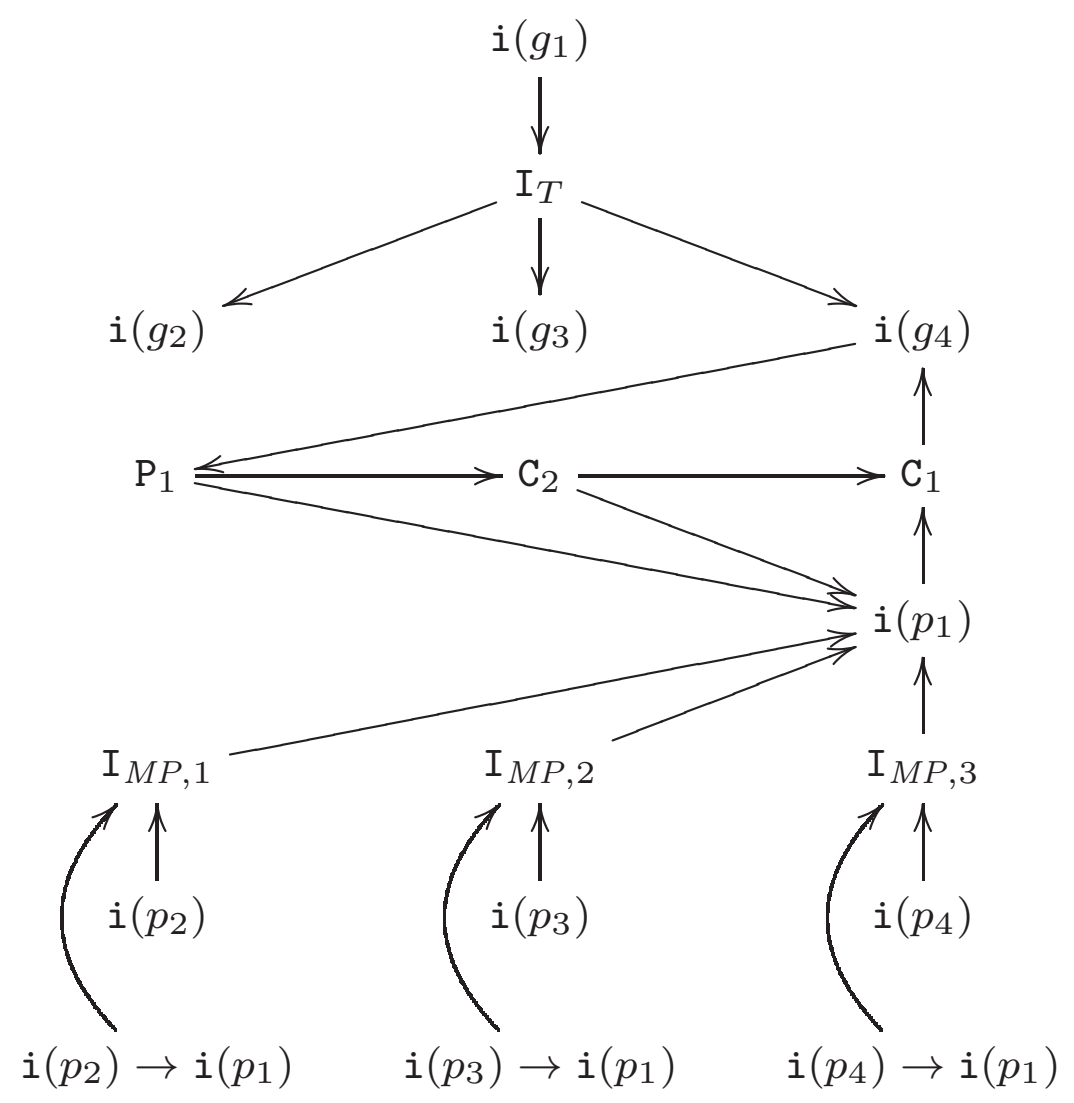

**Figure 1:** ACE graph capturing the application of the refinement method to the goal $g_1$ and the information offered in the discussion of that application of refinement.

An information vertex (i) can be involved in the application of a conflict (C) or of a preference rule (P). Suppose that a stakeholder indicates the following:

(Ex.6) $p_1$: *Revenue can be generated by charging subscriptions to users*

A vertex labeled C indicates an application of a *conflict rule*, that is, the application of criteria giving rise to a conflict between two or more other vertices in the graph. Since it is clear that not all features of the player are free when a paid subscription is available, we add the conflict vertex $\mathtt{C}_1(\mathtt{i}(p_1), \mathtt{i}(g_4))$ to the graph.

(Ex.7) $\mathtt{C}_1$: *Subscription and advertising revenue models should not be combined on this system.*

Additional information can be found to elaborate on the subscription revenue model:

(Ex.8) $p_2$: *Part of the music database can be restricted, so that the player only plays 30 seconds of some songs, until the user buys a subscription to listen full songs.*

(Ex.9) $p_3$: *According to competitors' services, some users are willing to pay to choose a different graphical layout for the online audio player; users can be allowed to choose among different graphical layouts and pay for each.*

(Ex.10) $p_4$: *Two versions of the player can be offered, one with basic and free features, and another with advanced features requiring subscription.*

We choose to relate each of $p_2$, $p_3$, $p_4$ via the modus ponens inference rule to $p_1$. Say that a survey concludes that users strictly prefer a free music on-demand service to one based on subscription. We capture this strict preference by the preference vertex $\mathtt{P}_1(\mathtt{i}(g_4), \mathtt{i}(p_1))$ and lines from $\mathtt{i}(g_4)$ to $\mathtt{P}_1(\mathtt{i}(g_4), \mathtt{i}(p_1))$, and from $\mathtt{P}_1(\mathtt{i}(g_4), \mathtt{i}(p_1))$ to $\mathtt{i}(p_1)$ in accordance to the direction of preference. A preference vertex represents the application of a *preference rule*, that is, the application of criteria defining a strict preference order between the conditions described in two or more other vertices; we can write that rule as follows:

(Ex.11) $\texttt{P}_1$: *Users strictly prefer a free music on-demand service to one based on subscription.*

If the stakeholders agree that $\texttt{P}_1(\texttt{i}(g_4), \texttt{i}(p_1))$ resolves the conflict $\texttt{C}_1(\texttt{i}(p_1), \texttt{i}(g_4))$, then an application of a conflict rule will be added, $\texttt{C}_2(\texttt{P}_1, \{\texttt{C}_1, \texttt{i}(p_2)\})$, from $\texttt{P}_1(\texttt{i}(g_4), \texttt{i}(p_1))$ to $\texttt{C}_1(\texttt{i}(p_1), \texttt{i}(g_4))$ and $\texttt{i}(p_2)$.

(Ex.12) $\texttt{C}_2$: *Users' preference of free over subscription services should be satisfied.*

Figure 1 summarizes this discussion; applications of inference, conflict, and preference rules are given in the abbreviated form therein (i.e., $\texttt{P}_1$ is written in place of $\texttt{P}_1(\texttt{i}(g_4), \texttt{i}(p_1))$), and each application of the modus ponens inference rule is indexed differently as it takes different inputs, i.e.:

- $\texttt{I}_{MP,1}(\{\texttt{i}(p_2) \rightarrow \texttt{i}(p_1), \texttt{i}(p_2)\}, \texttt{i}(p_1))$,
- $\texttt{I}_{MP,2}(\{\texttt{i}(p_3) \rightarrow \texttt{i}(p_1), \texttt{i}(p_3)\}, \texttt{i}(p_1))$, and
- $\texttt{I}_{MP,3}(\{\texttt{i}(p_4) \rightarrow \texttt{i}(p_1), \texttt{i}(p_4)\}, \texttt{i}(p_1))$.

There are three constraints (1)–(3) imposed by the meaning of the `To` relationhship. (1) Any two vertices in $G$ can be connected by at most one line. (2) No two information vertices can be connected; any information vertex must be connected to an inference, conflict, or preference vertex, for it is these vertices that establish the use to which the information vertices are put in $G$. (3) Any inference, conflict, or preference vertex must have at least one line that enters it, and another that exits it. There are no restrictions on the label of vertices to which an inference, conflict, or preference vertex can be connected. This makes the language rather versatile, as some forms of meta-reasoning can be captured. A preference may be given between other preferences (e.g., $\texttt{P}_1 \longrightarrow \texttt{P}_3(\texttt{P}_1, \texttt{P}_2) \longrightarrow \texttt{P}_2$) to capture the priority among preferences. Inference rules can be compared in terms of preference (e.g., $\texttt{I}_1 \longrightarrow \texttt{P}(\texttt{I}_1, \texttt{I}_2) \longrightarrow \texttt{I}_2$). Conflicts between preferences can be described, along with conflicts between conflicts, and conflicts between applications of inference rules.

## 4.2 Definition of the Language

We start from a terminology with four concepts. An ACE graphs captures the use of the instances of these concepts.

**Definition 4.1.** ACE ***terminology.*** *The* ACE *terminology is the quadruple:*

$$\langle \textsf{\textit{Proposition}}, \textsf{\textit{Inference rule}}, \textsf{\textit{Conflict rule}}, \textsf{\textit{Preference rule}} \rangle \tag{3}$$

The concepts in the ACE terminology obtain informal meaning as follows.

**Definition 4.2.** ***Proposition [16].*** *Any instance of* Proposition *is a shareable object of attitude (i.e., objects of beliefs, desires, intentions) and is a primary bearer of truth and falsity.*

*Remark* 4.3. Instances of proposition are not restricted to a particular langage or class(es) of language(s). As usual then, a proposition can be a sentence in natural language, or a sentence in a predicate logic.

**Definition 4.4.** ***Inference Rule.*** *Any instance of* Inference rule *is a particular rule of deductive or ampliative inference that is applied to a nonempty set of propositions in order conclude another nonempty set of propositions.*

*Example* 4.5. We applied modus ponens in our earlier example (cf., §4.1) to conclude $\texttt{i}(p_1)$ from $\texttt{i}(p_2)$ and $\texttt{i}(p_2) \rightarrow \texttt{i}(p_1)$. The inference rule in that case is modus ponens. For an example of an ampliative inference rule, suppose that we have the following proposition:

(Ex.13) *Main competing service for on-demand music had less than a hundred thousand users in the first month of operation in a country with the same population.*

This may lead us to conclude the proposition below:

(Ex.14) *The on-demand music service that is being built will not have more than a hundred thousand users in the first month of operation.*

To conclude this proposition, we applied reasoning by analogy, which is an ampliative inference rule.

**Definition 4.6.** ***Conflict Rule.*** *Any instance of* Conflict rule *are criteria indicating that a nonempty set of propositions opposes another nonempty set of propositions.*

*Remark* 4.7. Conflict rules may, but need not be project- and domain-independent. Logical inconsistency is an example of a domain- and project-independent rule of coflict. A project-dependent rule of conflict may indicate that conditions described by two propositions are alternative (hence, conflicting) not because they alone are logically inconsistent, but because together, they violate the conflict rule (e.g., meeting the conditions in both propositions together would overrun a deadline or budget).

**Definition 4.8.** ***Preference Rule.*** *Any instance of* Preference rule *are criteria indicating that the truth of a nonempty set of propositions is stritly preferred to the truth of another nonempty set of propositions.*

*Remark* 4.9. Any instance of Comparison rule equates to a binary relation on the universe of propositions. Let $\succ_i$ symbolize a generic $i$th instance of Comparison rule. The intuitive reading of $p \succ p'$ is "the truth of $p$ is strictly better than the truth of $v'$". A particular instance of Comparison rule, i.e., some $\succ_i$ is not property neutral, whereby its properties are identified when collecting its applications to propositions. E.g., some $\succ_i$ will be transitive, while others will not. What does it mean to say that an instance of Preference rule "are criteria"? Recall the earlier example (cf., §4.1), where we captured by a strict preference the conclusion of a survey, which say that users strictly prefer a free music on-demand service to one based on subscription. This users' preference is a criterion that we use to compare propositions pertaining to the preference. This criterion gives us an instance of a Preference rule. Now, if we subsequently establish that users strictly prefer a subscription-based music service to a per-song payment (i.e., if the user listens ten songs, she pays ten times the unit price of a song), and that the free service is strictly preferred to a per-song payment, then our instance of Preference rule is transitive for the three given propositions.

A discussion of the acceptability of the application of an RE method involves the application of inference rules, conflict rules, and preference rules. Instances of Inference rule are are applied to premises in order to draw conclusions. Instances of Conflict rule are used to highlight opposition between the conditions described by propositions. Instances of Preference rule are employed to indicate relative desirability of what the relevant propositions describe. It is therefore not the instances of inference, conflict, and preference rules that are captured for analysis in ACE, but the *application* of these instances to propositions.

**Definition 4.10.** ACE ***Graph (G).*** *An* ACE *graph $G$ is a directed labeled graph:*

$$G = \langle V(G), L(G), V_\lambda, L_\lambda, \iota, \lambda_V, \lambda_L \rangle \tag{4}$$

*where: $V(G)$ is a finite set of vertices (i.e., nodes, points); $L(G)$ is a finite set of lines (i.e., edges); $V_\lambda$ is the set of labels for vertices; $L_\lambda$ is the set of labels for lines; $\iota$, the incidence function, is a function from $L(G)$ to $(V(G))^2$; $\lambda_V$, the vertice labeling function, is a function from $V(G)$ to $V_\lambda$, associating with each vertex in $V(G)$ a label from $V_\lambda$; $\lambda_L$, the line labeling function, is a function from $L(G)$ to $\lambda_L$ that associates with each line in $L(G)$ a label from $\lambda_L$.*

*Example* 4.11. Figure 1 shows an ACE graph.

**Definition 4.12.** ***Labels for Vertices ($V_\lambda$).*** $V_\lambda = \{\mathtt{i}, \mathtt{I}, \mathtt{C}, \mathtt{P}\}$.

*Remark* 4.13. A vertex labeled `i` is called an *information* vertex, or simply *information*, and `I`, `C`, `P` are called, respectively, *inference*, *conflict*, and *preference*.

**Definition 4.14.** ***Vertex Labeling Function ($\lambda_V$) and the meaning of vertex labels.*** *For some vertex $v \in V(G)$:*

- $\lambda_V(v) = \mathtt{I}$ *iff $v$ represents the application of an instance of* Inference rule *to specific propositions;*
- $\lambda_V(v) = \mathtt{C}$ *iff $v$ represents the application of an instance of* Conflict rule *to specific propositions;*

- $\lambda_V(v) = \texttt{P}$ *iff* $v$ *represents the application of an instance of* Preference rule *to specific propositions; and*
- $\lambda_V(v) = \texttt{i}$ *iff* $v$ *is a proposition that is neither the application of an instance of* Inference rule*, nor* Conflict rule*, nor* Preference rule*.*

*Remark* 4.15. The case $\lambda_V(v) = \texttt{i}$ is defined above in contrast to the other three cases. This is because any application of an instance of Inference rule is evidently a proposition: if we denote the inference rule with a predicate, the application of the inference rule gives us a predicate with no free variables, which thereby carries a truth value and is a proposition. Same applies for instances of Conflict rule and Preference rule: any application of a specific conflict or preference rule is a proposition. Not all instances of propositions are applications of inference, conflict, or preference rules. Consequently, if we have an instance of Proposition that is not itself the application of an inference, conflict, or preference rule to other propositions, then that proposition is called an information. By allowing an inference, conflict, or preference rule to apply to propositions, and as these applications are themselves propositions, we allow forms of meta-reasoning in ACE, which we mentioned earlier (cf., §4.1).

**Definition 4.16.** ***Labels for Lines ($L_\lambda$).*** $L_\lambda = \{\texttt{To}\}$.

**Definition 4.17.** ***Line Labeling Function ($\lambda_L$).*** $\forall l \in L(G),\ \lambda_L(l) = \texttt{To}$.

*Remark* 4.18. The set of line labels is a singleton, so that all lines carry the same label and we chose above to omit this label from the visualization of the graphs (§4.1).

**Definition 4.19.** ***Incidence Function ($\iota$).*** *The incidence function obeys the following two constraints:*

1. *for any* ACE *graph* $G$ *and any two distinct vertices* $\{v, v'\} \subseteq V(G)$*, if there is a line from* $v$ *to* $v'$ *in* $L(G)$*, then there can be no line from* $v'$ *to* $v$ *in* $L(G)$*, that is:*

$$\forall G,\ \forall \{v, v'\} \subseteq V(G),\ \textit{if } \iota(v'v) \neq \emptyset \textit{ then } \iota(vv') = \emptyset \tag{5}$$

2. *no line can connect two information vertices, that is:*

$$\forall G,\ \forall \{v, v'\} \subseteq V(G),\ \nexists l \in L(G) \textit{ s.t. } \iota(l) \in \{v'v, vv'\} \textit{ and } \lambda_V(v) = \lambda_V(v') = \texttt{i} \tag{6}$$

Definitions of the ACE terminology together with the vertex labeling function provide the informal meaning of the vertices in an ACE graph. The meaning of a line in an ACE graph are determined from the vertices that the line connects and the direction of the line.

**Definition 4.20.** ***Informal Meaning of the Lines.*** *The meaning of the lines are given in Table 1 as a function of the labels on vertices that the line connects and the direction of the line.*

# 5 Algorithms

The graph in Figure 1 is a summary of the information offered in favor of and against the application of the AND-refinement method in Ex.1. Given such a graph, two tasks are relevant. The first, *retrieval* task is to search for particular subgraphs $G$ in order to retrieve information that may be of relevance for further discussion among the participants and the evaluation of acceptability. The second, *evaluation* task is to determine if some specific application of an RE method is acceptable. We discuss the retrieval (§5.1) and evaluation (§5.2) tasks in turn below.

## 5.1 Retrieval

An ACE graph is built by incrementally adding information and/or the applications of inference, conflict, and preference rules offered by the participants in the application of the RE methods. The subgraphs an ACE graph are retrieved to inform further debate and to act as the input to the evaluation of acceptability.

Equation 2 indicates that the application of an RE method is acceptable if and only if each proposition describing the inputs (i.e., $In(I_D)$), the application of the method to these inputs (i.e., $In(T(I_D))$), and the

**Table 1:** Informal meaning of the lines in an ACE graph. Columns indicate the label of the vertex, in which the line starts, while the corresponding column indicates the label of the vertex, in which the line ends. The intersection of a row and a column provide the informal meaning of the line. We use the following notational convention: in $\mathtt{I}_e(\mathit{Col},\cdot)$, $\mathtt{C}_e(\mathit{Col},\cdot)$, and $\mathtt{P}_e(\mathit{Col},\cdot)$, *Col* is replaced by the column head; e.g., at the intersection of the column $\mathtt{i}_s$ and $\mathtt{I}_e(\mathit{Col},\cdot)$, $\mathit{Col} = \mathtt{i}_s$, so that $\mathtt{I}_e(\mathit{Col},\cdot) = \mathtt{I}_e(\mathtt{i}_s,\cdot)$.

| | $\mathtt{i}_s$ | $\mathtt{I}_s$ | $\mathtt{C}_s$ | $\mathtt{P}_s$ |
|---|---|---|---|---|
| $\mathtt{i}_e$ | *Not allowed*, because it is unclear why they are linked. | $\mathtt{I}_s(\cdot,\mathtt{i}_e) \longrightarrow \mathtt{i}_e$: The inference rule application $\mathtt{I}_s$ concludes the information $\mathtt{i}_e$. | $\mathtt{C}_s(\cdot,\mathtt{i}_e) \longrightarrow \mathtt{i}_e$: The conflict rule application $\mathtt{C}_s$ makes the information $\mathtt{i}_e$ attacked by some other vertices. | $\mathtt{P}_s(\cdot,\mathtt{i}_e) \longrightarrow \mathtt{i}_e$: The preference rule application $\mathtt{P}_s$ makes the information $\mathtt{i}_e$ strictly less preferred than some other vertices. |
| $\mathtt{I}_e(\cdot,\cdot)$ | *Not allowed*, because an information vertex must be either a premise or a conclusion of an inference rule application, and $\mathtt{i}_s$ is not mentioned in $\mathtt{I}_e(\cdot,\cdot)$. | $\mathtt{I}_s(\cdot,\mathtt{I}_e) \longrightarrow \mathtt{I}_e(\cdot,\cdot)$: The inference rule application $\mathtt{I}_e(\cdot,\cdot)$ is the conclusion of the inference rule application $\mathtt{I}_s(\cdot,\cdot)$. | $\mathtt{C}_s(\cdot,\mathtt{I}_e) \longrightarrow \mathtt{I}_e(\cdot,\cdot)$: The conflict rule application $\mathtt{C}_s$ makes the inference rule application $\mathtt{I}_e(\cdot,\cdot)$ attacked by some other vertices. | $\mathtt{P}_s(\cdot,\mathtt{I}_e) \longrightarrow \mathtt{I}_e(\cdot,\cdot)$: The preference rule application $\mathtt{P}_s$ makes the inference rule application $\mathtt{I}_e(\cdot,\cdot)$ strictly less preferred than some other vertices. |
| $\mathtt{I}_e(\mathit{Col},\cdot)$ | $\mathtt{i}_s \longrightarrow \mathtt{I}_e(\mathtt{i}_s,\cdot)$: Information $\mathtt{i}_s$ is a premise in the inference rule application $I_e$. | $\mathtt{I}_s \longrightarrow \mathtt{I}_e(\mathtt{I}_s,\cdot)$: Inference rule application $\mathtt{I}_s$ is a premise in the inference rule application $I_e$. | $\mathtt{C}_s \longrightarrow \mathtt{I}_e(\mathtt{C}_s,\cdot)$: Conflict rule application $\mathtt{I}_s$ is a premise in the inference rule application $I_e$. | $\mathtt{P}_s \longrightarrow \mathtt{I}_e(\mathtt{P}_s,\cdot)$: Preference rule application $\mathtt{I}_s$ is a premise in the inference rule application $I_e$. |
| $\mathtt{C}_e(\cdot,\cdot)$ | *Not allowed*, because an information vertex linked to a conflict rule application must be subjected to that conflict, and $\mathtt{i}_s$ is not mentioned in $\mathtt{C}_e(\cdot,\cdot)$. | $\mathtt{I}_s(\cdot,\mathtt{C}_e) \longrightarrow \mathtt{C}_e(\cdot,\cdot)$: The conflict rule application $\mathtt{C}_e(\cdot,\cdot)$ is the conclusion of the inference rule application $\mathtt{I}_s(\cdot,\cdot)$. | $\mathtt{C}_s(\cdot,\mathtt{C}_e) \longrightarrow \mathtt{C}_e(\cdot,\cdot)$: The conflict rule application $\mathtt{C}_s$ makes the conflict rule application $\mathtt{C}_e(\cdot,\cdot)$ attacked by some other vertices. | $\mathtt{P}_s(\cdot,\mathtt{C}_e) \longrightarrow \mathtt{C}_e(\cdot,\cdot)$: The preference rule application $\mathtt{P}_s$ makes the conflict rule application $\mathtt{C}_e(\cdot,\cdot)$ strictly less preferred than some other vertices. |
| $\mathtt{C}_e(\mathit{Col},\cdot)$ | $\mathtt{i}_s \longrightarrow \mathtt{C}_e(\mathtt{i}_s,\cdot)$: Information $\mathtt{i}_s$ attacks some other vertices by the conflict rule application $C_e$. | $\mathtt{I}_s \longrightarrow \mathtt{C}_e(\mathtt{I}_s,\cdot)$: Inference rule application $\mathtt{I}_s$ attacks some other vertices by the conflict rule application $C_e$. | $\mathtt{C}_s \longrightarrow \mathtt{C}_e(\mathtt{C}_s,\cdot)$: Conflict rule application $\mathtt{I}_s$ attacks some other vertices by the conflict rule application $C_e$. | $\mathtt{P}_s \longrightarrow \mathtt{C}_e(\mathtt{P}_s,\cdot)$: Preference rule application $\mathtt{I}_s$ attacks some other vertices by the conflict rule application $C_e$. |
| $\mathtt{P}_e(\cdot,\cdot)$ | *Not allowed*, because an information vertex linked to a preference rule application must be subjected to that preference, and $\mathtt{i}_s$ is not mentioned in $\mathtt{P}_e(\cdot,\cdot)$. | $\mathtt{I}_s(\cdot,\mathtt{P}_e) \longrightarrow \mathtt{P}_e(\cdot,\cdot)$: The preference rule application $\mathtt{C}_e(\cdot,\cdot)$ is the conclusion of the inference rule application $\mathtt{I}_s(\cdot,\cdot)$. | $\mathtt{C}_s(\cdot,\mathtt{P}_e) \longrightarrow \mathtt{P}_e(\cdot,\cdot)$: The conflict rule application $\mathtt{C}_s$ makes the preference rule application $\mathtt{P}_e(\cdot,\cdot)$ attacked by some other vertices. | $\mathtt{P}_s(\cdot,\mathtt{P}_e) \longrightarrow \mathtt{P}_e(\cdot,\cdot)$: The preference rule application $\mathtt{P}_s$ makes the preference rule application $\mathtt{P}_e(\cdot,\cdot)$ strictly less preferred than some other vertices. |
| $\mathtt{P}_e(\mathit{Col},\cdot)$ | $\mathtt{i}_s \longrightarrow \mathtt{P}_e(\mathtt{i}_s,\cdot)$: Preference rule application $P_e$ makes the information $\mathtt{i}_s$ strictly preferred to some other vertices. | $\mathtt{I}_s \longrightarrow \mathtt{P}_e(\mathtt{I}_s,\cdot)$: Preference rule application $P_e$ makes the inference rule application $\mathtt{I}_s$ strictly preferred to some other vertices. | $\mathtt{C}_s \longrightarrow \mathtt{P}_e(\mathtt{C}_s,\cdot)$: Preference rule application $P_e$ makes the conflict rule application $\mathtt{C}_s$ strictly preferred to some other vertices. | $\mathtt{P}_s \longrightarrow \mathtt{P}_e(\mathtt{P}_s,\cdot)$: Preference rule application $P_e$ makes the preference rule application $\mathtt{P}_s$ strictly preferred to some other vertices. |

outputs is acceptable (i.e., $In(O_D)$). To inform the participants of some specific proposition, or evaluate the acceptability of that proposition, we retrieve the vertex in the ACE graph that captures this proposition, along with all vertices *relevant to the acceptability of that vertex*.

**Definition 5.1.** ***AC-Relevant vertex.*** *A vertex $v'$ is relevant to the acceptability of another vertex $v$ in the same* ace *graph if and only if $v'$ is directly or indirectly, in favor or against $v$.*

*Example* 5.2. Return to Figure 1 and consider the vertex $\mathtt{i}(g_4)$. The vertex $\mathtt{I}_T$ is *in favor* of $\mathtt{i}(g_4)$, as $\mathtt{i}(g_4)$ is a conclusion of the inference application $\mathtt{I}_T$. Moreover, $\mathtt{I}_T$ is *directly* in favor of $\mathtt{i}(g_4)$ given that there is a line from $\mathtt{I}_T$ to $\mathtt{i}(g_4)$. The vertex $\mathtt{i}(g_1)$ is a premise to the inference application $\mathtt{I}_T$, and is therefore directly in favor of $\mathtt{I}_T$. Since $\mathtt{i}(g_1)$ is directly in favor of $\mathtt{I}_T$, $\mathtt{I}_T$ is directly in favor of $\mathtt{i}(g_4)$, and there is no line from $\mathtt{i}(g_1)$ to $\mathtt{i}(g_4)$, the vertex $\mathtt{i}(g_1)$ is *indirectly* in favor of $\mathtt{i}(g_4)$. It is clear that the conflict application $\mathtt{C}_1$ is *against* $\mathtt{i}(g_4)$, as the conflict application opposes $\mathtt{i}(p_1)$ to $\mathtt{i}(g_4)$. The line $\mathtt{C}_1\mathtt{i}(g_4)$ makes $\mathtt{C}_1$ *directly* against $\mathtt{i}(g_4)$.

There are two ways for a vertex $v'$ to be directly against another vertex $v$: $v'$ may make $v$ attacked, or it may make $v$ dominated.

**Definition 5.3.** ***Attacked Vertex.*** *A vertex $v \in V(G)$ is attacked iff there is a line $l \in L(G)$ from a conflict vertex to $v$, i.e., $\exists l = v'v \in L(G)$ s.t. $\lambda_V(v') = \mathtt{C}$.*

*Remark* 5.4. In a subgraph pattern $v' \longrightarrow \mathtt{C}(v', v) \longrightarrow v$, we say that $v'$ attacks $v$ via the conflict rule application $\mathtt{C}(v', v)$; or equivalently, $v$ is the attacked vertex, $v'$ is the attacker vertex, and $\mathtt{C}(v', v)$ is the attack.

*Example* 5.5. In the attack $\mathtt{C}_2$, $\mathtt{P}_1$ is the attacker vertex, while and $\mathtt{C}_2$ and $\mathtt{i}(p_1)$ the attacked vertices. In the attack $\mathtt{C}_1$, $\mathtt{i}(g_4)$ is attacked by $\mathtt{i}(p_1)$.

**Definition 5.6.** ***Dominated Vertex.*** *A vertex $v \in V(G)$ is dominated iff there is a line from a preference vertex to $v$, i.e., $\exists l = v'v \in L(G)$ s.t. $\lambda_V(v') = \mathtt{P}$.*

*Remark* 5.7. In a subgraph pattern $v' \longrightarrow \mathtt{P}(v', v) \longrightarrow v$, we say that $v'$ dominates $v$ via the preference rule application $\mathtt{P}(v', v)$, or equivalently, that $v'$ is strictly better than $v$.

*Example* 5.8. The preference rule application $\mathtt{P}_1(\mathtt{i}(g_4), \mathtt{i}(p_1))$ in Figure 1 indicates that $\mathtt{i}(g_4)$ is strictly better than (i.e., dominates) $\mathtt{i}(p_1)$.

Both the conflict rule application that is an attack to a vertex, and the preference rule application making a vertex dominated, are directly against a vertex. While the attack on a vertex $v$ is directly against $v$ in $v' \longrightarrow \mathtt{C}(v', v) \longrightarrow v$, the attacker $v'$ is *indirectly* related to $v$. Since $v'$ attacks $v$ via the conflict rule application, $v'$ is indirectly *against* $v$. Similarly, $v'$ in $v' \longrightarrow \mathtt{P}(v', v) \longrightarrow v$ is indirectly against $v$. It is clear that whether $v$ is acceptable in $v' \longrightarrow \mathtt{C}(v', v) \longrightarrow v$ will depend on whether $\mathtt{C}(v', v)$ is acceptable, which in turn depends on the acceptability of $v'$: if the attacker is not acceptable, then the attack is irrelevant, and the attacked vertex is acceptable. Definition 5.1 and the examples above lead us to the following practical result.

**Proposition 5.9.** *If there is a path from a vertex $v'$ to $v$ in an* ACE *graph, then $v'$ is relevant to the acceptability of $v$.*

*Proof.* Cf., Appendix A.1. ☐

**Corollary 5.10.** *In any given* ACE *graph $G$, all vertices on all paths that end in $v$ are relevant to the acceptability of $v$.*

The question at this point is if all vertices on all paths that end in $v$ are *the only* vertices in $G$ that are relevant to the acceptability of $v$.

**Proposition 5.11.** *In any given* ACE *graph, if there is no path from a $v^{prime}$ to another vertex $v$, then $v'$ is not relevant to the acceptability of $v$.*

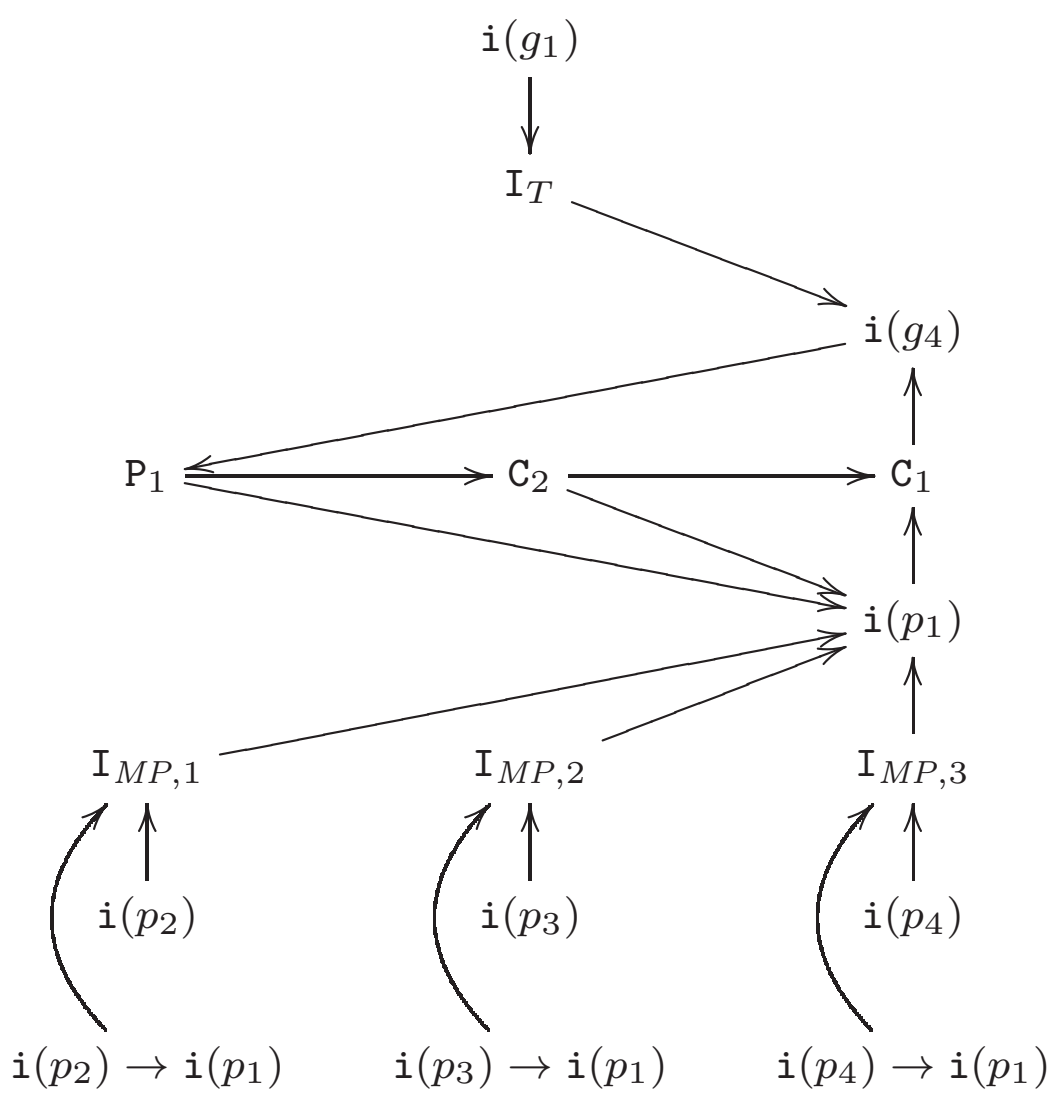

**Figure 2:** The discussion $D[\mathtt{i}(g_4)]$ from the ACE graph in Figure 1.

*Proof.* We prove Proposition 5.11 by contradiction. Suppose that $v'$ is relevant to the acceptability of $v$, but that there is no path from $v'$ to $v$. We know from Definition 5.1 that $v'$ is relevant to the acceptability of $v$ if and only if $v'$ is directly or indirectly, in favor or against $v$. If there is no path from $v'$ to $v$, then there is obviously no path from $v'$ to any vertex that is on any path that ends in $v$. Equivalently, there is no line in $L(G)$ that connects $v'$ to any vertex on any path that ends in $v$. This leads to three observations: (i) $v'$ is not a premise to an inference rule application concluding a vertex on a path that ends in $v$; (ii) $v'$ does not attack a vertex on a path that ends in $v$; and (iii) $v'$ does not dominate a vertex on a path that ends in $v$. This leads to a contradiction, as $v'$ must stand in at least one of the said three relationships to $v$ (and/or any vertex that is relevant to the acceptability of $v$) if it is to be relevant to the acceptability of $v$. □

*Example* 5.12. There are no paths between $\mathtt{i}(g_2)$ and $\mathtt{i}(g_3)$ in Figure 1, so that neither of these vertices is relevant for the acceptability of the other.

*Remark* 5.13. Observe that if $v'$ is not relevant to the acceptability of $v$, then $v$ may be relevant to the acceptability of $v'$. This occurs when there is a path from $v$ to $v'$, but no path from $v'$ to $v$.

We are interested in retrieving all parts of a given ACE graph that are sufficient to evaluate the acceptability of some specific vertex. Corollary 5.10 and Proposition 5.11 indicate that we must retrieve all distinct paths in $G$ that end in $v$, in order to evaluate the acceptability of $v$. This leads us to introduce the notion of *discussion*, which, for a given vertex $v \in V(G)$ carries all parts of the graph $G$ that are necessary to determine whether $\mathbf{AC}(v)$.

**Definition 5.14.** ***Discussion of*** $v$ **(**$D[v]$**).** *The discussion of a vertex $v$ in an* ACE *graph $G$ is the subgraph of $G$, which contains exactly all distinct paths in $G$ that end in $v$.*

*Example* 5.15. Figure 2 shows the discussion of $\mathtt{i}(g_4)$, i.e., $D[\mathtt{i}(g_4)]$. This discussion contains all distinct paths from the ACE graph in Figure 1 that end in $\mathtt{i}(g_4)$.

The acceptability of a vertex $v$ in an ACE graph is evaluated by the analysis of the discussion of $v$, i.e., $D[v]$. To perform the evaluation of acceptability of a vertex $v \in V(G)$ in an ACE graph $G$, we must first retrieve $D[v]$ from a given $G$. Algorithm 1 retrieves retrieves a *discussion* of the given vertex.

**Proposition 5.16.** *Algorithm 1 applied to a vertex $v$ in an* ACE *graph (i) does not loop indefinetly, (ii) returns all direct and indirect vertices in favor of or against the starting vertex $v$ (i.e., returns the discussion of $v$, $D[v]$), and (iii) has the running time in $O(|V(D[v])| + |L(D[v])|)$.*

**Algorithm 1** Find Discussion

**Require:** A nonempty ACE graph $G$, a starting vertex $First \in V(G)$;
**Ensure:** A nonempty graph $D[First]$, which is a subgraph of $G$;

```
procedure FindDiscussion(G, First)
    Empty the queue Q; V(D[First]) ← ∅; L(D[First]) ← ∅
    Add First to Q
    while Q is not empty do
        for each vertex v in Q do
            Add v to V(D[First])
            for each v′ ∈ V(G) s.t. ∃v′v ∈ L(G) do
                if v′v ∉ L(D[First]) then
                    Add v′v to L(D[First])
                end if
                if v′ ∉ V(D[First]) then
                    Add v′ to V(D[First])
                    Add v′ to Q
                end if
            end for
            Delete v from Q
        end for
    end while
end procedure
```

*Proof.* Cf., Appendix A.2. □

Algorithm 1 is an adaptation of the usual breadth first search algorithm (e.g., [13]). When applied to an ACE graph $G$, Algorithm 1 enqueues the starting vertex *First* and visits it. It then enqueues all vertices on lines incoming to the visited node, and dequeues the visited starting node. All vertices in the queue are visited in the same manner and are added to the discussion $D[First]$. The discussion also receives all lines from $G$ connecting the vertices in that discussion.

**Definition 5.17.** ***Subdiscussion.*** *$D[v']$ is a subdiscussion of $D[v]$ if and only if $V(D[v']) \subseteq V(D[v])$ and $L(D[v']) \subseteq L(D[v])$.*

**Proposition 5.18.** *If $v' \in D[v]$, then $D[v']$ is a subdiscussion of $D[v]$.*

*Proof.* We give a trivial proof by contradiction. A discussion $D[v \in V(G)]$ contains, by Definition 5.14, all distinct paths in $G$ that end in $v$. Suppose that there is a vertex $v' \in V(D[v])$ such that $V(D[v']) \cap V(D[v]) \neq V(D[v'])$. There is thus a vertex $v''$ that is on a path that ends in $v'$, but there is no path from $v''$ to $v$. This is a contradiction: if there is a path from $v''$ to $v'$, and a path from $v'$ to $v$, there there must be a path from $v''$ to $v'$. □

Proposition 5.18 is a useful result, since retrieving $D[v]$ also retrieves the discussions of all vertices in $V(D[v])$, that is, all subdiscussions of $D[v]$.

## 5.2 Evaluation

The acceptability of the vertex $v$ is evaluated by traversing and computing the labels on vertices in the discussion of $v$, $D[v]$. The *computed* label of a vertex is a secondary label, and is different from that assigned by $\lambda_V$. The computed label indicates whether a vertex is acceptable. We first give an informal overview below, of how a discussion is traversed and labels computed (§5.2.1), then provide the details of the acceptability evaluation algorithm (§5.2.2).

### 5.2.1 Overview of Evaluation

The computed label of any vertex in $D[v]$ is either **A** for *accepted*, **AD** for *accepted and dominated*, or **R** for *rejected*. We illustrate the computation of labels by simple examples first, then go on to label the discussion $D[\mathtt{i}(g_4)]$ obtained from the graph in Figure 1, and finally give an informal outline of the algorithm that computes the labels of any discussion.

Consider an ACE graph $G'$ with only a single vertex $V(G') = \{v\}$ and $L(G') = \emptyset$, so that $D[v] = G'$. Being alone in $D[v]$, $v$ is neither attacked nor dominated; we therefore say that $v$ is acceptable, and label it **A**, denoted ${}_{\mathbf{A}}v$. Consider now a graph $G''$ with three vertices and two lines between them. To compute the labels in $G''$, we need to know the direction of the lines and the primary labels on the vertices:

- if $G''$ is $\mathtt{i}_1 \longrightarrow \mathtt{I}(\mathtt{i}_1, \mathtt{i}_2) \longrightarrow \mathtt{i}_2$, then all vertices are neither attacked nor dominated, and they all take the label **A**, i.e., ${}_{\mathbf{A}}\mathtt{i}_1 \longrightarrow {}_{\mathbf{A}}\mathtt{I}(\mathtt{i}_1, \mathtt{i}_2) \longrightarrow {}_{\mathbf{A}}\mathtt{i}_2$;
- if $G''$ is $\mathtt{i}_1 \longrightarrow \mathtt{C}(\mathtt{i}_1, \mathtt{i}_2) \longrightarrow \mathtt{i}_2$, then $\mathtt{i}_2$ is attacked; we see that $\mathtt{i}_1$ and $\mathtt{C}$ are not attacked, and conclude that $\mathtt{i}_2$ is rejected: ${}_{\mathbf{A}}\mathtt{i}_1 \longrightarrow {}_{\mathbf{A}}\mathtt{C}(\mathtt{i}_1, \mathtt{i}_2) \longrightarrow {}_{\mathbf{R}}\mathtt{i}_2$;
- if $G''$ is $\mathtt{i}_1 \longrightarrow \mathtt{P}(\mathtt{i}_1, \mathtt{i}_2) \longrightarrow \mathtt{i}_2$, then $\mathtt{i}_2$ is dominated. To be dominated alone is not enough for rejection, so that $\mathtt{i}_2$ is accepted and dominated, that is ${}_{\mathbf{A}}\mathtt{i}_1 \longrightarrow {}_{\mathbf{A}}\mathtt{P}(\mathtt{i}_1, \mathtt{i}_2) \longrightarrow {}_{\mathbf{AD}}\mathtt{i}_2$.

These three cases illustrate the first important principle used in computing the labels of a discussion: the label on a vertex $v$ depends on the labels of all vertices $v_1, \ldots, v_n$ adjacent to $v$ by lines $v_1v, v_2v, \ldots, v_nv$. This alone is not enough, as we must know how the labels interact – consider the hypothetical discussion $D[\mathtt{i}_1]$ in Ex.15.

(Ex.15)

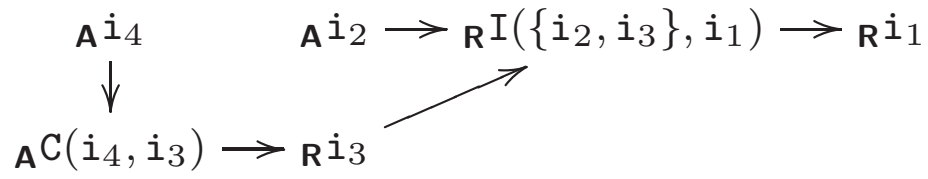

The inference $\mathtt{I}(\{\mathtt{i}_2, \mathtt{i}_3\}, \mathtt{i}_1)$ uses two inputs, one accepted $\mathtt{i}_2$ and another $\mathtt{i}_3$, which is attacked by the acccepted $\mathtt{i}_4$ (hence the rejection of $\mathtt{i}_3$). The inference itself cannot be accepted, since one of its inputs is rejected. Given that the application of the inference rule is rejected, the conclusion of the inference, $\mathtt{i}_1$, must be rejected as well. Ex.15 illustrates the choice that the label **R** has priority over **A**. We choose to be cautious in computing the labels, meaning that **R** has priority over **AD** and **A**, and **AD** has priority over **A**. In addition to this second principle employed in the computation of the labels, we have a third and final one, illustrated via Ex.16.

(Ex.16)

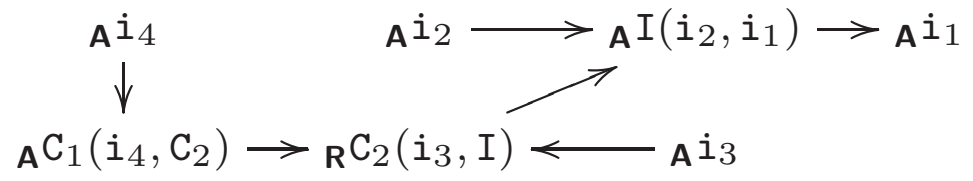

Suppose that no computed labels are given in Ex.16. We see immediately that $\mathtt{i}_4$, $\mathtt{i}_3$ and $\mathtt{i}_2$ should be accepted as they are not attacked, along with $\mathtt{C}_1(\mathtt{i}_4, \mathtt{C}_2)$. $\mathtt{C}_2(\mathtt{i}_3, \mathtt{I})$ must then be rejected, as it is attacked via $\mathtt{C}_1(\mathtt{i}_4, \mathtt{C}_2)$. Observe then that $\mathtt{I}(\mathtt{i}_2, \mathtt{i}_1)$ has two incoming lines, one from the accepted $\mathtt{i}_2$ and another from the rejected conflict $\mathtt{C}_2(\mathtt{i}_3, \mathtt{I})$. While it is true that **R** has priority over **A**, we conclude that $\mathtt{I}(\mathtt{i}_2, \mathtt{i}_1)$ is accepted, because the rejected conflict is not an input to the inference $\mathtt{I}(\mathtt{i}_2, \mathtt{i}_1)$. We have noted earlier that the meaning of a line in an ACE graph is determined from the labels that $\lambda_V$ assigns to the vertices connected by the line. The conclusion that $\mathtt{I}(\mathtt{i}_2, \mathtt{i}_1)$ is accepted cannot be reached without determining the meaning of each line ending in $\mathtt{I}(\mathtt{i}_2, \mathtt{i}_1)$. By reading these lines, we see that $\mathtt{I}(\mathtt{i}_2, \mathtt{i}_1)$ does not take the conflict $\mathtt{C}_2(\mathtt{i}_3, \mathtt{I})$ as its input, but that this conflict attacks $\mathtt{I}(\mathtt{i}_2, \mathtt{i}_1)$. If the conflict is accepted, $\mathtt{I}(\mathtt{i}_2, \mathtt{i}_1)$ should be rejected; however, the conflict is rejected, so that $\mathtt{I}(\mathtt{i}_2, \mathtt{i}_1)$ is accepted. More generally, the third principle we use in computing the label on a vertex $v$ is that the meaning of each line that ends in $v$ must be determined. This leads us to define a number of deduction rules for labels, which account for the meaning of the relevant line.

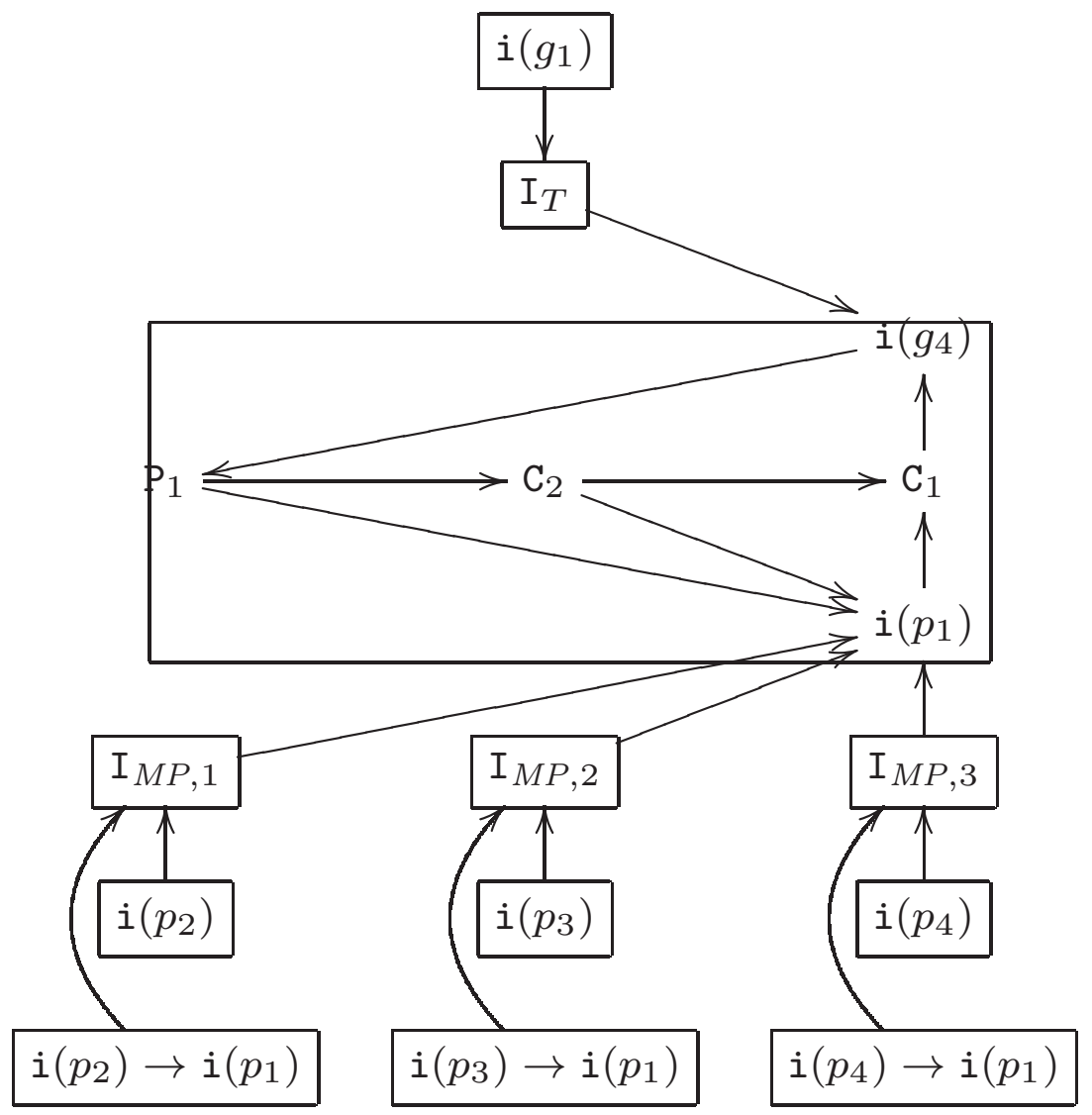

**Figure 3:** Stongly connected components in the discussion $D[\mathtt{i}(g_4)]$ from Figure 1

To see how these rules are used, consider again the vertex $\mathtt{I}(\mathtt{i}_2, \mathtt{i}_1)$. Since it has lines incoming from two different vertices, we use the following two *label deduction rules*:[1]

- from $_{\mathbf{A}}\mathtt{i} \longrightarrow \mathtt{I}(\mathtt{i}, \cdot)$, conclude that the inference $\mathtt{I}(\mathtt{i}, \cdot)$ should be acccepted (where $_{\mathbf{A}}\mathtt{i} \longrightarrow \mathtt{I}(\mathtt{i}, \cdot)$ means that the inference $\mathtt{I}(\mathtt{i}, \cdot)$ uses the accepted $\mathtt{i}$ as its input and concludes something else, i.e., "$\cdot$"); and
- from $_{\mathbf{R}}\mathtt{C}(\cdot, \mathtt{I}) \longrightarrow \mathtt{I}(\cdot, \cdot)$, conclude that the inference $\mathtt{I}(\cdot, \cdot)$ should be accepted (where $_{\mathbf{R}}\mathtt{C}(\cdot, \mathtt{I}) \longrightarrow \mathtt{I}(\cdot, \cdot)$ means that the inference $\mathtt{I}(\cdot, \cdot)$ is attacked by the rejected conflict $\mathtt{C}(\cdot, \mathtt{I})$).

The application of two rules, as above, gives us two labels, both **A**; it is thus clear that $\mathtt{I}(\mathtt{i}_2, \mathtt{i}_1)$ will bear the label **A**. More generally, if $v$ has $n$ incoming lines $v_1v, v_2v, \ldots, v_nv$, then we will apply $n$ rules, selected depending on the label of each $v_i \in \{v_1, v_2, \ldots, v_n\}$. This will result in $n$ labels. The one label that we will assign to $v$ will be that of the $n$ labels, which has the priority over others, according to the principle of how the labels interact, and given above. E.g., if we have the set of $n$ labels, in which there is at least one label **R**, we will conclude $_{\mathbf{R}}v$; if each label is **A**, then $_{\mathbf{A}}v$; if there are no **R** labels, but only **A** and **AD** labels, then $_{\mathbf{AD}}v$. Seventy two label deduction rules cover all cases allowed by the meaning of the $\mathtt{To}$ line in any ACE graph. We provide their definitions later on (cf., §5.2.2).

We now ask if $\mathtt{i}(g_4)$ is acceptable. The answer can be given once we compute the labels on the discussion $D[\mathtt{i}(g_4)]$. We find the discussion $D[\mathtt{i}(g_4)]$ via Algorithm 1 and then proceed as follows:

1. If there are preferences in the discussion that are transitive, the steps below are performed on the transitive closure of these transitive preferences on the discussion of choice. Regardless of whether $\mathtt{P}_1$ is transitive, the transitive closure of $\mathtt{P}_1$ on $D[\mathtt{i}(g_4)]$ is the same as $D[\mathtt{i}(g_4)]$.
2. We now need to find a topological sort of the strongly connected components of the discussion $D[\mathtt{i}(g_4)]$. A discussion can contain cycles, which is why we must first identify the strongly connected components.[2] Figure 3 shows the discussion $D[\mathtt{i}(g_4)]$, where each strongly connected component is delimited by a rectangle. The largest strongly connected component contains cycles, while the others contain no cycles. Once we have a topological components, we need their topological sort. It is well known that contracting each strongly connected component in a directed graph gives a directed acyclic graph, where each vertex is a contracted strongly connected component. A topological sort of that directed acyclic graph is a linear ordering of its vertices, in which a vertex comes before all vertices, to which it has outcoming lines. To see why we need a topological sort of the strongly connected components, consider the problem of labeling $\mathtt{I}_{MP,1}$ (same applies to the problem of labeling $\mathtt{I}_{MP,2}$ and $\mathtt{I}_{MP,3}$): the label of $\mathtt{I}_{MP,1}$ depends on the labels of $\mathtt{i}(p_2)$ and $\mathtt{i}(p_2) \rightarrow \mathtt{i}(p_1)$. Consequently, we must label $\mathtt{i}(p_2)$ and $\mathtt{i}(p_2) \rightarrow \mathtt{i}(p_1)$ before we label $\mathtt{I}_{MP,1}$. In a topological sort, $\mathtt{I}_{MP,1}$ comes after both $\mathtt{i}(p_2)$ and $\mathtt{i}(p_2) \rightarrow \mathtt{i}(p_1)$. A topological sort therefore gives the order, in which the strongly connected components should be labeled.

[1] As a notational convention, we write "$\cdot$" for the parameter that is not important for the application of the given rule.

[2] As usual, a strongly connected component is a graph, in which there is a path from any vertex to any other vertex.

3. Given a topological sort of the strongly connected components of the discussion $D[\mathtt{i}(g_4)]$, we label all elements in the sort that have no incoming lines. We consequently label as accepted the following vertices: $_{\mathsf{A}}\mathtt{i}(g_1)$, $_{\mathsf{A}}\mathtt{i}(p_2)$, $_{\mathsf{A}}\mathtt{i}(p_3)$, $_{\mathsf{A}}\mathtt{i}(p_4)$, $_{\mathsf{A}}(\mathtt{i}(p_2) \rightarrow \mathtt{i}(p_1))$, $_{\mathsf{A}}(\mathtt{i}(p_3) \rightarrow \mathtt{i}(p_1))$, and $_{\mathsf{A}}(\mathtt{i}(p_4) \rightarrow \mathtt{i}(p_1))$. Once these are labeled, the next elements in the sort are the three applications of modus ponens and $\mathtt{I}_T$. They are not attacked and their inputs are accepted, so that $_{\mathsf{A}}\mathtt{I}_{MP,1}$, $_{\mathsf{A}}\mathtt{I}_{MP,2}$, $_{\mathsf{A}}\mathtt{I}_{MP,3}$, and $_{\mathsf{A}}\mathtt{I}_T$. Labeling $\mathtt{i}(p_1)$ is more difficult and requires a different strategy. This is because $\mathtt{i}(p_1)$ lies on at least one simple cycle.[3] To label a strongly connected component with cycles, we proceed as follows:

   (a) The count, say $C$ of simple cycles in the strongly connected component is computed. $C = 3$ in the strongly connected component containing $\mathtt{i}(p_1)$.

   (b) We add an empty sequence of labels on each vertex in the strongly connected component, and add the label **A** to the sequence of each vertex.

   (c) We choose a vertex according to a specific heuristic (namely, we take the last added vertex in the strongly conncted component, as discussed in detail later on – cf., §5.2.2) and call it the *First* vertex. In $D[\mathtt{i}(g_4)]$, $\mathtt{P}_1$ is that vertex. In doing so, we in fact merely hypothesize that $\mathtt{P}_1$ is accepted. To understand intuitively what happens next, suppose that there are $C$ walkers stationed at $\mathtt{P}_1$. Each walker obeys the following: (i) it takes equal time to traverse a vertex; (ii) it can only go forward (i.e., over lines that start in a vertex); and (iii) no two walkers will start from *First* and return to *First* along the exact same path. In the strongly connected component with $\mathtt{i}(p_1)$, we place three walkers at the vertex $\mathtt{P}_1$ and send them along the lines outgoing from $\mathtt{P}_1$. After the first step, two walkers will reach $\mathtt{C}_2$ and one will reach $\mathtt{i}(p_1)$. Once they reach a vertex, they compute the label on that vertex by using the label deduction rules we explained earlier. The computed label is appended to the sequence of labels on the vertex. For $_{\langle \mathsf{A} \rangle}\mathtt{C}_2$, we append the sequence of labels with **A**, and obtain $_{\langle \mathsf{A},\mathsf{A} \rangle}\mathtt{C}_2$. Since all three applications of modus ponens are accepted and $_{\langle \mathsf{A} \rangle}\mathtt{P}_1$, we get $_{\langle \mathsf{A},\mathsf{AD} \rangle}\mathtt{i}(p_1)$. After the second step, two walkers are at $\mathtt{C}_1$, and the third is at $\mathtt{i}(p_1)$, so that $_{\langle \mathsf{A},\mathsf{A},\mathsf{R},\mathsf{R} \rangle}\mathtt{C}_1$ and $_{\langle \mathsf{A},\mathsf{AD},\mathsf{R} \rangle}\mathtt{i}(p_1)$. The fourth step results in $_{\langle \mathsf{A},\mathsf{A} \rangle}\mathtt{i}(g_4)$. After the fourth step, two walkers arrive simultaneously at the first vertex $\mathtt{P}_1$. However, the first vertex obtains its second label (i.e., $_{\langle \mathsf{A} \rangle}\mathtt{P}_1$ becomes $_{\langle \mathsf{A},\mathsf{A} \rangle}\mathtt{P}_1$) only after the slowest walker, i.e., the one traversing the longest path back to the first vertex, arrives at that vertex.[4]

   The stopping criterion for the walkers is as follows. If the last two labels in the sequence of labels on the first vertex are identical, the walkers are not sent to traverse the strongly connected component any further. This is the case in the example, where $_{\langle \mathsf{A},\mathsf{A} \rangle}\mathtt{P}_1$. The last label in the sequence of labels on any vertex is the label that indicates the acceptability of that vertex: if **A**, the vertex is acceptable, if **AD**, the vertex is acceptable and dominated, if **R**, the vertex is not acceptable. In case the first two labels in the sequence of the first vertex are not identical, the walkers will be sent out in the same way as described above, until the first vertex has four labels. When the first vertex has four labels and its last two labels are not identical, then the given discussion is inconclusive with regards to acceptability: more vertices need to be added (i.e., the discussion should continue) before the discussion is evaluated again.

The sequence of steps exemplified above is the informal outline of the algorithm that labels any discussion. The algorithm, called EVALUATEDISCUSSION is formally introduced and discussed below.

[3]As usual, a simple cycle is a cycle that passes once through all vertices except its starting vertex, which the cycle passes twice (as the cycle starts and ends in that vertex).

[4]Observe that the number of labels in the sequence of labels of a vertex, in a strongly connected component with cycles, equals 1 plus the number of times a walker traversed that vertex. This is valid for all vertices other than the *First* vertex.

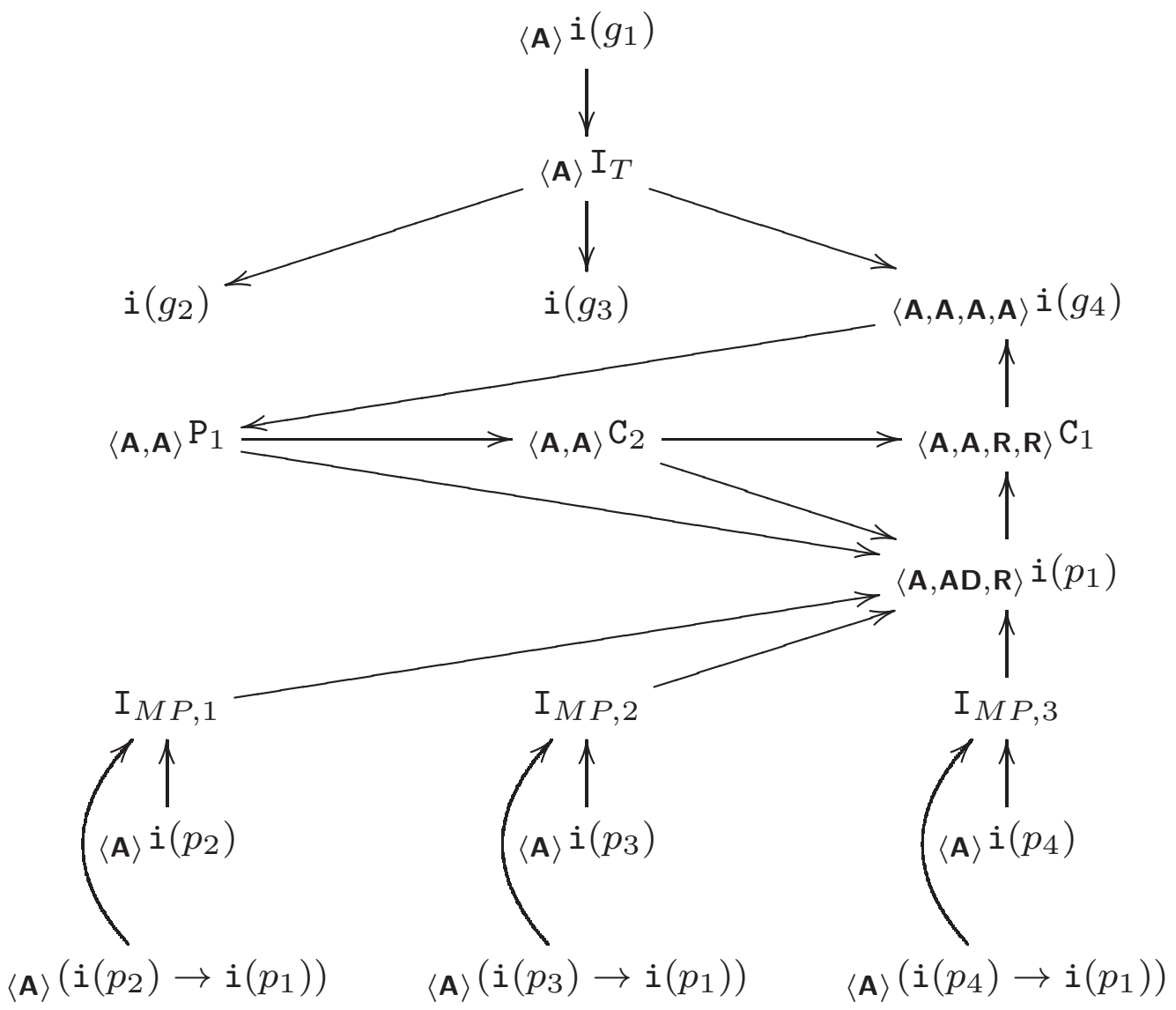

**Figure 4:** Result of evaluating the discussion $D[\mathtt{i}(g_4)]$ for acceptability.

### 5.2.2 Definition of Evaluation

We now present the algorithm that was informally outlined above. We start by defining the three computed labels and the rules governing their interaction.

**Definition 5.19.** ***C-label.*** *A computed label, or c-label, on any vertex $v$ in an* ACE *graph $G$ is either* **A** *for accepted (equivalently, acceptable),* **AD** *for accepted and dominated (i.e., acceptable and dominated), or* **R** *for rejected (i.e., not acceptable).*

*Remark* 5.20. An **A** c-label on $v$ indicates that $v$ is acceptable, that is, $\mathbf{AC}(v)$ holds. If $v$ carries the c-label **R**, then $\mathbf{AC}(v)$ does not hold. The c-label **AD** gives two indications: one is that $\mathbf{AC}(v)$ holds, while the **D** in **AD** says that there are acceptable vertices in $G$ that are strictly more preferred to $v$.

The c-label of a vertex $v \in V(G)$ is computed by taking into account the c-labels on vertices $v_1, v_2, \ldots, v_n$ such that $\exists\{v_1v, v_2v, \ldots, v_nv\} \subseteq L(G)$. We call each of $v_1, v_2, \ldots, v_n$ an *in-adjacent* vertex. The c-labels on the various in-adjacent vertices can be different, so that it becomes necessary to know how to combine them when computing the c-label on $v$. Consider the graph pattern below.

(Ex.17)

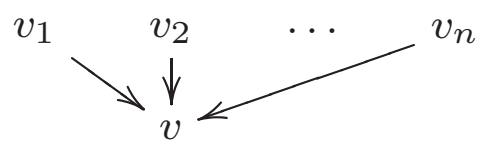

Suppose that vertices $v_1$ to $v_{n-2}$ are all accepted, $v_{n-1}$ is accepted and dominated, and $v_n$ is rejected. To compute the c-label on $v$, we obviously need to know the labels assigned by $\lambda_V$ to each vertex $v, v_1, \ldots, v_n$. E.g., if $\lambda_V(v_1) = \mathtt{C}$, then the c-label **AD** will be propagated from $v_1$ to $v$. We therefore require the rules for how c-labels are sent between vertices in an ACE graphs, depending on the labels of vertices (i.e., `i`, `I`, `C`, and `P`) and the direction of the `To` line between a given pair of vertices.

*Remark* 5.21. The rules for the propagation of c-labels across lines are used by the evaluation algorithm when it traverses a vertex. Since the application of such rules will propagate as many c-labels on $v$ as there are in-adjacent vertices to $v$, we associate a set, called *t-set* to a vertex. Each member of a t-set if a c-label. If $v$ has $n$ in-adjacent vertices, the t-set of $v$ will receive $n$ members each time the evaluation algorithm traverses $v$. Because the evaluation algorithm mey traverse a vertex more than once, we also add a sequence of c-labels, called *c-sequence* on each vertex in an ACE graph. The evaluation algorithm appends a c-label to the c-sequence of a vertex, by inferring that c-label from the c-labels in the t-set of that vertex. Each member of a c-sequence is a c-label. The use of both the t-set and the c-sequence of a vertex will be clarified below. When it is relevant to make explicit the c-sequence of $v$, we write $_{\langle\ldots\rangle}v$. When it is relevant to make the t-set of $v$ explicit, we write $^{\{\cdots\}}v$.

**Definition 5.22.** ***propagateLabel Function.*** *propagateLabel* : $V(G) \times V(G) \longrightarrow \{\mathsf{A}, \mathsf{AD}, \mathsf{R}\}$. *For a line $v'v \in L(G)$ in an* ACE *graph, such that $v'$ has a c-label, the propagateLabel function returns a c-label for the vertex $v$, based on $\lambda_V(v')$ and the direction of the line from $v'$ to $v$. Table 2 completely defines the function propagateLabels.*

*Remark* 5.23. Table 2 has a similar structure as Table 1. For each case in Table 1, Table 2 studies three cases, one per allowed label. E.g., for the case $\mathtt{I}_s(\cdot, \mathtt{i}_e) \longrightarrow \mathtt{i}_e$ (second column, first row in Table 1), Table 2 gives three rules: (i) if $\mathtt{I}_s(\cdot, \mathtt{i}_e)$ has the c-label $\mathsf{A}$, then $\mathtt{i}_e$ obtains the c-label $\mathsf{A}$ (rule 17 in Table 2); (ii) if $\mathtt{I}_s(\cdot, \mathtt{i}_e)$ has the c-label $\mathsf{AD}$, then $\mathtt{i}_e$ obtains $\mathsf{A}$ (rule 18 in Table 2); and (iii) if $\mathtt{I}_s(\cdot, \mathtt{i}_e)$ carries $\mathsf{R}$, then $\mathtt{i}_e$ obtains $\mathsf{R}$ (rule 19 in Table 2).

*Example* 5.24. In Figure 2:

- *propagateLabel*($_{\mathsf{A}}\mathtt{i}(g_1), \mathtt{I}_T$) = $\mathsf{A}$;
- *propagateLabel*($_{\mathsf{A}}\mathtt{I}_T, \mathtt{i}(g_4)$) = $\mathsf{A}$;
- *propagateLabel*($_{\mathsf{A}}\mathtt{C}_1, \mathtt{i}(g_4)$) = $\mathsf{R}$; and so on.

To compute the c-label on a vertex $v$, which has in-adjacent vertices $v_1, \ldots, v_n$, we propagate a c-label from each of $v_1, \ldots, v_n$. We consequently end up with $n$ c-labels in the t-set of $v$, so that it becomes necessary to determine which of the $n$ c-labels to infer from the t-set of $v$. We do so by applying the inference rules for multiple labels.

**Definition 5.25.** ***Inference Rules for Multiple C-Labels.*** *When there are $n$ alternative c-labels for a vertex $v$, then the c-vertex on $v$ is determined by the application of the following rules:*

1. *Rejection* $\mathsf{R}$ *overrules acceptance* $\mathsf{A}$*: from* $\mathsf{R}$ *and* $\mathsf{A}$*, conclude* $\mathsf{R}$.
2. *Rejection* $\mathsf{R}$ *overrules dominated acceptance* $\mathsf{AD}$*: from* $\mathsf{R}$ *and* $\mathsf{AD}$*, conclude* $\mathsf{R}$.
3. *Dominated acceptance* $\mathsf{AD}$ *overrules acceptance* $\mathsf{A}$*: from* $\mathsf{AD}$ *and* $\mathsf{A}$*, conclude* $\mathsf{AD}$.

*Example* 5.26. In Figure 2, suppose that both $\mathtt{I}_T$ and $\mathtt{C}_1$ carry the c-label $\mathsf{A}$. To determine the c-label on $\mathtt{i}(g_4)$, we call the function *propagateLabel* on both vertices that are in-adjacent to $\mathtt{i}(g_4)$. We thus have *propagateLabel*($_{\mathsf{A}}\mathtt{I}_T, \mathtt{i}(g_4)$) = $\mathsf{A}$ and *propagateLabel*($_{\mathsf{A}}\mathtt{C}_1, \mathtt{i}(g_4)$) = $\mathsf{R}$. We thus have $^{\{\mathsf{A},\mathsf{R}\}}\mathtt{i}(g_4)$. We follow the first rule in Definition 5.25 and conclude that the c-label on $\mathtt{i}(g_4)$ is $\mathsf{R}$. We append the c-sequence of $v$ with the c-label inferred from the t-set (and empty the t-set), so that $_{\langle\ldots,\mathsf{R}\rangle}\mathtt{i}(g_4)$.

We combine the *propagateLabel* function (cf., Definition 5.22) and the inference rules for multiple labels in Algorithm 2. The procedure ComputeLabel is called by the evaluation algorithm whenever that algorithm visits a vertex in the discussion that is being labeled.

Algorithm 2 is given a vertex $v_i$ from a discussion $D[v]$. The algorithm first empties the t-set of $v_i$ (cf., Line 2 in Algorithm 2). The **for each** loop then considers all in-adjacent vertices of $v_i$ in the dicussion $D[v]$ (cf., Lines 3–9). For each considered vertex $v'$, the *propagateLabel* function is called: if its result is an error, then the line $v'v_i$ is not one allowed by the meaning of the To link – the algorithm stops and reports the error. If *propagateLabels*($v', v_i$) returns a c-label instead of an error, that c-label is added to the t-set of $v_i$. The **for each** loop finishes after it has considered all vertices in $D[v]$ that are in-adjacent to $v_i$. Let *inDegree*($v_i, G$) = $|\{v' \mid \exists v'v_i \in L(G)\}|$, so that *inDegree*($v_i, D[v]$) is the number of vertices in-adjacent to $v_i$ in $D[v]$. Observe that the cardinality of the t-set of $v_i$ equals *inDegree*($v_i, D[v]$). Once the t-set has *inDegree*($v_i, D[v]$) elements, Definition 5.25 is applied via the last **if** block (cf., Lines 10–16) to determine the overruling c-label in the t-set.

**Proposition 5.27.** *Algorithm 2 applied to a vertex $v_i$ in a discussion $D[v]$ (i) does not loop indefinetly, (ii) returns the overruling c-label among the c-labels propagated from all in-adjacent vertices to $v_i$ in $D[v]$, and (iii) has the running time in $O(inDegree(v_i, D[v]))$.*

**Table 2:** Complete definition of the *propagateLabel* function. To simplify the presentation below, the notation $[_{\langle\ldots,\mathbf{A}\rangle}\mathtt{i},\mathtt{I}(\mathtt{i},\cdot)] \Rightarrow [\mathbf{A}]$ is an abbreviation of *propagateLabel*$(_{\langle\ldots,\mathbf{A}\rangle}\mathtt{i},\mathtt{I}(\mathtt{i},\cdot)) = \mathbf{A}$, where $\mathbf{A}$ is the c-label added to the t-set of $\mathtt{I}(\mathtt{i},\cdot)$. As in Table 1, we write $\mathtt{I}(a,b)$ to indicate that $a$ is used to infer $b$ by the application of $\mathtt{I}$; $\mathtt{P}(a,b)$ indicates that $a$ attacks $b$ by the conflict rule application $\mathtt{P}$; $\mathtt{P}(a,b)$ indicates that $a$ is strictly better than $b$ according to the preference rule application $P$. An additional convention is that if we write *propagateLabel*$(_{\langle\ldots,\mathbf{A}\rangle}\mathtt{i},\mathtt{I}(\mathtt{i},\cdot))$, then the labeled $\mathtt{i}$ is the parameter $\mathtt{i}$ in $\mathtt{I}(\mathtt{i},\cdot)$; if we write *propagateLabel*$(_{\langle\ldots,\mathbf{A}\rangle}\mathtt{i},\mathtt{I})$ then the labeled $\mathtt{i}$ is not a parameter in $\mathtt{I}$ (i.e., the parameters of $\mathtt{I}$ are some vertices other than the given information vertex $\mathtt{i}$).

| | | | |
|---|---|---|---|
| In any case other than (8)–(79), *getLabel* returns error. | $[_{\langle\ldots,\mathbf{A}\rangle}\mathtt{i},\mathtt{I}(\mathtt{i},\cdot)] \Rightarrow [\mathbf{A}]$ (7) | $[_{\langle\ldots,\mathbf{A}\rangle}\mathtt{i},\mathtt{C}(\mathtt{i},\cdot)] \Rightarrow [\mathbf{A}]$ (10) | $[_{\langle\ldots,\mathbf{A}\rangle}\mathtt{i},\mathtt{P}(\mathtt{i},\cdot)] \Rightarrow [\mathbf{A}]$ (13) |
| | $[_{\langle\ldots,\mathbf{AD}\rangle}\mathtt{i},\mathtt{I}(\mathtt{i},\cdot)] \Rightarrow [\mathbf{A}]$ (8) | $[_{\langle\ldots,\mathbf{AD}\rangle}\mathtt{i},\mathtt{C}(\mathtt{i},\cdot)] \Rightarrow [\mathbf{A}]$ (11) | $[_{\langle\ldots,\mathbf{AD}\rangle}\mathtt{i},\mathtt{P}(\mathtt{i},\cdot)] \Rightarrow [\mathbf{A}]$ (14) |
| | $[_{\langle\ldots,\mathbf{R}\rangle}\mathtt{i},\mathtt{I}(\mathtt{i},\cdot)] \Rightarrow [\mathbf{R}]$ (9) | $[_{\langle\ldots,\mathbf{R}\rangle}\mathtt{i},\mathtt{C}(\mathtt{i},\cdot)] \Rightarrow [\mathbf{R}]$ (12) | $[_{\langle\ldots,\mathbf{R}\rangle}\mathtt{i},\mathtt{P}(\mathtt{i},\cdot)] \Rightarrow [\mathbf{R}]$ (15) |
| $[_{\langle\ldots,\mathbf{A}\rangle}\mathtt{I}(\cdot,\mathtt{i}),\mathtt{i}] \Rightarrow [\mathbf{A}]$ (16) | $[_{\langle\ldots,\mathbf{A}\rangle}\mathtt{I}(\cdot,\mathtt{I}'),\mathtt{I}'] \Rightarrow [\mathbf{A}]$ (19) | $[_{\langle\ldots,\mathbf{A}\rangle}\mathtt{I}(\cdot,\mathtt{C}),\mathtt{C}] \Rightarrow [\mathbf{A}]$ (25) | $[_{\langle\ldots,\mathbf{A}\rangle}\mathtt{I}(\cdot,\mathtt{P}),\mathtt{P}] \Rightarrow [\mathbf{A}]$ (31) |
| $[_{\langle\ldots,\mathbf{AD}\rangle}\mathtt{I}(\cdot,\mathtt{i}),\mathtt{i}] \Rightarrow [\mathbf{A}]$ (17) | $[_{\langle\ldots,\mathbf{A}\rangle}\mathtt{I},\mathtt{I}'(\mathtt{I},\cdot)] \Rightarrow [\mathbf{A}]$ (20) | $[_{\langle\ldots,\mathbf{A}\rangle}\mathtt{I},\mathtt{C}(\mathtt{I},\cdot)] \Rightarrow [\mathbf{A}]$ (26) | $[_{\langle\ldots,\mathbf{A}\rangle}\mathtt{I},\mathtt{P}(\mathtt{I},\cdot)] \Rightarrow [\mathbf{A}]$ (32) |
| $[_{\langle\ldots,\mathbf{R}\rangle}\mathtt{I}(\cdot,\mathtt{i}),\mathtt{i}] \Rightarrow [\mathbf{R}]$ (18) | $[_{\langle\ldots,\mathbf{AD}\rangle}\mathtt{I}(\cdot,\mathtt{I}'),\mathtt{I}'] \Rightarrow [\mathbf{A}]$ (21) | $[_{\langle\ldots,\mathbf{AD}\rangle}\mathtt{I}(\cdot,\mathtt{C}),\mathtt{C}] \Rightarrow [\mathbf{A}]$ (27) | $[_{\langle\ldots,\mathbf{AD}\rangle}\mathtt{I}(\cdot,\mathtt{P}),\mathtt{P}] \Rightarrow [\mathbf{A}]$ (33) |
| | $[_{\langle\ldots,\mathbf{AD}\rangle}\mathtt{I},\mathtt{I}'(\mathtt{I},\cdot)] \Rightarrow [\mathbf{A}]$ (22) | $[_{\langle\ldots,\mathbf{AD}\rangle}\mathtt{I},\mathtt{C}(\mathtt{I},\cdot)] \Rightarrow [\mathbf{A}]$ (28) | $[_{\langle\ldots,\mathbf{AD}\rangle}\mathtt{I},\mathtt{P}(\mathtt{I},\cdot)] \Rightarrow [\mathbf{A}]$ (34) |
| | $[_{\langle\ldots,\mathbf{R}\rangle}\mathtt{I}(\cdot,\mathtt{I}'),\mathtt{I}'] \Rightarrow [\mathbf{R}]$ (23) | $[_{\langle\ldots,\mathbf{R}\rangle}\mathtt{I}(\cdot,\mathtt{C}),\mathtt{C}] \Rightarrow [\mathbf{R}]$ (29) | $[_{\langle\ldots,\mathbf{R}\rangle}\mathtt{I}(\cdot,\mathtt{P}),\mathtt{P}] \Rightarrow [\mathbf{R}]$ (35) |
| | $[_{\langle\ldots,\mathbf{R}\rangle}\mathtt{I},\mathtt{I}'(\mathtt{I},\cdot)] \Rightarrow [\mathbf{R}]$ (24) | $[_{\langle\ldots,\mathbf{R}\rangle}\mathtt{I},\mathtt{C}(\mathtt{I},\cdot)] \Rightarrow [\mathbf{R}]$ (30) | $[_{\langle\ldots,\mathbf{R}\rangle}\mathtt{I},\mathtt{P}(\mathtt{I},\cdot)] \Rightarrow [\mathbf{R}]$ (36) |
| $[_{\langle\ldots,\mathbf{A}\rangle}\mathtt{C}(\cdot,\mathtt{i}),\mathtt{i}] \Rightarrow [\mathbf{R}]$ (37) | $[_{\langle\ldots,\mathbf{A}\rangle}\mathtt{C}(\cdot,\mathtt{I}),\mathtt{I}] \Rightarrow [\mathbf{R}]$ (40) | $[_{\langle\ldots,\mathbf{A}\rangle}\mathtt{C}(\cdot,\mathtt{C}'),\mathtt{C}'] \Rightarrow [\mathbf{R}]$ (46) | $[_{\langle\ldots,\mathbf{A}\rangle}\mathtt{C}(\cdot,\mathtt{P}),\mathtt{P}] \Rightarrow [\mathbf{R}]$ (52) |
| $[_{\langle\ldots,\mathbf{AD}\rangle}\mathtt{C}(\cdot,\mathtt{i}),\mathtt{i}] \Rightarrow [\mathbf{R}]$ (38) | $[_{\langle\ldots,\mathbf{A}\rangle}\mathtt{C},\mathtt{I}(\mathtt{C},\cdot)] \Rightarrow [\mathbf{A}]$ (41) | $[_{\langle\ldots,\mathbf{A}\rangle}\mathtt{C},\mathtt{C}'(\mathtt{C},\cdot)] \Rightarrow [\mathbf{A}]$ (47) | $[_{\langle\ldots,\mathbf{A}\rangle}\mathtt{C},\mathtt{P}(\mathtt{C},\cdot)] \Rightarrow [\mathbf{A}]$ (53) |
| $[_{\langle\ldots,\mathbf{R}\rangle}\mathtt{C}(\cdot,\mathtt{i}),\mathtt{i}] \Rightarrow [\mathbf{A}]$ (39) | $[_{\langle\ldots,\mathbf{AD}\rangle}\mathtt{C}(\cdot,\mathtt{I}),\mathtt{I}] \Rightarrow [\mathbf{R}]$ (42) | $[_{\langle\ldots,\mathbf{AD}\rangle}\mathtt{C}(\cdot,\mathtt{C}'),\mathtt{C}'] \Rightarrow [\mathbf{R}]$ (48) | $[_{\langle\ldots,\mathbf{AD}\rangle}\mathtt{C}(\cdot,\mathtt{P}),\mathtt{P}] \Rightarrow [\mathbf{R}]$ (54) |
| | $[_{\langle\ldots,\mathbf{AD}\rangle}\mathtt{C},\mathtt{I}(\mathtt{C},\cdot)] \Rightarrow [\mathbf{A}]$ (43) | $[_{\langle\ldots,\mathbf{AD}\rangle}\mathtt{C},\mathtt{C}'(\mathtt{C},\cdot)] \Rightarrow [\mathbf{A}]$ (49) | $[_{\langle\ldots,\mathbf{AD}\rangle}\mathtt{C},\mathtt{P}(\mathtt{C},\cdot)] \Rightarrow [\mathbf{A}]$ (55) |
| | $[_{\langle\ldots,\mathbf{R}\rangle}\mathtt{C}(\cdot,\mathtt{I}),\mathtt{I}] \Rightarrow [\mathbf{A}]$ (44) | $[_{\langle\ldots,\mathbf{R}\rangle}\mathtt{C}(\cdot,\mathtt{C}'),\mathtt{C}'] \Rightarrow [\mathbf{A}]$ (50) | $[_{\langle\ldots,\mathbf{R}\rangle}\mathtt{C}(\cdot,\mathtt{P}),\mathtt{P}] \Rightarrow [\mathbf{A}]$ (56) |
| | $[_{\langle\ldots,\mathbf{R}\rangle}\mathtt{C},\mathtt{I}(\mathtt{C},\cdot)] \Rightarrow [\mathbf{R}]$ (45) | $[_{\langle\ldots,\mathbf{R}\rangle}\mathtt{C},\mathtt{C}'(\mathtt{C},\cdot)] \Rightarrow [\mathbf{R}]$ (51) | $[_{\langle\ldots,\mathbf{R}\rangle}\mathtt{C},\mathtt{P}(\mathtt{C},\cdot)] \Rightarrow [\mathbf{R}]$ (57) |
| $[_{\langle\ldots,\mathbf{A}\rangle}\mathtt{P}(\cdot,\mathtt{i}),\mathtt{i}] \Rightarrow [\mathbf{AD}]$ (58) | $[_{\langle\ldots,\mathbf{A}\rangle}\mathtt{P}(\cdot,\mathtt{I}),\mathtt{I}] \Rightarrow [\mathbf{AD}]$ (61) | $[_{\langle\ldots,\mathbf{A}\rangle}\mathtt{P}(\cdot,\mathtt{C}),\mathtt{C}] \Rightarrow [\mathbf{AD}]$ (67) | $[_{\langle\ldots,\mathbf{A}\rangle}\mathtt{P}(\cdot,\mathtt{P}'),\mathtt{P}'] \Rightarrow [\mathbf{AD}]$ (73) |
| $[_{\langle\ldots,\mathbf{AD}\rangle}\mathtt{P}(\cdot,\mathtt{i}),\mathtt{i}] \Rightarrow [\mathbf{AD}]$ (59) | $[_{\langle\ldots,\mathbf{A}\rangle}\mathtt{P},\mathtt{I}(\mathtt{P},\cdot)] \Rightarrow [\mathbf{A}]$ (62) | $[_{\langle\ldots,\mathbf{A}\rangle}\mathtt{P},\mathtt{C}(\mathtt{P},\cdot)] \Rightarrow [\mathbf{A}]$ (68) | $[_{\langle\ldots,\mathbf{A}\rangle}\mathtt{P},\mathtt{P}'(\mathtt{P},\cdot)] \Rightarrow [\mathbf{A}]$ (74) |
| $[_{\langle\ldots,\mathbf{R}\rangle}\mathtt{P}(\cdot,\mathtt{i}),\mathtt{i}] \Rightarrow [\mathbf{A}]$ (60) | $[_{\langle\ldots,\mathbf{AD}\rangle}\mathtt{P}(\cdot,\mathtt{I}),\mathtt{I}] \Rightarrow [\mathbf{AD}]$ (63) | $[_{\langle\ldots,\mathbf{AD}\rangle}\mathtt{P}(\cdot,\mathtt{C}),\mathtt{C}] \Rightarrow [\mathbf{AD}]$ (69) | $[_{\langle\ldots,\mathbf{AD}\rangle}\mathtt{P}(\cdot,\mathtt{P}'),\mathtt{P}'] \Rightarrow [\mathbf{AD}]$ (75) |
| | $[_{\langle\ldots,\mathbf{AD}\rangle}\mathtt{P},\mathtt{I}(\mathtt{P},\cdot)] \Rightarrow [\mathbf{A}]$ (64) | $[_{\langle\ldots,\mathbf{AD}\rangle}\mathtt{P},\mathtt{C}(\mathtt{P},\cdot)] \Rightarrow [\mathbf{A}]$ (70) | $[_{\langle\ldots,\mathbf{AD}\rangle}\mathtt{P},\mathtt{P}'(\mathtt{P},\cdot)] \Rightarrow [\mathbf{A}]$ (76) |
| | $[_{\langle\ldots,\mathbf{R}\rangle}\mathtt{P}(\cdot,\mathtt{I}),\mathtt{I}] \Rightarrow [\mathbf{A}]$ (65) | $[_{\langle\ldots,\mathbf{R}\rangle}\mathtt{P}(\cdot,\mathtt{C}),\mathtt{C}] \Rightarrow [\mathbf{A}]$ (71) | $[_{\langle\ldots,\mathbf{R}\rangle}\mathtt{P}(\cdot,\mathtt{P}'),\mathtt{P}'] \Rightarrow [\mathbf{A}]$ (77) |
| | $[_{\langle\ldots,\mathbf{R}\rangle}\mathtt{P},\mathtt{I}(\mathtt{P},\cdot)] \Rightarrow [\mathbf{R}]$ (66) | $[_{\langle\ldots,\mathbf{R}\rangle}\mathtt{P},\mathtt{C}(\mathtt{P},\cdot)] \Rightarrow [\mathbf{R}]$ (72) | $[_{\langle\ldots,\mathbf{R}\rangle}\mathtt{P},\mathtt{P}'(\mathtt{P},\cdot)] \Rightarrow [\mathbf{R}]$ (78) |

**Algorithm 2** Compute a C-Label

**Require:** A discussion $D[v]$ and a vertex $v_i \in V(D[v])$;
**Ensure:** A c-label for $v_i$;

**procedure** ComputeLabel($v_i, D[v]$)
  *Initialize:*
  Empty the t-set of $v_i$
  *Fill the t-set of $v_i$:*
  **for each** $\exists v' v_i \in L(D[v])$ **do**
  *Apply Definition 5.22 to propagate a c-label to $v_i$ from the vertex $v'$:*
    **if** *propagateLabel*$(v', v_i)$ is not an error **then**
      Add the result of *propagateLabel*$(v', v_i)$ to the t-set of $v_i$
    **else**
      Stop and return error: disallowed graph structure encountered.
    **end if**
  **end for**
  *Apply Definition 5.25 to infer the overruling label in the t-set of $v_i$:*
  **if** there is at least one **R** in the t-set of $v_i$ **then**
    Stop and return **R**.
  **else if** there is at least one **AD** in the t-set of $v_i$ **then**
    Stop and return **AD**.
  **else**
    Stop and return **A**.
  **end if**
**end procedure**

*Proof.* Trivial; cf., Appendix A.3. □

The ComputeLabel procedure is called within the evaluation algorithm, of which an informal outline was given earlier (cf., §5.2.1). Three steps were identified: (1) the algorithm builds the transitive closure of the transitive preference rules, the applications of which appear in the discussion being evaluated; (2) the algorithm builds a topological sort of the strongly connected components of the transitive closure of the discussion; and (3) the algorithm labels the strongly connected components, in the order of their topological sort. These three steps are performed in the given sequence by Algorithm 3, called EvaluateDiscussion. Given a discussion $D[v]$, EvaluateDiscussion will compute, if they exist, stable c-labels on each vertex in the discussion. Errors will be returned if the c-labels are not stable, or if the algorithm encounters a line that violates the meaning of lines in an ace graph, as established in Definition 4.20.

**Definition 5.28.** ***Stable C-Label.*** *The $m$th c-label in the c-sequence of a vertex $v$ is a stable label for that vertex if and only if for any integer $i > 0$, any $(m+i)$th c-label, computed at the $(m+i)$th traversal of $v$, equals the $m$th c-label in the c-sequence of $v$.*

If the algorithm finds stable c-labels, we can immediately answer the question of whether each vertex in the discussion is acceptable. If EvaluateDiscussion returns an error, then the discussion must be revised: new vertices may need to be added if stable labels are absent, while all lines violating the meaning of the lines in an ace graphs should be removed.

**Proposition 5.29.** *Agorithm 3 applied to a discusssion $D[v]$ (i) does not loop indefinetly, (ii) returns stable c-labels for all vertices in $D[v]$ if they exist, an error otherwise, and (iii) has the running time in $O(C(D^c[v])(|L(D^c[v])| + 2|V(D^c[v])|))$, where $C(D^c[v])$ is the number of simple cycles in $D^c[v]$ and $D^c[v]$ is the transitive closure of the transitive preference rules in $D[v]$ on $D[v]$.*

*Proof.* Cf., Appendix A.5. □

EvaluateDiscussion applied to $D[v]$ starts by building the transitive closure on transitive preference rule applications in $D[v]$ to obtain $D^c[v]$. $P^T$ is a set of sets; each element in $P^T$ is a set of preference rule applications that are transitive. Figure 5(a) gives a hypothetical discussion $D[\mathtt{i}_1]$, where $\mathtt{C}_{1,1}$, $\mathtt{C}_{1,2}$, and $\mathtt{C}_{1,3}$ are assumed to be three applications of the same transitive preference rule $\mathtt{C}_1$. Figure 5(b) shows the discussion $D^c[\mathtt{i}_1]$, obtained by computing the transitive closure of $\mathtt{C}_1$ on the discussion $D[\mathtt{i}_1]$. $D^c[\mathtt{i}_1]$ is the result of BuildTransitiveClosure($D[\mathtt{i}_1], P^T$), where $P^T$ has only one element, which is the set $\{\mathtt{P}_{1,1}, \mathtt{P}_{1,2}, \mathtt{P}_{1,3}\}$. The procedure BuildTransitiveClosure is described in detail in Appendix A.4.

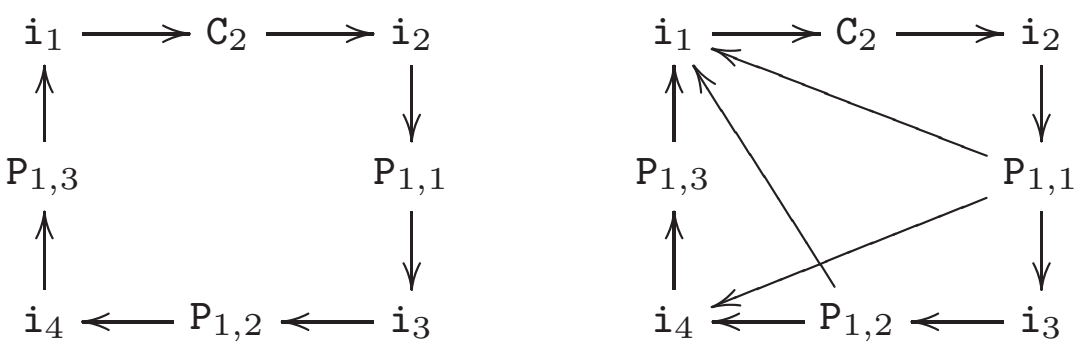

**Figure 5:** Figure 5(a) shows a hypothetical discussion, in which all preference rule applications are different applications $\mathtt{P}_{1,1}, \mathtt{P}_{1,2}, \mathtt{P}_{1,3}$ of the same comparison rule $\mathtt{P}_1$ and $\mathtt{P}_1$ is assumed to be transitive. Figure 5(b) shows the same discussion complex smallest justification updated for the transitive closure of $\mathtt{P}_1$ on the discussion in Figure 5(a).

The next step (cf., Line 4 in Algorithm 3) is to find all strongly connected compontens in $D^c[v]$. $D^c[\mathtt{i}_1]$ in Figure 5 has a single strongly connected component, while $D^c[\mathtt{i}(g_4)]$ in Figure 2 has twelve strongly connected components, as illustrated in Figure 3. We do not discuss the detail of EnumerateSCC, as it amounts to Tarjan's strongly connected components algorithm [19], or an improved variant thereof (cf., e.g., [18]).

The strongly connected components of a $D^c[v]$ returned by EnumerateSCC($D^c[v]$) are recorded in the set $\mathcal{C}$. Each element $c_i \in \mathcal{C}$ is a strongly connected component of $D^c[v]$, and thereby a subgraph of $D^c[v]$. The c-labels are computed by traversing each strongly connected component separately. Since the labels on the vertices $V(c_i)$ in a strongly connected component $c_i$ may depend on the labels on vertices $V(c_j)$ of another strongly connected component $c_j$, an order must be established, in which the strongly connected comoponents should be traversed to compute c-labels on their vertices. It is well-known that if we contract each strongly connected component in a directed graph down to a single vertex, the result is a directed acyclic graph. The procedure ContractSCC takes a $D^c[v]$ and the set of its strongly connected components, and returns a directed acyclic graph $D_C$. Each vertex $w_i \in V(V_C)$ represents exactly one strongly connected component in $\mathcal{C}$. ContractSCC starts by adding as many vertices $w_i$ to $V(D_C)$ as there are strongly connected components in $\mathcal{C}$. The first **for each** loop then considers each strongly connected component $c_i \in \mathcal{C}$, and defines the function *standsFor* : $V(D_C) \longrightarrow \mathcal{C}$. The *standsFor* function associates to each $w_i \in V(D_C)$ a strongly connected component $c_i \in \mathcal{C}$ that $w_i$ represents; one can understand a vertex $w_i$ as being the result of shrinking $c_i$ down to a single vertex. We shall say that the function *standsFor* returns the strongly connected component in $\mathcal{C}$ that is represented by a given vertex in $D_C$. After the current $c_i$ is associated to a $w_i$ in Line 13, the **for each** loop in Lines 15–23 considers for each $w_i \in V(D_C)$, all other $w_j \neq w_i$ in order to check if there is a line in $D^c[v]$ from a vertex in $V(\mathit{standsFor}(w_i))$ (i.e., $V(c_i)$) to a vertex in $V(\mathit{standsFor}(w_j))$ (i.e., $c_j$). If such a line exists, a line between $w_i$ and $w_j$ is added to $L(D_C)$. ContractSCC thereby returns a graph, in which each vertex stands for exactly one strongly connected component of $D^c[v]$, and in which any two distinct vertices $w_i$ and $w_j$ are connected by a line $w_iw_j$ if and only if there is a line from a vertex in $V(\mathit{standsFor}(w_i))$ to a vertex in $V(\mathit{standsFor}(w_j))$.

Given a directed acyclic graph $D_C$ and the function *standsFor*, both from ContractSCC($D^c[v], \mathcal{C}$), we can determine the order, in which to traverse the strongly connected components of $D^c[v]$. A topological order of the vertices in $D_C$ is a sequence of vertices, in which (i) the vertices without incoming lines occupy the first places, (ii) vertices in-adjacent to any of the first vertices (i) occupy the next places, (iii) vertices in-adjacent to the vertices on second places (ii) are next, and so on; the last vertices in the sequence have no outoing lines. TopologicalSort($D_C$) returns $\mathcal{S}_C$, which is a topological sort of the vertices in $V(D_C)$. As the elements in $\mathcal{S}_C$ are the vertices of $D_C$, the procedure ExpandSCC is called on $\mathcal{S}_C$ in order to obtain a topological sort of the strongly connected components of $D^c[v]$. ExpandSCC will first set $\mathcal{S}$ to equate $\mathcal{S}_C$, then replace each element $w_i$ of $\mathcal{S}$ with the result of *standsFor*($w_i$). Each element $w_i$ will thereby be replaced with the strongly connected component $c_i$, which *standsFor* relates to $c_i$. As soon as we have a topological sort $\mathcal{S}$ of the strongly connected components of $D[v]$, the traversal of the strongly connected compontents can be performed and c-labels computed on the vertices therein.

**Algorithm 3** Evaluate a Discussion

**Require:** A discussion $D[v]$;
**Ensure:** If there are stable c-labels for all vertices in $D[v]$, then a function $\Lambda : V(D[v]) \longrightarrow \{\mathsf{A}, \mathsf{AD}, \mathsf{R}\}$, which returns the stable c-label for each vertex in $D[v]$; if there are no stable labels, then an error;

**procedure** EvaluateDiscussion($D[v], P^T$)
*Initialize:*
    Empty $D^c[v]$, $\mathcal{C}$, $D_C$, $\mathcal{S}_C$, and $\mathcal{S}$
*Build the transitive closure of the transitive preference rules applied in $D[v]$:*
    $D^c[v] \leftarrow$ BuildTransitiveClosure($D[v], P^T$)
*Find all strongly connected compontents of $D^c[v]$ and obtain their topological sort:*
    $\mathcal{C} \leftarrow$ EnumerateSCC($D^c[v]$) ▷ $\mathcal{C}$ is the set of all strongly connected components of $D^c[v]$.
    $D_C \leftarrow$ ContractSSC($D^c[v], \mathcal{C}$)
    $\mathcal{S}_C \leftarrow$ TopologicalSort($D_C$)
    $\mathcal{S} \leftarrow$ ExpandSCC($\mathcal{S}_C$) ▷ An element of $\mathcal{S}$ is a strongly connected component from $\mathcal{C}$.
*Label all strongly connected components from the first to the last in their topological sort:*
    $\Lambda \leftarrow$ LabelSCC($D^c[v], \mathcal{S}$)
**end procedure**

*Obtain a directed acyclic graph $D_C$ by contracting the strongly connected components in $D^c[v]$:*
**procedure** ContractSCC($D^c[v], \mathcal{C}$)
    $V(D_C) \leftarrow \{w_1, w_2, \ldots, w_{|\mathcal{C}|}\}$ ▷ $w_1, \ldots, w_{|\mathcal{C}|}$ are vertices of $D_C$.
    **for each** $c_i \in \mathcal{C}$ **do** ▷ $c_i$ is a strongly connected component of $D^c[v]$.
        *standsFor*($w_i$) $\leftarrow c_i$
    **end for**
    **for each** $w_i \in V(D_C)$ **do**
        **for each** $w_j \in V(D_C)$ s.t. $w_i \neq w_j$ **do**
            **for each** $v \in V($*standsFor*$(w_i))$ **do**
                **if** $\exists v'v \in L(D^c[v])$ s.t. $v' \in V($*standsFor*$(w_j))$ **then**
                    Add the line from $w_j$ to $w_i$ to $L(D_C)$
                **end if**
            **end for**
        **end for**
    **end for**
**end procedure**

*Obtain a topological sort $\mathcal{S}$ of the strongly connected components of $D^c[v]$:*
**procedure** ExpandSCC($\mathcal{S}_C$)
    $\mathcal{S} \leftarrow \mathcal{S}_C$
    **for each** $w_i$ in $\mathcal{S}$ **do**
        Replace $w_i$ with the result of *standsFor*($w_i$)
    **end for**
**end procedure**

**Algorithm 4** Label Strongly Connected Components

**Require:** $D^c[v]$ and $\mathcal{S}$, a topological sort of the strongly connected components of $D^c[v]$;
**Ensure:** If there are stable c-labels for all vertices in $D[v]$, then a function $\Lambda : V(D[v]) \longrightarrow \{\mathsf{A}, \mathsf{AD}, \mathsf{R}\}$, which returns the stable c-label for each vertex in $D[v]$; if there are no stable labels, then an error;

**procedure** LabelSCC($D^c[v], \mathcal{S}$)
  *Initialization:*
  Add an empty c-sequence and an empty t-set on each vertex in $V(D^c[v])$
  Let $i$ in $c_i$ be the position of the strongly connected component $c$ in $\mathcal{S}$, i.e., $i \equiv \mathit{position}(c, \mathcal{S})$
  $i \leftarrow 1$

  *Label each strongly connected component of* $D^c[v]$*:*
  **while** $i \leq \mathit{length}(\mathcal{S})$ **do**
    **if** $|V(c_i)| = 1$ **then**
      **if** $\nexists v''v' \in L(D^c[v])$ s.t. $v' \in V(c_i),\ v'' \in \bigcup_{j=1}^{i-1} V(c_j)$ **then**
        Append the c-sequence of the only vertex $v' \in V(c_i)$ with the label $\mathsf{A}$
      **else**
        ComputeLabel($v' \in V(c_i), D^c[v]$)
      **end if**
    **else**
      Let $v'$ be the newest vertex in $V(c_i)$
      LabelComplexSCC($c_i, v'$)
    **end if**
    $i \leftarrow i + 1$
  **end while**

  *Define the function* $\Lambda : V(D[v]) \longrightarrow \{\mathsf{A}, \mathsf{AD}, \mathsf{R}\}$
  **for each** vertex $v' \in V(D^c[v])$ **do**
    $\Lambda(v') \leftarrow$ (the last c-label in the c-sequence of $v$)
  **end for**
  Stop and return message: Stable labels found.
**end procedure**

To label the vertices in the strongly connected components of $D^c[v]$, Algorithm 3 calls the procedure `LabelSCC` on $D^c[v]$ and a topological sort $\mathcal{S}$ of the strongly connected components of $D^c[v]$. Detail of the LabelSCC is given in Algorithm 4.

LabelSCC($D^c[v], \mathcal{S}$) initializes by adding an empty t-set and an empty c-sequence to each vertex of $VD^c[v]$. As discussed earlier in relation to the procedure ComputeLabel in Algorithm 2, the t-sets and c-sequences are needed to compute c-labels at each traversal of a vertex. We then introduce the function *position*, which returns the positive integer denoting the position of a strongly connected component $c$ in the topological sort $\mathcal{S}$ of the strongly connected components. $i$ in $c_i$ abbreviates $\mathit{position}(c, \mathcal{S})$. Initialization ends by setting $i = 1$, so that the **while** loop moves from the first $c$ in $\mathcal{S}$. At each iteration, the **while** loop considers the strongly connected component $c$ at the $i$th position in $\mathcal{S}$, i.e., $c_i$. The **while** loop stops after it has processed the last element in the sequence $\mathcal{S}$. $\mathit{length}(\mathcal{S})$ is the number of elements in $\mathcal{S}$. For a given $c_i$, the outer **if** block in the **while** loop checks if the $c_i$ has one or more vertices:

- If $c_i$ has a single vertex, i.e., $|V(c_i)| = 1$, then we must determine whether that vertex has incoming lines in $D^c[v]$:
  - If the only vertex in $c_i$ has no incoming lines, there are no vertices in $D^c[v]$, which are relevant to its acceptability (cf., Definition 5.1 and Proposition 5.9) and that vertex can be given the c-label $\mathsf{A}$;
  - If the only vertex in $c_i$ has incoming lines, then there are vertices that are relevant to its acceptability. We therefore call the procedure ComputeLabel on that vertex and the graph $D^c[v]$;

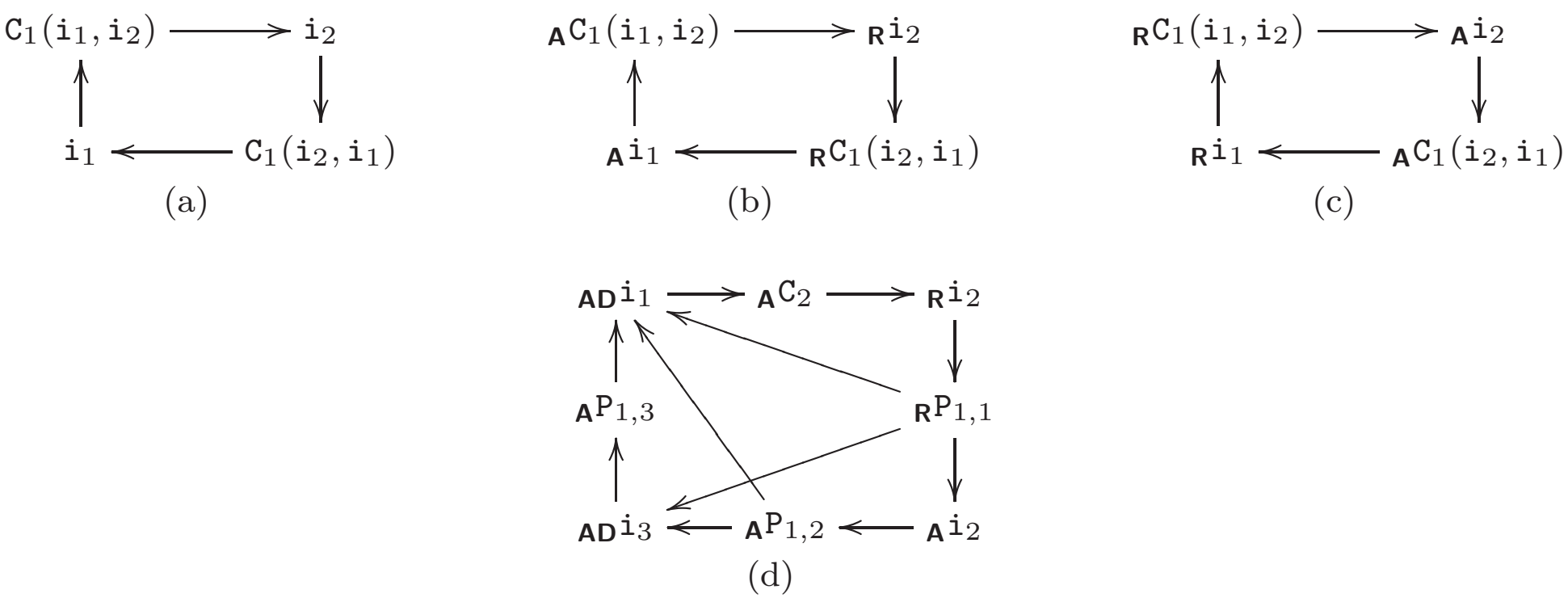

**Figure 6:** Figure 6(a) shows a strongly connected component, say $c_1$ from the transitive closure $D^c[v]$ of some discussion $D[v]$. Figure 6(b) shows the stable c-labels obtained by calling LabelComplexSCC($c_1$, $\mathtt{C}_1$). Figure 6(b) shows the stable c-labels obtained by calling LabelComplexSCC($c_1$, $\mathtt{i}_2$). We assume for simplicity that no vertex in $c_1$ has an incoming line in $D^c[v]$ that is not in $c_1$. Observe from Figures 6(b) and 6(c) that the strongly connected component $c_1$ has no unique stable labels. Figure 6(d) shows the unique stable labels on the strongly connected component from Figure 5(b).

- If $c_i$ has more than one vertex, then each vertex in $c_i$ is on at least one cycle. To label the vertices on cycles, one of the vertices must be chosen as a starting vertex. The problem with this choice is that an absolute criterion for choosing a starting vertex is abesent. To see why, consider the strongly connected component in Figure 6(a). We observe that calling LabelComplexSCC on two different vertices gives different stable labels. This same problem arises when LabelComplexSCC is called on the largest strongly connected component in Figure 3, once on the vertex $\mathtt{P}_1$, and in another call on $\mathtt{C}_1$. In other words, we can obtain stable labels, but there are patterns of strongly connected components, in which the choice of the starting vertex for the labeling procedure influences the stable labels that are obtained. Not all complex (i.e., $|V(c)| > 1$) strongly connected components suffer from this problem: an example is the strongly connected component shown in Figure 5(b). If we assume that none of its vertices has incoming lines other than those shown in Figure 5(b), then unique stable labels exist for all of its vertices: whatever the chosen starting vertex for the labeling procedure, the stable labels will always be identical for each of its vertices. We discuss the uniqueness of stable c-labels in detail below.

If no errors are reported, the **while** loop will terminate after all strongly connected components in $\mathcal{S}$ have been traversed, and all vertices therein assigned stable c-labels. We then proceed to define the $\Lambda$ function via the last **for each** loop. Each vertex $v' \in D^c[v]$ is considered, and $\Lambda(v')$ is given the last c-label of the c-sequence of $v'$.

We have noted above that stable labels may not be unique. We now consider this problem in more detail.

**Definition 5.30.** ***Unique Stable C-Label.*** *Let $c$ be a strongly connected component of the transitive closure of a discussion, and such that $|V(c)| > 1$. A stable c-label on a vertex $v \in V(c)$ is unique if and only if* LabelComplexSCC *always assigns the same stable c-label on $v$, regardless on which vertex in $c$ it is called.*

Suppose that there are in fact stable c-labels in a given strongly connected component $c$ and that there are no lines in $c$ that violate the line meaning (cf., §4.2), so that LabelComplexSCC will not return an error. We then see four strategies to attack the question of whether the returned stable c-labels of the vertices in $c$ are unique:

1. *Acyclical discussion.* If there are no cycles in the transitive closure $D^c[v]$ of a discussion $D[v]$, the problem of uniqueness of stable c-labels disappears. The absence of cycles in $D^c[v]$ makes each vertex of $D^c[v]$ a strongly connected component, so that the **if** condition in Line 6 of LabelSCC will verify for each vertex in $D^c[v]$ and LabelComplexSCC will not be called at all. There are two principal shortcomings of imposing acyclical discussions. Part of the expressive power would be eliminated, as various intuitively valid structures, such as the one in our earlier example (cf., §4.1), cannot be captured.

The other difficulty is that constraints must be placed on the process of building a discussion, if cycles are to be avoided at discussion-time.

2. *Brute force uniqueness check.* The brute force uniqueness check would amount to call, for a given $c$, LabelComplexSCC on each of the vertices in $c$. In case the stable labels are not unique, the part of the discussion that the given strongly connected component captures should be revisited; after changes are made, EvaluateDiscussion would be run again. The proof of Proposition 5.29 points out that the running time of LabelComplexSCC on a strongly connected component $c$ is $O((|V(c)|+|L(c)|)(C(c)+1)+C(c)|V(c)|)$, where $C(c)$ is the number of simple cycles in $c$. The part $O((|V(c)|+|L(c)|)(C(c)+1))$ is due to the computation of the number of simple cycles in $c$, so that it need be executed only once in the brute force uniqueness check. It is then apparent that the brute force uniqueness check will increase the complexity from $O((|V(c)| + |L(c)|)(C(c) + 1) + C(c)|V(c)|)$ to $O((|V(c)| + |L(c)|)(C(c) + 1) + C(c)(|V(c)|)^2)$. The principal disadvantage in the brute force approach is the increase in complexity.

3. *Uniqueness check by random sample.* Instead of running LabelComplexSCC on each vertex in $c_i$ and comparing results, a random sample of the vertices could be chosen, and LabelComplexSCC would be called on each of the vertices in the sample. If LabelComplexSCC returns the same c-labels in each call on a vertex from the sample, the probability of these labels being the unique ones can be computed. Instead then of being certain that the c-labels are unique (as in the brute force approach), the computed probability would give a level of confidence in the c-labels being unique. If $X$ is the sample of vertices, and $S \subseteq V(c)$, then the complexity of this approach is $O((|V(c)| + |L(c)|)(C(c) + 1) + C(c)|X||V(c)|)$.

4. *Choose the first vertex via a heuristic.* The idea here is to offer a heuristic for choosing the starting vertex in $c$, on which to call LabelComplexSCC. One possible heuristic comes from the saying "The one who has the last word laughs best." In other words, the starting vertex for labeling could be the last vertex among those in a given strongly connected component. In Figure 6(a), if $\mathtt{C}_1(\mathtt{i}_1, \mathtt{i}_2)$ was the last vertex added among the four vertices in that figure, then LabelComplexSCC($c_1, \mathtt{C}_1$) is called and the resulting c-labels are given in Figure 6(a). This heuristic is meaningful in the example used throughout the paper (cf., §4.1): $\mathtt{P}_1$ and $\mathtt{C}_2$ were added last to the part of the graph, which encompasses the largest strongly connected component of $D[\mathtt{i}(g_4)]$. Calling LabelComplexSCC on either $\mathtt{P}_1$ or $\mathtt{C}_2$ gives the same stable labels. If LabelComplexSCC is called on a vertex older than $\mathtt{P}_1$ or $\mathtt{C}_2$, different c-labels will be obtained. We adopt this heuristic at present and include in the evaluation algorithm. As Dung [6] observes, the saying "he who has the last word laughs best" illustrates a very simple principle, on which people seem to often base the exchange of arguments. Given that LabelComplexSCC is called only once in this strategy, LabelComplexSCC will have the running time in $O((|V(c)| + |L(c)|)(C(c) + 1) + C(c)|V(c)|)$.

Observe that the first strategy is the simplest, while the second is the most complex. The third strategy is between the fourht and the second in terms of complexity. We use the heuristic given above in order to choose the newest vertex in the strongly connected component. This is reflected in Line 13 of LabelSCC, in Algorithm 4, where the newest vertex is chosen among the vertices of a given strongly connected component, and is input to LabelComplexSCC. It is clear that other relevant heuristics may be found, but we leave such discussions for future work.

LabelComplexSCC is defined in Algorithm 5. LabelSCC calls LabelComplexSCC on each strongly connected component $c$ having more than one vertex. In addition to $c$, LabelComplexSCC is supplied with the vertex of $c$, say $v$, from which to traverse $c$. We gave earlier (cf., §5.2.1) an informal outline of how $c$ is traversed from $v$ in order to compute the c-labels on vertices in $c$. We now discuss LabelComplexSCC in more detail via an example.

The initialization of LabelComplexSCC starts by counting the simple cycles in $c$, which is stored in $C(c)$. The queue $Q$ is emptied, and the starting vertex $v$ is added to $Q$. $Q$ will carry the vertices that the outer **for each** loop in the **while** loop will traverse. All vertices are assumed acceptable at the outset, so that the c-label **A** is added to the c-sequence of the each vertex in $c$. The c-label **A** is taken as the initial hypothetical label because it is the weakest of the three labels (cf., Definition 5.25). Two counters $p$ and $q$ are initialized. The starting vertex $v$ is called *First*. To see how LabelComplexSCC traverses $c$, observe first that the number of cycles $C(c)$ gives the upper bound on the number of simple paths from any vertex to any other vertex in $c$. Assume then that we have a $C(c)$ number of walkers for the given $c$. Recall from a

**Algorithm 5** Label a Complex Strongly Connected Component

**Require:** A strongly connected component $c$ of the transitive closure $D^c[v_i]$, s.t., $|V(c)| > 1$, and a vertex $v \in V(c)$;

**Ensure:** If there are stable c-labels for all vertices in $c$, then at least two c-labels in the c-sequence of each vertex in $V(\mathcal{J}[v])$; an error otherwise;

```
procedure LabelComplexSCC(c, v)
Initialization:
    C(c) ← CountSimpleCycles(c)                    ▷ C(c) is the number of simple cycles in c.
    Empty Q
    Add v to Q
    Add the c-label A to the c-sequence of each vertex in c
    Call First the vertex v in Q
    q ← 0
    p ← 1                      ▷ p counts the number of c-labels in the c-sequence of First.
Traversal:
    while Q is not empty do
        for each vertex v in Q do
            Delete v from the queue Q
            for each v′ ∈ V(c) s.t. ∃vv′ ∈ L(c) do
                if the result of ComputeLabel(v′, D^c[v_i]) is not an error then
                    if v′ ≠ First then
                        Append the c-sequence of v′ with the result of ComputeLabel(v′, D^c[v_i])
                    else
                        if q < C(c) then
                            q ← q + 1
                        else if q = C(c) then
                            Empty Q
                            Append the c-sequence of v′ with the result of ComputeLabel(v′, D^c[v_i])
                            q ← 0
                            p ← p + 1
                        end if
                    end if
                    Add v′ to Q
                    if the last two labels in the c-sequence of First are identical then
                        Empty Q and return message: Stable labels found.
                        For each vertex in V(c), keep only the last label in its c-sequence, delete others.
                    else if p = 4 then
                        Empty Q and return error: c has unstable labels.
                    end if
                else
                    Empty Q and return error: Disallowed graph structure encountered.
                end if
            end for
        end for
    end while
end procedure
```

discussion above (cf., §5.2.1) that tach walker obeys the following rules: (i) it takes equal time to traverse a vertex; (ii) it can only go forward (i.e., over lines that start in a vertex); and (iii) no two walkers will start from *First* and return to *First* along the exact same path.

Consider now again the strongly connected component $c$ in Figure 5(b) having four simple cycles, i.e., $C(c) = 4$. We shall assume for simplicity the following: (i) $c$ is a strongly connected component in the transitive closure $D^c[v_i]$, of some given discussion $D[v_i]$; and (ii) none of the vertices in $c$ has an incoming line in $D^c[v_i]$ that is not in $c$. Let $\mathtt{C}_2$ be the newest vertex in $c$, and is therefore called *First* and added to the queue $Q$. With $\mathtt{C}_2$ in $Q$, LabelComplexSCC enters the **while** loop. The outer **for each** loop will delete *First* from $Q$, while the inner **for each** loop will consider each line that starts in $\mathtt{C}_2$. We put all four walkers on $\mathtt{C}_2$ and move them along the line $\mathtt{C}_2\mathtt{i}_2$ to $\mathtt{i}_2$. Since $\mathtt{i}_2 \neq \mathtt{C}_2$, the **if** condition in Line 14, in the inner **for each** loop verifies and ComputeLabel($\mathtt{i}_2, D^c[v_i]$) is called. The c-sequence of $\mathtt{i}_2$ thus becomes $\langle \mathsf{A}, \mathsf{R} \rangle$ and $\mathtt{i}_2$ is added to $Q$ and the walkers are at $\mathtt{i}_2$. The **while** loop then processes $\mathtt{i}_2$: the walkers move along the line $\mathtt{i}_2\mathtt{P}_{1,1}$ to $\mathtt{P}_{1,1}$, and the c-sequence of $\mathtt{P}_{1,1}$ becomes $\langle \mathsf{A}, \mathsf{R} \rangle$ and $Q$ contains only $\mathtt{P}_{1,1}$. Because there are three lines starting in $\mathtt{P}_{1,1}$ and none of the vertices on which these lines end is *First*, the inner **for each** loop will be adding three vertices to $Q$.

We consequently send the walkers along three different lines that start in $\mathtt{P}_{1,1}$: walker A will move to $\mathtt{i}_1$, B to $\mathtt{i}_4$, and the other two, C and D to $\mathtt{i}_3$. It is important to observe here that the walker A will reach the vertex *First* in the least number of steps, while one of the walkers C or D will reach *First* in the most steps. Suppose that D is the last to arrive to *First*. We can now explain the role of the counter $q$. When A arrives to *First*, it will apply the ComputeLabel procedure, which will return a c-label based on the last c-label in the c-sequence of $\mathtt{i}_1$, which in turn was computed from the last c-label in the c-sequences of vertices $\mathtt{P}_{1,1}$, $\mathtt{P}_{1,2}$ and $\mathtt{P}_{1,3}$. At the time when A computes the label on $\mathtt{C}_2$, the walker C will be computing the label on $\mathtt{P}_{1,2}$. Moreover, it is only after A has appended the c-sequence of $\mathtt{C}_2$ that D will be computing the label on $\mathtt{P}_{1,3}$. If D adds a label to $\mathtt{P}_{1,3}$ that is different from the label that A had on $\mathtt{P}_{1,3}$, D will may end up computing a different label on $\mathtt{C}_2$ from the label that A computed. The point is that it is not appropriate to let A append the c-sequence of *First* before the last walker returns to *First*, because the labels computed by the walkers before D do not account for all of the information in $c$. It is only after the last walker arrives to *First* that we know that all vertices of $c$ were traversed. The **else** block in Lines 16–25 in Algorithm 5 verifies when a walker arrives at *First*, and it is only when the last walker has arrived that the c-sequence of *First* is appended with a label. Observe thus that $q$ counts the number of walkers arriving at *First*, while $p$ counts the number of c-labels in the c-sequence of *First*. Figure 7 shows the result of applying LabelComplexSCC to $c$ in Figure 5(b).

It is on the basis of patterns of c-labels in the c-sequence of *First* that LabelComplexSCC detects the presence or absence of stable c-labels in a complex strongly connected component. When no stable c-labels can be found, no pair of same c-labels will be identified in the four first c-labels of the c-sequence of *First*. Otherwise, stable labels are present and the algorithm will return a stable label for each vertex. The proof of Proposition 5.29 discusses in detail the case of unstable labels. Figure 8(a) gives an example of a complex smallest justification that has no stable c-labels, while Figure 8(b) illustrates a partial result of the application of LabelComplexSCC to the strongly connected component in Figure 8(a).

# 6 Acceptability Condition Revisited

We wrote in Equation 2 above (cf., §3) that $\mathbf{AC}(I_D, T(I_D), O_D)$ holds if and only if $\forall p \in \mathit{In}(I_D) \cup \mathit{In}(O_D) \cup \mathit{In}(T(I_D))$, $\mathbf{AC}(p)$. After we have introduced the EvaluateDiscussion procedure (cf., §5.2.2 and Algorithm 3), we can provide the revision of the preliminary definition of the acceptability condition (cf., Definition 3.1).

**Definition 6.1.** ***AC* (revised).** *The application of the* RE *method* $T$ *to the input* $I_D$ *to produce the output* $O_D$ *is acceptable, denoted* $\boldsymbol{AC}(I_D, T(I_D), O_D)$ *if and only if:*

$$\forall p \in \mathit{In}(I_D) \cup \mathit{In}(O_D) \cup \mathit{In}(T(I_D)),\ \boldsymbol{AC}(p) \tag{79}$$

*where:*

$$\boldsymbol{AC}(p) \text{ iff } \Lambda(\lambda_V(p)) \in \{\mathsf{A}, \mathsf{AD}\} \tag{80}$$

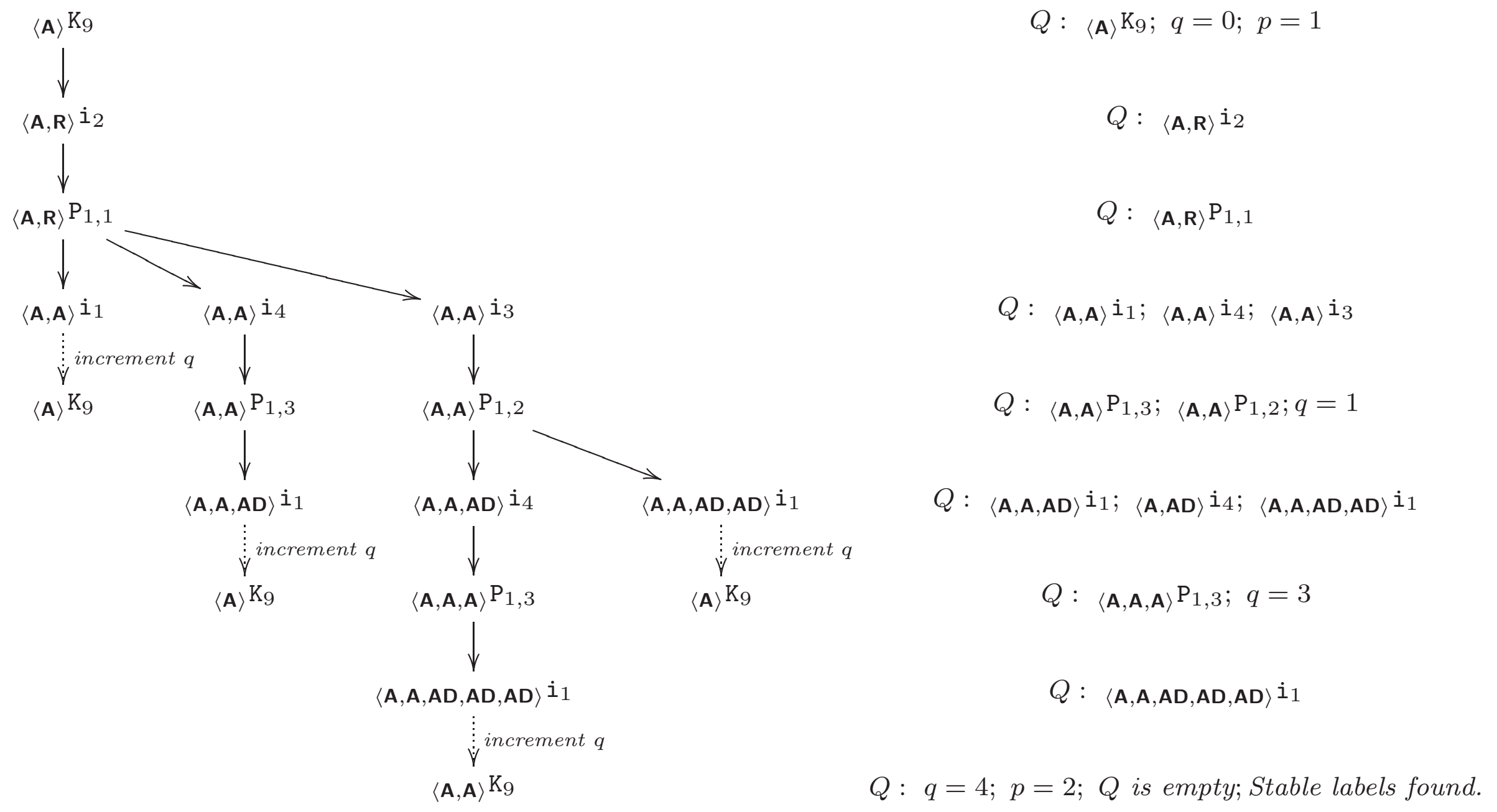

**Figure 7:** Traversal of complex strongly connected component in Figure 5(b) by the procedure LabelComplexSCC in Algorithm 5. The graph in Figure 5(b) has four simple cycles (i.e., $C(c) = 4$). The content of the queue $Q$ is given on the right hand side of the figure above. Stable c-labels are found in this example. The stable label for any given vertex is the last label in the sequence of labels for that vertex.

*Remark* 6.2. The definition above simply indicates that the stable c-label on the vertex $\lambda_V(p)$ that represents the proposition $p$, should either be **A** or **AD**. $\Lambda(\lambda_V(p))$ returns that stable c-label, if it exists. Recall from Algorithm 4 that the function $\Lambda$ is defined by calling EvaluateDiscussion on a discussion. We will therefore need to apply EvaluateDiscussion on all discussions that together have vertices representing all propositions in $\mathit{In}(I_D) \cup \mathit{In}(O_D) \cup \mathit{In}(T(I_D))$. The relationship between subdiscussions and discussions, highligthed in Proposition 5.17, means that we will not necessarily need to apply EvaluateDiscussion to the discussion of each vertex representing a proposition in $\mathit{In}(I_D) \cup \mathit{In}(O_D) \cup \mathit{In}(T(I_D))$, as some vertices may be within discussions of other vertices.

# 7 Notes on Implementation

The implementation of ace graphs and of the associated retrieval and evaluation algorithms is not available at the time of writing. The implementation in progress is based on the adaptation of standard open source internet forum software. This choice is based on two observations: (i) internet forums are popular means for discussion, and are a well known kind of software, having evolved from bulletin board systems of the early 1970s; and (ii) a discussion in ace amounts to a forum discussion performed according to a set of simple rules. By default, a discussion in an internet forum contains a collection of untyped posts. Any post may be a response to the original post (i.e., the root post), or a response to a later post. A forum discussion thus typically resembles an unlabeled tree, where each post is a vertex. To obtain an ace discussion instead of a standard forum discussion, the following rules must be followed by the participants:

1. Any post is either an information, inference, conflict, or preference post.
2. Any post that is not the first post in a discussion, must be related to another post according to the meaning of the `To` relationship.

Compared to a classical forum, an ace forum is thus one where a user chooses the label for a new post, and relates that post to others. The first rule above ensures that we have the label for any post, while the second rule guarantees that no post is disconnected from the others. A discussion in an ACE forum can thus be interpreted as an ACE graph (or ACE discussion), on which we can perform retrieval and evaluation operations via the algorithms discussed earlier. The evaluation algorithm provides to participants the indication on the acceptability of each post, so that they can, in case of rejection, intervene and respond in defense of their claims.

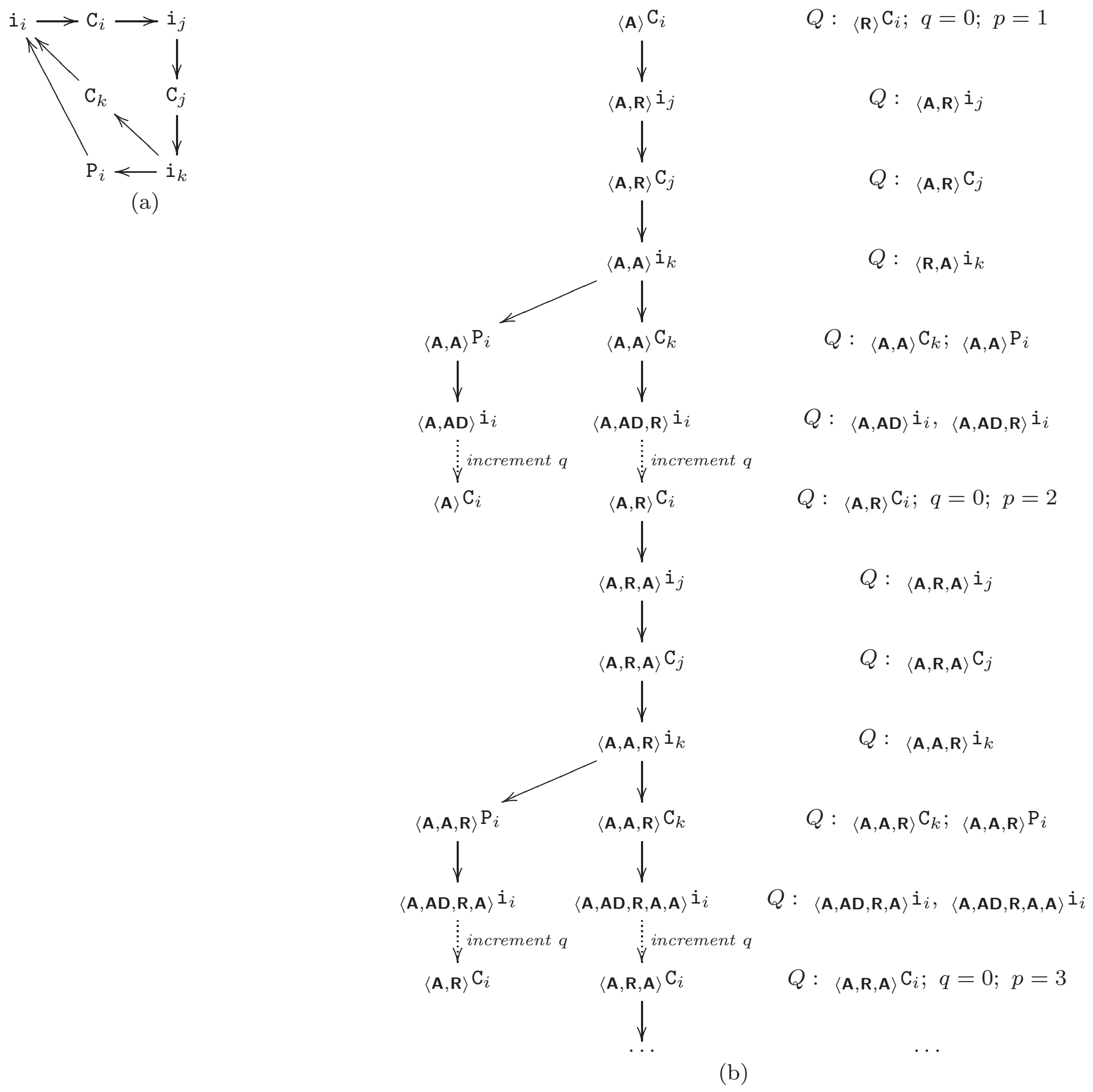

**Figure 8:** Figure 8(b) shows part of the traversal of the complex smallest justification in Figure 8(a) by the procedure LabelComplexSCC in Algorithm 5. Only part of the traversal is shown as the rest of it can be straightforwardly reconstructed by the reader. The graph in Figure 8(a) has two simple cycles (i.e., $C(c) = 2$). The content of the queue $Q$ is given on the right hand side in Figure 8(b). The strongly connected component $c$ in Figure 8(a) has no stable labels.

# 8 Discussion and Related Work

This work is the continuation of our efforts to advance the analysis of decision making in requirements engineering. At the 2006 edition of this conference, we discussed the problem of the acceptability of goal models [11] via their justification. We argued at the time that modeling choices – e.g., the inclusion or exclusion of some information about the system-to-be and its environment – should be *justified* for a goal model to be acceptable. A manual procedure for justifying and evaluating justifications was offered. The concept and procedure borrowed from the contributions on the analysis of arguments in artificial intelligence. A justification amounted to a tree, where any vertex can either support or attack other vertices. That work was subsequently extended [12] to analyze the clarity of information offered when justifying modeling choices; these manual techniques can identify the use of unclear information, by detecting, e.g., ambiguity and vagueness. This work was subsequently applied to the analysis of the acceptability of changes in Unified Modeling Language models [10]. The present work advances our prior results in several respects: (1) ACE allows participants to express preferences over information, and applications of inferences, conflicts, and other preferences in ACE graphs. Our prior analysis of acceptability via justification could not account for preferences, the set of inferences and conflict rules was closed (i.e., two inference rules were available and one conflict rule), and no forms of meta-reasoning could be captured (one could not express that the application of an inference rule is in conflict with the application of another inference rule, that a conflict may be in conflict with the application of an inference rule, and so on). (2) ACE offers algorithms for the retrieval of information relevant for the evaluation of acceptability; no such features were available for our justification graphs. (3) The evaluation of acceptability in ACE is automated via the evaluation algorithm outlined above; acceptability was evaluated manually in justification graphs. In conclusion, ACE is (i) more expressive (due to the presence of preferences and the support for forms of meta-reasoning), (ii) more "practical" (due to the presence of retrieval and evaluation algorithms), and (iii) more general (as argued throughout the paper) than the goal-oriented approaches, which we suggested previously.

The language in ACE, and more precisely, the choice of the four labels – information, inference, conflict, and preference – was influenced by the initiative towards a core ontology for argumentation in artificial intelligence, within the Argument Interchange Format (AIF) [3]. AIF has recently been suggested to facilitate the representation and exchange of data between various tools and agent-based applications that rely on arguments. The basis for AIF is the AIF Argument Network, which is "the core ontology for argument entities and relations between argument entities" [3]. The notion of argument is a construct in AIF, defined as a particular subgraph in an argument network. It is by placing restrictions on, or by specializing the concepts in the argument network that classical argumentation frameworks can be obtained. ACE thereby reuses the core ontology of AIF for the labels in ACE graphs. To the best of our knowledge, no framework based on AIF and comparable to ACE in terms of the language and the retrieval and evaluation algorithms has been suggested.

Validation is an old problem in RE, and has been raised in particular in relation to the very early requirements elicited from the parties involved in RE. Leite and Freeman argued in an important paper [14] that requirements should be elicited from different viewpoints, and "that examination of the differences resulting from them can be used as a way of assisting in the early validation of requirements". They suggested a language for capturing viewpoints, and heuristics for a syntacticly oriented analysis of views to the aim of resolving their inconsistencies. This approach provides inputs to the negotiation required to reconcile different opinions. An attractive characteristic of their proposal is that it is a means of validation applicable very early in the elicitation of requirements. We are very close to their work in this motivation, although our proposal differs in several important respects. ACE has a considerably simpler language than their framework. Since automated reasoning about acceptability happens in a propositional framework in ACE, we can accommodate various forms of RE artifacts. As we treat propositions as atoms, Leite and Freeman's approach can go into more detail, and study the structure of the propositional content. While they can point precisely the disrepancies between views, we can help by evaluating the outcomes of a discussion of these discrepancies. Nothing similar to the automated evaluation of acceptability is present in the viewpoint resolution method. Discussions are not studied in detail. ACE is complementary in this respect, as it can be used to record and evaluate discussions of views and their inconsistencies. Gervasi and Nuseibeh [7] check predefined properties on models generated from parsed text to identify nontrivial inconsistencies. While $\mathbf{AC}(I_D, T(I_D), O_D)$ indicates agreement about these artifacts, it certainly does not entail the internal consistency of the $I_D$, $T(I_D)$, and $O_D$; inconsistency can be present even if agreement is present: inconsistency in that case remains undetected, making ACE complementary to any approach tailored to detect nontrivial internal inconsistencies. Validation of late requirements is well illustrated in a recent paper from Uchitel et al. [21]. They establish the relation between scenarios and goals via "fluents that describe how the events of the operational descriptio change the state of the basic propositions from which goals are expressed." Graphical animations can be synthesized from scenarios and goal model checking over scenarios is enabled to guide animations through goal violation traces. Animations are subsequently presented to the relevant parties for discussion:

> "The users were given a view to interact with and asked to perform certain tasks. This initiated discussion as to how well the system supported their achieving their goals, and what might be changed in order to make it more effective. Suggestions about variations in the order in which activities were performed could be incorporated into the model with a few minutes work."

ACE can complement such an approach, especially when user-centered sessions involve the asynchronous participation of geographically distributed users, so that discussion via forum-like tools becomes relevant. Boehm's WinWin groupware supports negotiation of requirements via the general WinWin approach. It is usually defined as [1] "a set of principles, practices, and tools, which enable a set of interdependent stakeholders to work out a mutually satisfactory (winwin) set of shared commitments." WinWin differs from ACE in that it focuses on the negotiation context, which differs from discussions on which ACE focuses. ACE is not tailored to negotiation.

Absence of validity can reflect errors in the rationale of the decisions taken when a method is applied. The primary aim of design rationale (DR) research is to capture the *why* behind decisions in a design activity. The classical IBIS method [4] starts with a participant who posts the root *issue* of an IBIS tree. Others then post *positions* (i.e., ways of resolving issues) and *arguments* (to support or object to positions). Issues, positions, and arguments are related via some of the allowed relationships: generalize, specialize, object, support, replace, question. The process stops when consensus has been reached regarding the resolution of issues. Subsequent DR methods, as reviewed by Louridas and Loucopoulos [15], share many characteristics with IBIS. The principal novelty of ACE with regards to these other DR methods lies in the way it answers the *relationship expressivity question*: What are the relationships between concepts in the DR method, and how are they defined? At stake in this question is how to build a DR approach in the face of the multitude of potentially relevant relationships; Louridas and Loucopoulos highlight this problem in the following passage:

> "A fixed set of links [i.e., relationships] limits both the expressive and the functional capabilities of the [design rationale] model. Regarding the expressive weaknesses of any such attempt, the sheer number of the proposed relationships in the various [design rationale] approaches and the differences in their semantics from approach to approach indicate that it is difficult to arrive at a widely accepted set of predefined links. Each approach commits to a certain set, but there is no reason to believe that one of these sets is innately better than the others." ([15]: p.222–223)

This leads Louridas and Loucopoulos to leave out the definitions of the relationships in their Reasoning Loop Model, which is their synthetic proposal for reflective design. In doing so, they adopt the so-called *free-link* approach to the construction of a DR methods. This differs from the classical *fixed-link* approach, where a closed set of relationships is defined, and their meaning set once and for all. The benefit of free-links approach is that it leaves considerable freedom in use; its main downside is that it is hard to define, even manual, techniques for the analysis of its graphs. The fixed-link approach allows for the definition of analysis techniques, although the tendency to use many relationships renders the definition of alorithms for the analysis of graphs difficult. ACE adopts a third option, which has not been explored elsewhere to the best of our knowledge. ACE has a single relationship, the meaning of which are derived from the labels of vertices that it connects. An inference, conflict, and preference label does not designate a specific inference rule, conflict rule, or preference rule: one inference vertex may capture the application of modus ponens, another the application of defesible inference; one preference vertex may capture the application of a transitive preference order, while another may capture the application of an intransitive preference. Finally, and very importantly, ACE allows one to accept or reject the application of an inference, conflict, and/or preference rule, which is clearly impossible in the fixed-links approach to the construction of DR methods. If links are fixed, no meta-reasoning on them is allowed by the DR method itself. Making the links free is no better solution: what are then the criteria to accept or reject an arbitrary link? ACE is interesting because it leaves the freedom in the actual choice of an inference, conflict, or preference, while at the same time fixing how an inference, conflict, or preference relates, in terms of acceptability, to another information, inference, conflict, or preference vertex. E.g., $\mathtt{i}_1 \longrightarrow \mathtt{I} \longrightarrow \mathtt{i}_2$ tells us in terms of acceptability that $\mathtt{i}_1$ does not make $\mathtt{i}_2$ unacceptable, but is evidence to the acceptability of $\mathtt{i}_2$, and this regardless of the actual inference applied to conclude $\mathtt{i}_2$ from $\mathtt{i}_1$. By writing $\mathtt{i}_1 \longrightarrow \mathtt{I} \longrightarrow \mathtt{i}_2$, we may be abbreviating the expression "$\mathtt{i}_1$ supports $\mathrm{I}_2$", where "supports" has the same meaning as in IBIS. $\mathtt{I}$ in $\mathtt{i}_1 \longrightarrow \mathtt{I} \longrightarrow \mathtt{i}_2$ is thus locally defined (as opposed to fixed-lines approach), but we still have precise criteria for accepting or rejecting any local reading of $\mathtt{I}$ in $\mathtt{i}_1 \longrightarrow \mathtt{I} \longrightarrow \mathtt{i}_2$ (as opposed to the absence of these criteria in the free-links approach): we will reject it if the evaluation algorithm labels it **R**. Overall, ACE shows a noveal way to balance the tradeoff between freedom in use and the automated analysis of graphs, without falling into extremes of the usual fixed-links, or the recent free-links approach to the construction of DR methods.

# 9 Conclusions and Future Work

Perfectly valid RE artifacts capture *exactly* what the stakeholders *really* need. We distinguished this absolute validity from relative validity. The latter asks if the stakeholders agree on the content of an RE artifact, being thereby relative to the stakeholders. Checking relative validity inevitably leads to a discussion between the stakeholders and the requirements engineer. This paper offered the *acceptability condition* on an artifact as a proxy for relative validity, and the ACE framework for the evaluation of the acceptability condition via the analysis of discussions. If the acceptability condition holds, then this signals that the relative validity verifies for the given artifact and for the participants in a given discussion. The ACE framework incorporates a simple, but expressive language for the representation of the pieces of information exchanged in a discussion, and the inference, conflict, and preference relationships between these pieces of information. Discussions are represented via directed labeled graphs. We suggested an algorithm to retrieve subgraphs in order to inform the discussions between the participants. ACE incorporates another algorithm to evaluate the acceptability condition in these graphs. Data from actual use will expectedly open many questions regarding usability and relevance in practice.

# A Proofs

## A.1 Proof of Proposition 5.9

*Proof.* We prove Proposition 5.9 by induction on the length of a path $P$ in an ACE graph $G$. Suppose that $v \in V(G)$ and there are various paths in $G$ that end in $v$.

Take a path $P = v'' \longrightarrow v' \longrightarrow v$. Consider all possible combinations of labels on $v'$ and $v$. Of all possible combinations, twenty four are allowed by the meaning of the `To` link in an ACE graph. All twenty four are listed in Table 1. We discuss in turn the first three cases, then skip the others as they become obvious after the first three:

1. $P = v'' \longrightarrow \mathtt{I}_s(v'', \mathtt{i}_e) \longrightarrow \mathtt{i}_e$: Table 1 indicates that the inference rule application $\mathtt{I}_s(v'', \mathtt{i}_e)$ concludes $\mathtt{i}_e$. If the inference rule application is not acceptable, then its conclusion is not acceptable, so that $\mathtt{I}_s(v'', \mathtt{i}_e)$ is relevant to the acceptability of $\mathtt{i}_e$. $v''$ is the premise to the inference rule application $\mathtt{I}_s(v'', \mathtt{i}_e)$: if $v''$ is not acceptable, then the inference rule application will not be acceptable, so that $v''$ is **AC**-relevant to the inference rule application, and thereby to the conclusion of the inference rule application. Both $v''$ and $\mathtt{I}_s(v'', \mathtt{i}_e)$ are therefore relevant to the acceptability of $\mathtt{i}_e$.

2. $P = v'' \longrightarrow \mathtt{C}_s(v'', \mathtt{i}_e) \longrightarrow \mathtt{i}_e$: The conflict rule application $\mathtt{C}_s(v'', \mathtt{i}_e)$ makes $v''$ attack $\mathtt{i}_e$. If the conflict rule application is not accepted, then $\mathtt{i}_e$ will be accetable (at least as far as the current path is concerned). If instead the conflict rule application is accepted, but the dominant vertex $v''$ is not, then $\mathtt{i}_e$ still is accepted. It follows that both $v'$ and $\mathtt{C}_s(v'', \mathtt{i}_e)$ are relevant to the acceptablity of $\mathtt{i}_e$.

3. $P = v'' \longrightarrow \mathtt{P}_s(v'', \mathtt{i}_e) \longrightarrow \mathtt{i}_e$: The preference rule application $\mathtt{P}_s(v'', \mathtt{i}_e)$ makes $\mathtt{i}_e$ strictly less preferred than $v''$, that is, $\mathtt{i}_e$ is dominated by $v''$. If the preference rule application is not accepted, then $\mathtt{i}_e$ will not be dominated. If instead the conflict rule application is accepted, but the dominant vertex $v''$ is not, then $\mathtt{i}_e$ still is accepted and not dominated. Both $v'$ and $\mathtt{P}_s(v'', \mathtt{i}_e)$ are clearly relevant to the acceptablity of $\mathtt{i}_e$.

4. By applying the same reasoning as above for the remaining twenty one cases, observe that both $v''$ and $v'$ are relevant to the acceptability of $v$, and this regardless of the labels on all three vertices (and as long as the meaning of the `To` link are not violated).

From there on, we extend the path with an additional vertex $v'''$, so that $P' = v''' \longrightarrow v'' \longrightarrow v' \longrightarrow v$. If we consider separately the subpath $P^x = v''' \longrightarrow v'' \longrightarrow v'$, and study all combinations of labels on $v'''$, $v''$ and $v'$ that are in accordance with the meaning of the link `To`, we need to consider exactly the same cases as above for the initial path $P = v'' \longrightarrow v' \longrightarrow v$. For each case, the conclusion will be identical: $v'''$ and $v''$ are both relevant to the acceptability of $v'$. As $v'$ is relevant to the acceptability of $v$, we conclude that $v'''$ and $v''$ are both relevant to the acceptability of $v$. It is trivial to see that whatever the number of vertices on a path between a vertex $v^x$ and the vertex $v$, $v^x$ is relevant to the acceptability of $v$. □

## A.2 Proof of Proposition 5.16

*Proof.* We first prove that the algorithm *(i) does not loop indefinetly.* $V(G)$ and $L(G)$ are finite; the **while** loop explores only incoming lines to a given vertex, and never adds the same line or vertex twice to $D[v]$. It follows that the algorithm will never loop indefinetly.

We prove by contradiction that the algorithm *(ii) returns all direct and indirect vertices in favor of or against the starting vertex* $v$. Suppose that there is a vertex $v'$ that is either in favor or against the starting vertex $v$, and that is not visited by the algorithm. The inner **for each** loop (Lines 7–15) moves from the starting vertex along its incoming lines to its nonvisited adjacent vertices, adds these to the queue $Q$, and removes the starting vertex. The **while** loop guarantees that any vertex added to $Q$ is visited, along with its outgoing lines. The **while** loop thereby ensures that any vertex in $G$ having a path to the starting vertex is visited. If $v'$ was not visited by the algorithm, then $v'$ is not on a path that ends in $v$. It is therefore a contradiction that $v'$ is in favor or against $v$, but that it has not been found by the algorithm.

Finally, we prove that the algorithm *(iii) has the running time in* $O(|V(D[v])| + |L(D[v])|)$. The **if-then** blocks in the inner **for each** loop (Lines 7–15) guarantee that no vertex or line in $G$ will enter $Q$ more than once, and that all lines and vertices visited for the first time will be added to $D[v]$. It follows that the worst case arises when $D[v] = G$, so that the algorithm will traverse all lines and vertices of $G$, which gives the $O(|V(G)| + |L(G)|)$ as the upper bound on the time complexity, and $O(|V(D[v])| + |L(D[v])|)$ as the time complexity for the algorithm. □

## A.3 Proof of Proposition 5.27

*Proof.* We first prove that Algorithm 2 *(i) does not loop indefinetly.* The **for each** loop considers only the vertices that are in-adjacent to $v_i$. The number of vertices in $D[v]$ is finite, so that $inDegree(v_i, D[v])$ is finite. The algorithm therefore always terminates.

We now prove that Algorithm 2 *(ii) returns the overruling c-label among the c-labels propagated from all in-adjacent vertices to $v_i$ in $D[v]$.* The **for each** loop ensures that the t-set of $v_i$ receives a c-label from each vertex in $D[v]$ that is in-adjacent to $v_i$, so that all c-labels propagated from these vertices are included in the t-set. It is obvious that the last **if** block (cf., Lines 10–16) is simply a rewriting of the inference rules for multiple c-labels (cf., Definition 5.25), so that the appropriate overruling label is computed by the ComputeLabel procedure.

We finally prove that Algorithm 2 *(iii) has the running time in $O(inDegree(v_i, D[v]))$.* This is obvious, as the **for each** loop considers each vertex in $D[v]$ that is in-adjacent to $v_i$. □

## A.4 Build the Transitive Closure of a Discussion

Any discussion $D[v_i]$ can contain preference rule applications. All preference rule applications need not be the applications of the same preference rule. That is, different preference rules can be applied in a discussion. Some of these preference rules may be transitive, so that it becomes necessary to build the transitive closure $D^c[v_i]$ of these transitive preference rules on $D[v_i]$. This is accomplished via the procedure BuildTransitiveClosure in Algorithm 6. The procedure takes a discussion $D[v_i]$ and a set $P^T$. Each member of $P^T$ is a set of applications of one same transitive preference rule; e.g., $P^T = \{\{\mathtt{P}_{1,1}, \mathtt{P}_{1,2}, \ldots\}, \{\mathtt{P}_{2,1}, \mathtt{P}_{2,1}, \ldots\}, \ldots\}$, where $\{\mathtt{P}_{1,1}, \mathtt{P}_{1,2}, \ldots\}$ is the set of applications of the preference rule $\mathtt{P}_1$, $\{\mathtt{P}_{2,1}, \mathtt{P}_{2,2}, \ldots\}$ is the set of applications of the preference rule $\mathtt{P}_2$, and so on. The potential presence of more than one transitive relation in a discussion and the representation of the application of the transitive relations via preference vertices (and not direct lines between vertices) makes it impossible to reuse as-is the standard algorithms for the computation of the transitive closure of a directed graph.

BuildTransitiveClosure initialzes in two steps. The first step is to copy the vertices and lines of the $D[v_i]$ to the new graph $D^c[v_i]$, so that all original vertices and lines are kept. The second step is the **for each** loop in Lines 3–13. The loop considers each set $p_j \in P^T$ of applications of the same transitive preference rule. An empty graph $Z_j$ is created for each $p_j$. Recall that a preference rule application relates two sets of vertices, that is, any member of $p_j$ is of the form $\mathtt{P}(A, B)$, where $A$ is the set of vertices of $D[v]$, each of which is made strictly more preferred by the preference application $\mathtt{P}(A, B)$ than each of the vertices in the set $B$. I.e., any $v \in A$ is strictly more preferred than any vertex $v' \in B$. The purpose of $Z_j$ is to carry only all vertices, on which any member of $p_j$ is applied. E.g., if $\mathtt{P}(A, B) \in p_j$, then $Z_j$ will contain all vertices in $A \cup B$ and all lines from $v \in A$ to the vertex $\mathtt{P}(A, B)$, and all lines from the vertex $\mathtt{P}(A, B)$ to each vertex $v' \in B$. The iteration of the **for each** loop in Lines 5–12 on $\mathtt{P}(A, B)$ will first add to $Z_j$ (via the **for each** loop in Lines 6–8) the vertices of $A$ and lines from each vertex in $A$ to $\mathtt{P}(A, B)$, then add to $Z_j$ (via the **for each** loop in Lines 9–11) all vertices of $B$ and lines from $\mathtt{P}(A, B)$ to each vertex in $B$.

After the initialization step is performed, a set of graphs $Z_j$ is available. A graph $Z_j$ contains only all preference rule applications from $p_j$ and all vertices from $D[v]$, to which these preference rule applications were applied. Observe that a $Z_j$ need not be a connected graph. Consider the graph below.

(Ex.18)

$$\mathtt{i}_1 \longrightarrow \mathtt{P}_{1,1}(\mathtt{i}_1, \mathtt{i}_2) \longrightarrow \mathtt{i}_2 \longrightarrow \mathtt{C}_1(\mathtt{i}_2, \mathtt{i}_3) \longrightarrow \mathtt{i}_3 \longrightarrow \mathtt{P}_{1,2}(\mathtt{i}_3, \mathtt{i}_4) \longrightarrow \mathtt{i}_4$$

Suppose that the graph above is a fragment of a discussion, and $\mathtt{P}_{1,1}(\mathtt{i}_1, \mathtt{i}_2)$ and $\mathtt{P}_{1,2}(\mathtt{i}_3, \mathtt{i}_4)$ are the only applications in that discussion of the same transitive preference rule $\mathtt{P}_1$, then $p_1 = \{\mathtt{P}_{1,1}(\mathtt{i}_1, \mathtt{i}_2), \mathtt{P}_{1,2}(\mathtt{i}_2, \mathtt{i}_3)\}$. The initialization step will give $Z_1$ for $p_1$, whereby $Z_1$ is shown below.

**Algorithm 6** Build the Transitive Closure of a Discussion

**Require:** A discussion $D[v_i]$ and a set $P^T$. Each member $p_j$ of $P^T$ is the set of all applications of the same transitive preference rule;

**Ensure:** The transitive closure $D^c[v_i]$ of transitive preferences in $P^T$

**procedure** BuildTransitiveClosure($D[v_i], P^T$)
*Initialization:*
  $V(D^c[v_i]) \leftarrow V(D[v_i])$ and $L(D^c[v_i]) \leftarrow L(D[v_i])$
  **for each** set of transitive preference applications $p_j \in P^T$ **do**
    Create an empty graph $Z_j$
    **for each** $v \in p_j$ **do**
      **for each** $v'v \in L(D[v_i])$ and $v$ is a preference on $v'$ **do**
        Add $v'$ to $V(Z_j)$ and add $v'v$ to $L(Z_j)$
      **end for**
      **for each** $vv' \in L(D[v_i])$ and $v$ is a preference on $v'$ **do**
        Add $v'$ to $V(Z_j)$ and add $vv'$ to $L(Z_j)$
      **end for**
    **end for**
  **end for**

*Identify and add new lines to* $D^c[v_i]$
  **for each** $Z_j$ **do**
    **for each** connected subgraph $Z_{j,k}$ of $Z_j$ **do**
      Create an empty queue $W_{j,k}$
      Create an empty queue $X_{j,k}$
      **for each** $v \in V(Z_{j,k})$ s.t. $\nexists v'v \in L(Z_{j,k})$ **do**
        Add $v$ to $W_{j,k}$
      **end for**
      **if** $W_{j,k}$ is empty **then**
        Choose a random vertex $v$ in $Z_{j,k}$ s.t. $\exists vv' \in L(Z_{j,k})$
        Add $v$ to $W_{j,k}$
      **end if**
      **while** $W_{j,k}$ is not empty **do**
        **for each** vertex $v$ in $W_{j,k}$ **do**
          Delete $v$ from $W_{j,k}$
          **if** $v \in p_j$ **then**
            Append $v$ to $X_{j,k}$
            **for each** $v' \neq v$ in $X_{j,k}$ **do**
              **if** there is a path from $v'$ to $v$ in $Z_{j,k}$ **then**
                **for each** $v''' \in V(Z_{j,k})$ s.t. $\exists vv''' \in L(Z_{j,k})$ **do**
                  **if** $v'v''' \notin L(D^c[v_i])$ **then**
                    Add $v'v'''$ to $L(D^c[v_i])$
                  **end if**
                **end for**
              **end if**
            **end for**
          **end if**
          **for each** $v'' \in V(Z_{j,k})$ s.t. $v''$ was never before in $W_{j,k}$ and $\exists vv'' \in L(Z_{j,k})$ **do**
            Add $v''$ to $W_{j,k}$
          **end for**
        **end for**
      **end while**
    **end for**
  **end for**
**end procedure**

(Ex.19)

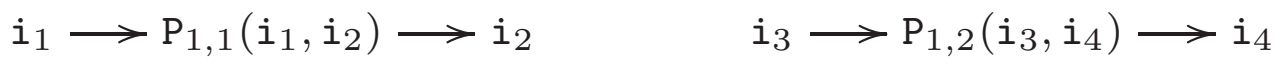

The graph $Z_1$ above is not connected and has two connected subgraphs. The **for each** loop in Lines 14–46 in Algorithm 6 considers each $Z_j$. For a given $Z_j$, each of its connected subgraphs $Z_{j,k}$ is processed in turn. For a given $Z_{j,k}$, two queues are created: $W_{j,k}$ will carry vertices that have not been traversed and thereby need to be traversed, while $X_{j,k}$ will contain preference rule appliactions from $p_j$, which were already traversed. Lines 18–24 add vertices to $W_{j,k}$: if there are vertices in $Z_{j,k}$ that have no incoming lines, then these vertices are added to $W_{j,k}$; if there are no such vertices, a random vertex with at least one outgoing line is added to $W_{j,k}$.

At Line 25 in Algorithm 6, there is at least one vertex in $W_{j,k}$ so that we can enter the **while** loop. Each vertex in $W_{j,k}$ will first be removed from $W_{j,k}$ in Line 27. The **if** block in Lines 28–39 will check if the vertex $v$ is a preference rule application from $p_j$. If so, $v$ will be added to the queue $X_{j,k}$, which contains members of $p_j$, which were already visited through the **while** loop. After $v$ is added to $X_{j,k}$, we consider each of the preference rule applications $v'$ already in $X_{j,k}$ and determine if there is a path in $Z_{j,k}$ from $v'$ to $v$. If such a path exists, a line is added in Line 34 from $v'$ to each of the vertices in $Z_{j,k}$, to which $v$ applies. For illustration, let the graph below be a $Z_{j,k}$.

(Ex.20)

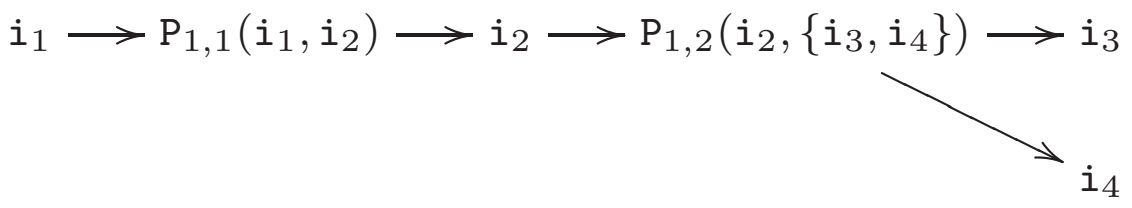

If we run the **while** loop on the graph $Z_{j,k}$ above, $\mathtt{i}_1$ will be the first vertex added to $W_{j,k}$. Given that $\mathtt{i}_1$ is not a preference rule application, the condition in Line 28 will not verify, so that the **for each** loop in Lines 40–42 will be executed, and result in adding $\mathtt{P}_{1,1}(\mathtt{i}_1, \mathtt{i}_2)$ to $W_{j,k}$. The **while** loop will then consider $\mathtt{P}_{1,1}(\mathtt{i}_1, \mathtt{i}_2)$, and the condition in Line 28 will verify. $\mathtt{P}_{1,1}(\mathtt{i}_1, \mathtt{i}_2)$ will be added to $X_{j,k}$, which is empty. The **for each** loop in Lines 40–42 will then add $\mathtt{i}_2$ to $W_{j,k}$. As for $\mathtt{i}_1$, no change will be made to $X_{j,k}$ when the **while** loop runs on $\mathtt{i}_2$. After $\mathtt{i}_2$, the **while** loop will consider $\mathtt{P}_{1,2}(\mathtt{i}_2, \{\mathtt{i}_3, \mathtt{i}_4\})$. This preference rule application will be added to $X_{j,k}$, which already contains $\mathtt{P}_{1,1}(\mathtt{i}_1, \mathtt{i}_2)$. As there is a path from $\mathtt{P}_{1,1}(\mathtt{i}_1, \mathtt{i}_2)$ to $\mathtt{P}_{1,2}(\mathtt{i}_2, \{\mathtt{i}_3, \mathtt{i}_4\})$, the condition in Line 31 will verify, and two lines will be added to $D^c[v_i]$: $\mathtt{P}_{1,1}(\mathtt{i}_1, \mathtt{i}_2)\mathtt{i}_3$ and $\mathtt{P}_{1,1}(\mathtt{i}_1, \mathtt{i}_2)\mathtt{i}_4$, as shown in the graph below.

(Ex.21)

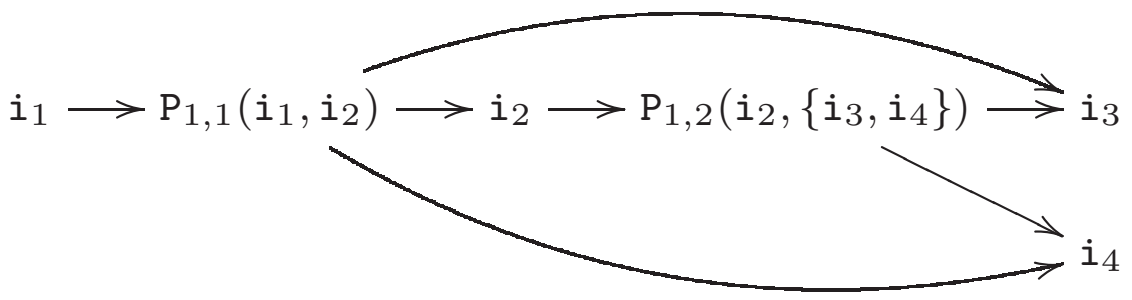

**Proposition A.1.** *The procedure* BuildTransitiveClosure *applied to a discussion $D[v_i]$ and a set $P^T$ (whereby each member of $P^T$ is a set of applications of one same transitive preference rule) (i) does not loop indefinetly, (ii) returns the transitive closure $D^c[v_i]$ of all transitive preference rule applications $P^T$ on $D[v_i]$, and (iii) has the running time in $O(|V(D^c[v_i])| + |L(D^c[v_i])|)$.*

*Proof.* We first prove that BuildTransitiveClosure *(i) does not loop indefinetly.* The **for each** loop in Lines 3–13 considers each $p_j \in P^T$. As the number of applications of transitive preference rules is below the number of vertices in $D[v_i]$, and $D[v_i]$ is finite, the first **for each** loop always terminates. The **for each** loop in Lines 14–46 considers each $Z_j$, the number of which equals to $|P^T|$, which is finite. The **for each** loop in Lines 18–20 always terminates, as $|V(Z_{j,k})|$ is finite. The **for each** loop in Lines 32–36 also always terminates because $|V(Z_{j,k})|$ is finite. $X_{j,k}$ is also finite, so that the **for each** loop in Lines 30–38 always terminates. The **for each** loops in Lines 15–45 and 14-46 will always terminate only if the **while** loop always terminates. The latter stops when $W_{j,k}$ empties. $W_{j,k}$ will empty after all vertices in $Z_{j,k}$ were visited once. The procedure BuildTransitiveClosure therefore does not loops indefinetly.

We now prove that BuildTransitiveClosure *(ii) returns the transitive closure $D^c[v_i]$ of all transitive preference rule applications $P^T$ on $D[v_i]$*. We prove this by contradiction. Remark first that all applications of transitive preference rules are input to the procedure via the set $P^T$. Each member $p_j$ of that set is a set of vertices in $D[v_i]$, and each of these vertices is a different application of one same transitive preference rule. We therefore know from the outset the preference rules that are transitive and their applications in $D[c_i]$. Suppose now that $D^c[v_i]$ is returned by BuildTransitiveClosure($D[v_i], P^T$), and that it is missing a line $\mathtt{P}_{1,1}v$ from a transitive preference rule application $\mathtt{P}_{1,1}$ to the vertex $v$, as shown in the graph below. We thus assume that the graph below is a subgraph of $D^c[v_i]$, and that there are no lines in $D^c[v_i]$ between the vertices shown below other than the lines shown below.

(Ex.22)

$$v'' \longrightarrow \mathtt{P}_{1,1}(v'', v') \longrightarrow v' \longrightarrow \mathtt{P}_{1,2}(v', v) \longrightarrow v$$

There are three ways for the procedure to miss to add $\mathtt{P}_{1,1}v$ to $D^c[v_i]$:

1. At initialization, the **for each** loop in Lines 3–13 does not add the line $v'\mathtt{P}_{1,2}$ to the relevant $Z_j$. This could only happen if that line was not in $L(D[v_i])$, which contradicts the premise that the line is in fact in $D[v_i]$.

2. At initialization, the **for each** loop in Lines 3–13 places $\mathtt{P}_{1,1}$ in some $Z_a$ and $\mathtt{P}_{1,2}$ in some $Z_b$, with $Z_a \neq Z_b$. This is only possible if $\mathtt{P}_{1,1}$ and $\mathtt{P}_{1,2}$ were in two different members $p_j$ of $P^T$. If that was the case, then $\mathtt{P}_{1,1}$ and $\mathtt{P}_{1,2}$ are applications of two different transitive preference rules, and the line $\mathtt{P}_{1,1}v$ should not be in $D^c[v_i]$ anyway.

3. Let $\mathtt{P}_{1,1}$ and $\mathtt{P}_{1,2}$ be in $V(Z_{1,1})$. If there is no path in $Z_{1,1}$ from $\mathtt{P}_{1,1}$ to $\mathtt{P}_{1,2}$, then the **while** loop will not add the line $\mathtt{P}_{1,1}v$ to $D^c[v_i]$. That path is absent is $\mathtt{P}_{1,1}$ and $\mathtt{P}_{1,2}$ are in two disconnected subgraphs of $Z_{1,1}$, which is a contradiction, as $Z_{1,1}$ is by definition a connected subgraph of $Z_1$ (cf., Line 15). That path can also be absent if $Z_{1,1}$ is a connected graph. This, however, is again a contradiction, as we assumed that a a path from $\mathtt{P}_{1,1}$ to $\mathtt{P}_{1,2}$ is present in $D[v_i]$.

We conclude that BuildTransitiveClosure returns the transitive closure $D^c[v_i]$ of all transitive preference rule applications $P^T$ on $D[v_i]$.

We finally prove that BuildTransitiveClosure *(iii) has the running time in $O(V(D^c[v]) + L(D^c[v]))$*. The upper bound on running time in the initialization step is $O(|V(D[v_i])| + |L(D[v_i])|)$. This is a pessimistic estimate, as it assumes that all vertices in $D[v_i]$ are linked to at least one application of a transitive preference rule. The pessimistic estimate for the main **for each** loop is $O(|V(D^c[v_i])| + |L(D^c[v_i])|)$, because it will consider each $Z_j$, and each $Z_{j,k}$ only once, and the **while** loop will visit each vertex in $Z_{j,k}$ only once. We conclude that $O(|V(D^c[v_i])| + |L(D^c[v_i])|)$ is the upper bound on the running time of BuildTransitiveClosure. □

## A.5 Proof of Proposition 5.29

*Proof.* The proof of Proposition 5.29 is a combination of propositions on the following procedures called in Algorithm 3:

- Proposition A.1 on the BuildTransitiveClosure procedure, which is called on a given discussion $D[v]$ and given a set of transitive preferences $P^T$;

- Proposition A.2 on the EnumerateSCC procedure, which is called on the transitive closure $D^c[v]$ of a discussion $D[v]$;

- Proposition A.3 on the ContractSSC procedure, which is called on the set $\mathcal{C}$ of strongly connected components of $D^c[v]$;

- Proposition A.4 on the TopologicalSort procedure, which is called on the directed acyclic graph $D_C$, in which each vertex represents a strongly connected component from $\mathcal{C}$;

- Proposition A.5 on the ExpandSCC procedure, which is called on the topological sort $\mathcal{S}_C$ of the vertices in $D_C$; and
- Proposition A.6 on the LabelSCC procedure, which is called on $D^c[v]$ and the topological sort of the strongly connected components of $D^c[v]$.

We first prove that Agorithm 3 applied to a discusssion $D[v]$ *(i) does not loop indefinetly.* Given that the procedures BuildTransitiveClosure, EnumerateSCC, ContractSCC, TopologicalSort, ExpandSCC, and LabelSCC all always terminate (cf., respectively, Propositions A.1, A.2, A.3, A.4, A.5, and A.6) when EvaluateDiscussion is called on a finite $D[v]$ and a finite $P^T$, EvaluateDiscussion always terminates if it is called on a finite $D[v]$ and a finite $P^T$.

The proof that Agorithm 3 applied to a discusssion $D[v]$ *(ii) returns stable c-labels for all vertices in* $D[v]$ *if they exist, an error otherwise* follows trivially from Propositions A.1, A.2, A.3, A.4, A.5, and A.6.

Finally, we prove that Agorithm 3 applied to a discusssion $D[v]$ *has the running time in* $O(C(D^c[v])(|L(D^c[v])|+2|V(D^c[v])|))$*, where* $C(D^c[v])$ *is the number of simple cycles in* $D^c[v]$. This follows trivially from Propositions A.1, A.2, A.3, A.4, A.5, and A.6, as $O(C(D^c[v])(|L(D^c[v])| + 2|V(D^c[v])|))$ is the worst of the running times of the procedures called by the procedure EvaluateDiscussion in Algorithm 3. □

**Proposition A.2.** *The procedure* EnumerateSCC *applied to the transitive closure* $D^c[v]$ *of a discussion* $D[v]$ *(i) does not loop indefinetly, (ii) returns the set of all strongly connected components of* $D^c[v]$*, and (iii) has the running time in* $O(V(D^c[v]) + L(D^c[v]))$*.*

*Proof.* We do not discuss the detail of the proof as Tarjan's strongly connected components algorithm [19] can be reused as is, or an improved variant thereof can be employed instead (cf., e.g., [18]). □

**Proposition A.3.** *The procedure* ContractSCC *applied to the set* $\mathcal{C}$ *of strongly connected components of* $D^c[v]$ *(i) does not loop indefinetly, (ii) returns an acyclic directed graph* $D_C$*, in which each vertex represents exactly one strongly connected component from* $\mathcal{C}$ *and* $D_C$ *corresponds to the graph that would be obtained by contracting down to a single vertex each strongly connected component in* $D^c[v]$*, and (iii) has the running time in* $O(|\mathcal{C}|(|\mathcal{C}| - 1)|L(D^c[v])|)$*.*

*Proof.* ContractSCC always terminates, as the number of strongly connected components is finite, and the number of vertices in each of these components is finite.

We now prove that ContractSCC *(ii) returns an acyclic directed graph* $D_C$*, in which each vertex represents exactly one strongly connected component from* $\mathcal{C}$ *and* $D_C$ *corresponds to the graph that would be obtained by contracting down to a single vertex each strongly connected component in* $D^c[v]$. The procedure starts by adding as many vertices to $D_C$ as there are strongly connected components in $\mathcal{C}$. As the rest of the procedure does not add or remove vertices, $|V(D_C)|$ equals the number of elements in $\mathcal{C}$. The first **for each** loop will relate, via the function *standsFor* each vertex in $D_C$ to exactly one strongly connected component in $\mathcal{C}$. The second **for each** loop, and its first inner **for each** loop will together consider each pair of vertices in $D_C$. For each such pair, the innermost **for each** loop (cf., Lines 17–21 in Algorithm 3) will add to $D_C$ a line between these two vertices only if there is a line in $D^c[v]$ between these two vertices, and in the relevant direction. The **for each** loops in Lines 15 and 16 together with the *standsFor* ensure that all lines in $D^c[v]$, which are not within strongly connected components, will be considered. For each such line between two strongly connected components, a line will be added to $D_C$ between the vertices that represent the two strongly connected components. It follows that $D_C$ corresponds to the graph that would be obtained by contracting down to a single vertex each strongly connected component in $D^c[v]$.

We finally prove that ContractSCC *(iii) has the running time in* $O(|\mathcal{C}|(|\mathcal{C}|-1)|L(D^c[v])|)$. The first **for each** loop has $|\mathcal{C}|$ cases to consider. The second **for each** loop will consider, for each of the $|\mathcal{C}|$ cases, $|\mathcal{C}|-1$ other case, then at most $|L(c)|$ lines. The upper bound on the number of operations that the procedure will perform is thus $|\mathcal{C}| + |\mathcal{C}|(|\mathcal{C}| - 1)|L(D^c[v]|)$. It follows that the upper bound on running time is in $O(|\mathcal{C}|(|\mathcal{C}| - 1)|L(D^c[v])|)$. □

**Proposition A.4.** *The procedure* TopologicalSort *applied to the directed acyclic graph* $D_C$ *(i) does not loop indefinetly, (ii) returns the topological sort of the vertices of* $D_C$*, and (iii) has the running time in* $O(V(D_C) + L(D_C))$*.*

*Proof.* We do not discuss the detail of the proof as Tarjan's depth-first search algorithm [20] can be reused as is. □

**Proposition A.5.** *The procedure* EXPANDSCC *applied to the sequence* $\mathcal{S}_C$ *(i) does not loop indefinetly, (ii) returns the sequence* $\mathcal{S}$*, which is the topological sort of the strongly connected components of* $D^c[v]$*, and (iii) has the running time in* $O(|\mathcal{C}|)$.

*Proof.* $\mathcal{S}_C$ has $|V(D_C)|$ elements, and procedure CONTRACTSCC makes $V(D_C)$ equal to the number of strongly connected components in $D^c[v]$. As $D^c[v]$ has a finite number of strongly connected components, EXPANDSCC alsways terminates.

It is trivial to see that $\mathcal{S}$, returned by EXPANDSCC is the topological sort of the strongly connected components of $D^c[v]$. EXPANDSCC first makes $\mathcal{S}$ the replica of $\mathcal{S}_C$, then replaces via the function *standsFor* (defined by CONTRACTSCC) each element of $\mathcal{S}$ by the corresponding strongly connected component of $D^c[v]$.

Finally, the **for each** loop considers each element in $\mathcal{S}$. The number of elements in $\mathcal{S}$ equals the number of strongly connected components, so that EXPANDSCC has the running time in $O(\mathcal{C})$ □

**Proposition A.6.** *The procedure* LABELSCC *applied to the transitive closure* $D^c[v]$ *of a discussion* $D[v]$ *and the topological sort* $\mathcal{S}$ *of all strongly connected components of* $D^c[v]$ *(i) does not loop indefinetly, (ii) if there are stable c-labels for all vertices in* $D[v]$*, then a function* $\Lambda : V(D[v]) \longrightarrow \{\mathsf{A}, \mathsf{AD}, \mathsf{R}\}$*, which returns the stable c-label for each vertex in* $D[v]$*; if there are no stable c-labels, then an error, and (iii) has the running time in* $O(C(D^c[v])(|L(D^c[v])| + 2|V(D^c[v])|))$*, where* $C(D^c[v])$ *is the number of simple cycles in* $D^c[v]$.

*Proof.* It is straightforward to see that the **while** loop will terminate for any finite $\mathcal{S}$ and finite $D^c[v]$, since it considers each strongly connected component only once. For each strongly connected component, the **while** loop will call either COMPUTELABEL or LABELCOMPLEXSCC. Proposition 5.27 indicates that COMPUTELABEL always terminates. Proposition A.7 indicates that LABELCOMPLEXSCC always terminates. Consequently, LABELSCC will never loop indefinetly, provided that $D^c[v]$ is finite.

We now prove that *if there are stable c-labels for all vertices in* $D[v]$*, then* LABELSCC *returns a function* $\Lambda : V(D[v]) \longrightarrow \{\mathsf{A}, \mathsf{AD}, \mathsf{R}\}$*, which returns the stable c-label for each vertex in* $D[v]$*; if there are no stable c-labels, then an error.* The **while** loop distinguishes simple from complex strongly connected components. A strongly connected component $c$ is simple if $|V(c)| = 1$, and complex otherwise. Given that the **while** loop considers each $c$ according to the topological sort of strongly connected components of $D^c[v]$, the procedure COMPUTELABEL will be called on a simple $c$ only after all strongly connected components that precede it obtained stable c-labels for all of their vertices. Consequently, all vertices that are relevant for the acceptability of the vertex in the simple $c$ already have stable c-labels when COMPUTELABEL is called on a simple $c$. Consequently, if COMPUTELABEL does not return an error when called on a simple $c$, then the only vertex in $c$ will obtain its stable c-label. When the **while** loop encounters a complex $c$, it will call LABELCOMPLEXSCC. Proposition A.7 indicates that LABELCOMPLEXSCC will return at least two c-labels in the c-sequence of each vertex in $c$, provided that there are stable c-labels for the vertices in $c$, whereby the last c-label of the c-sequence of each vertex in $c$ is its stable c-label. If the calls of COMPUTELABEL and LABELCOMPLEXSCC do not return an error, and after all strongly connected components obtained their c-labels, the function $\Lambda : V(D[v]) \longrightarrow \{\mathsf{A}, \mathsf{AD}, \mathsf{R}\}$ is defined in the **for each** loop in Lines 18–20.

Finally, we prove that LABELSCC *has the running time in* $O(C(D^c[v])(|L(D^c[v])| + 2|V(D^c[v])|))$*, where* $C(D^c[v])$ *is the number of simple cycles in* $D^c[v]$. COMPUTELABEL$(v_j, D^c[v])$ is in $O(\mathit{inDegree}(v_j, D^c[v]))$ (cf., Proposition 5.27), and the running time of LABELCOMPLEXSCC$(c_i, v_j)$ is $O((|L(c_i)| + |V(c_i)|)(C(c_i) + 1) + C(c_i)|V(c_i)|)$ (cf., Proposition A.7). The running time will be governed by the labeling of complex strongly connected components of $D^c[v]$, so that the running time of LABELSCC will be in:

$$O\left(\sum_{c_i \text{ s.t. } |c_i|>1} [(|L(c_i)| + |V(c_i)|)(C(c_i) + 1) + C(c_i)|V(c_i)|]\right)$$

In the worst case, every strongly connected component in $D^c[v]$ is complex. It follows that the upper bound on the time complexity for LABELSCC is $O(C(D^c[v])(|L(D^c[v])| + 2|V(D^c[v])|))$, where $C(D^c[v])$ is the number of simple cycles in $D^c[v]$. □

**Proposition A.7.** *The procedure* LabelComplexSCC *applied to a complex strongly connected component c (i) does not loop indefinetly, (ii) if there are stable c-labels for all vertices in c, then at least two c-labels in the c-sequence of each vertex in c, with the last c-label of the c-sequence of each vertex being its stable c-label, an error if there are no stable c-labels, and (iii) has the running time in* $O((|L(c)|+|V(c)|)(C(c)+1)+C(c)|V(c)|)$, *where* $V(c)$ *is the number of vertices,* $L(c)$ *the number of lines, and* $C(c)$ *the number of simple cycles in the strongly connected component c of the transitive closure* $D^c[v]$ *of the discussion* $D[v]$.

*Proof.* We first prove that LabelComplexSCC *(i) does not loop indefinetly.* The procedure terminates as soon as $Q$ empties. There are three ways to empty $Q$, so that we consider three cases:

1. ComputeLabel *returns an error.* This occurs if *propagateLabel* encounters a disallowed input (cf., Line 7 in Algorithm 2). Lines 34–36 in Algorithm 5 ensure that $Q$ empties if ComputeLabel returns an error.

2. *The last two c-labels on the c-sequence of First are identical (cf., Line 27 in Algorithm 5).* This case is possible because LabelComplexSCC will traverse *First* at least once after initialization, and thereby add a second c-label. Condition in Line 27 verifies and $Q$ empties, so that the procedure terminates. If the last two c-labels in the c-sequence of *First* are not identical, then the next case applies.

3. *The last two c-labels in the c-sequence of First are not identical (Line 30).* The procedure will not empty $Q$ until $p = 4$. $Q$ will therefore be emptied after a finite number of traversals of $c$, so that the procedure cannot loop indefinetly.

We now prove that the procedure LabelComplexSCC *(ii) if there are stable c-labels for all vertices in c, then at least two c-labels in the c-sequence of each vertex in c, with the last c-label of the c-sequence of each vertex being its stable c-label, an error if there are no stable c-labels.* Three cases must be considered:

1. *c contains a disallowed graph structure.* The procedure detects a disallowed graph structure via the procedure ComputeLabel. Lines 33–35 in Algorithm 5 ensure that the procedure returns an error when the ComputeLabel procedure returns an error. ComputeLabel returns an error if the function *propagateLabel* returns an error, that is, when *propagateLabel* is given an input other than those listed in Table 1. If $c$ contains a disallowed graph structure, then there is no interest in traversing the graph any further. The error must be resolved before the labeling is attempted again.

2. *The last two c-labels on the c-sequence of First are identical (Line 27 in Algorithm 5).* We must prove that *the labels returned after the condition in Line 28 verifies indeed are the stable labels for all vertices in c*. We prove this in three steps:

   (a) First, we prove that *if there are two c-labels in the c-sequence of First, then each vertex in c has at least two c-labels in its c-sequence.* We prove this by contradiction. When the procedure initializes, it adds the label **A** to the c-sequence of each vertex in $c$ (cf., Line 5 in Algorithm 5). The procedure adds *First* to the queue $Q$, and enters the **while** loop. The inner **for each** loop (cf., Lines 12–36) adds c-labels to all vertices on lines that start in *First* and adds these vertices to $Q$. The outer **for each** loop (cf., Lines 10–37) guarantees that each of the vertices in $Q$ will be traversed by the inner **for each** loop. The only way for some vertex $v^x$ not to obtain a second c-label in its c-sequence before *First* obtains its second c-label is if the inner **for each** loop misses $v^x$, which can only occur if there are no lines in $c$ that end in $v^x$. We know that, in any strongly connected component, any vertex is on at least one cycle. It is consequently a contradiction for $c$ to be a strongly connected component *and* have at least one vertex $v^x$ that has one c-label in its c-sequence when *First* has two c-labels in its c-sequence. LabelComplexSCC therefore ensures that all vertices have at least two c-labels if *First* has two labels.[5]

[5] It may be relevant to highlight why we need at least two c-labels in the c-sequence of each vertex after *First* has two c-labels. During initialization, the procedure assigns the c-label **A** to all vertices. **A** is chosen because it is the weakest label, being overruled by both **AD** and **R** (cf., Definition 5.25). If $Q$ empties and only a single c-label is assigned to some vertices, then we would be assuming that the stable c-label for those vertices is the label **A**, which was assigned when the procedure initialized. We clearly cannot assume that the first c-label in a c-sequence is the stable c-label, because the it is added without considering the c-labels on vertices adjecent via incoming lines.

(b) It follows from the above that the second c-label is added to the c-sequence of *First* after all vertices and lines in $c$ were visited at least once. The next, third c-label is added to the c-sequence of *First*, as soon as the procedure traverses all other vertices and all lines in $c$ at least once after the second c-label was added to the c-sequence of *First*. And so on. In other words, $p$ increments (cf., Line 23 in Algorithm 5) exactly when the procedure returns to *First*, having traversed all other vertices and all lines in $c$ at least once. It is apparent that *if there are stable labels in c and are discovered as soon as the the pth c-label is added to the c-sequence of First, then the c-labels returned by the procedure for all vertices when the pth c-label is added to the c-sequence of First must be identical to all c-labels returned by the procedure when the* $(p+i)$*th c-label is added to the c-sequence of First, where* $i > 0$ *is a positive integer.* We prove this claim by induction on $i$ in $p+i$ and the number $C(c)$ of simple cycles in $c$.

A preliminary remark is in order: The procedure LabelComplexSCC will return a c-label for each vertex as soon as the $(p-1)$th and $p$th c-labels in the c-sequence of *First* are identical. In order to add the $(p+1)$th label when $(p-1)$th and $p$th c-labels are identical, we will assume that Lines 28–29 do not execute when the $p$th label is added, which is to say that *First* will remain in $Q$ when the $p$th label is added.

i. Suppose that $C(c) = 1$, i.e., $c$ has a single simple cycle. (Note that, since $c$ is a strongly connected component, it must have at least one cycle, so that $c$ cannot have CountSimpleCycles$(c) = 0$.) If we assume $C(c) = 1$, all of $c$'s vertices are on that simple cycle (i.e., $c$ has a directed Hamiltonian cycle[6]). There is consequently exactly one *simple path* from any vertex in $c$ to any other vertex, and of course, exactly one simple cycle that starts and ends in a vertex. Equivalently, there is in $c$ exactly one line, say $v'v$, that ends in any one vertex, say $v$. Since there are three possible c-labels **A**, **AD**, and **R** on any $v'$, the result of *propagateLabel*$(v', v)$ will be any one of these three labels. We can picture $c$ as shown below:

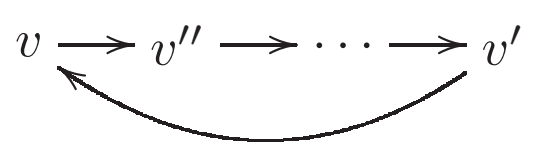

Given that all vertices are on the same simple cycle, and all vertices other than *First* must be traversed at least once (cf., Point 2.a above) before LabelComplexSCC returns to *First*, we know that as soon as *First* obtains its $p$th c-label, all other vertices will have $p$ c-labels in their c-sequences. Suppose that $v = First$ in the graph above, and its $(p-1)$th and $p$th labels are identical, and observe the following:

A. *propagateLabel*$(v, v'')$ when $v$ has $p$ labels is identical to *propagateLabel*$(v, v'')$ when $v$ has $p+1$ labels in its c-sequence; *propagateLabel*$(v'', \ldots)$ when $v''$ has $p$ labels is identical to *propagateLabel*$(v'', \ldots)$ when $v''$ has $(p+1)$ labels in its c-sequence; and so on. It ensues that *propagateLabel*$(v', v)$ when $v'$ has $p$ labels is identical to *propagateLabel*$(v', v)$ when $v$ has $p+1$ labels in its c-sequence. Now, a vertex in $c$ may have other lines that end in that vertex, but which are in $D^c[v]$, but not in $c$. They do not influence the observations here, beccause the vertices outside $c$, in which these lines originate, have obtained stable c-labels before $c$ is submitted to LabelComplexSCC. This is true by definition of the procedure EvaluateDiscussion, as these vertices outside $c$ are in strongly connected components other than $c$; such strongly connected components precede $c$ in the topological sort of the strongly connected components of $D^c[v]$.

B. We obtain the exact same conclusions as above if we replace $(p+1)$ with $(p+2)$ and $p$ with $(p+1)$. It is trivial to see that we will obtain the same conclusions when we replace $(p+i+1)$ with $(p+i+2)$ and $(p+i)$ with $(p+i+1)$, for any positive integer $i > 0$.

It follows then that if $c$ has exactly one simple cycle and has stable c-labels, and they are discovered as soon as the the $p$th c-label is added to the c-sequence of *First*, then the c-labels returned by the procedure LabelComplexSCC for all vertices when the $p$th c-label is added to the c-sequence of *First* must be identical to all c-labels returned by the procedure when the $(p+i)$th c-label is added to the c-sequence of *First*, where $i > 0$ is a positive integer.

[6] A directed Hamiltonian cycle is a cycle in a directed graph that traverses each vertex in the graph exactly once, except for the vertex in which the cycle starts and ends (which is visited twice).

ii. $C(c) = 2$, i.e., $c$ has two simple cycles. In order to have the two distinct simple cycles, there must be exactly one vertex, say $v^{out}$ in $c$ such that there are exactly two lines that start in $v^{out}$, and there must be exactly one other vertex, say $v^{in}$ in $c$ and exactly two other lines that end in $v^{in}$. $c$ therefore has some variant of the structure shown below:[7]

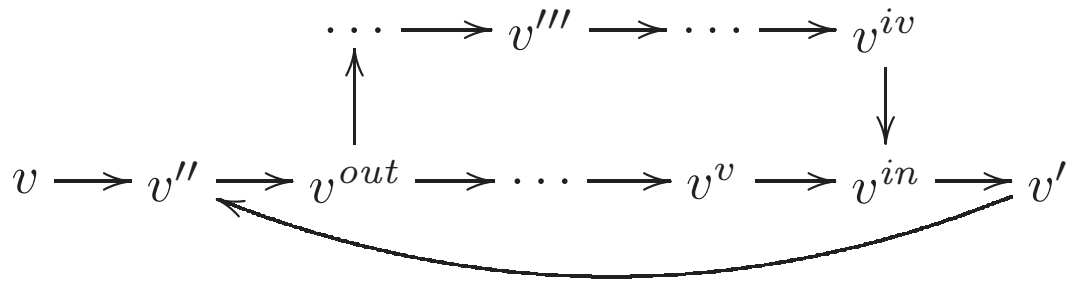

Suppose that $v = First$. To add the $(p+1)$th c-label to the c-sequence of *First*, the procedure must traverse at least once all other vertices in $c$. This is ensured by the **if–then** block in Lines 14–25. To see how, suppose that there are more vertices on the path from $v^{out}$ to $v^{in}$ that does not pass through $v'''$, than there are on the path from $v^{out}$ to $v^{in}$ that does pass through $v'''$. When the procedure reaches $v^{out}$, it will consider separately each branch outgoing from $v^{out}$. We can picture this in the following way. When the procedure computes the number of simple cycles, it obtains the upper bound on the number of simple paths from any vertex to any other vertex in $c$. As we have two simple cycles in the graph shown above, we shall place two walkers X and Y at the vertex *First* and assume that each walker takes equal time to traverse a vertex and can only go forward in steps of one vertex. The two walkers will reach $v^{out}$ at the same time, and take different paths along the branches from $v^{out}$. One, say X is thus sent along the shorter path and the other Y along the longer path. Clearly, X will reach $v'$ before Y. Once X reaches $v'$, its next step is *First*. We do not allow X to add a label to the c-sequence of *First* because some of the information in the graph, i.e., $c$, is not taken into account. Namely, when X entered $v^{in}$, it computed the c-label on $v^{in}$ by taking into account the last c-label in the c-sequences of the vertices $v^{iv}$ and $v^{v}$. This can be seen directly from the ComputeLabel procedure. At that same moment, Y was somewhere on the path that ends in $v^{in}$. Since Y had yet some vertices to traverse, and thereby their new c-labels to compute, Y will add a label to $v^{iv}$. Since that c-label may be different from that assumed by X when X labeled $v^{in}$, Y may label $v^{in}$ with a different c-label when it gets to $v^{in}$. The procedure is designed so that it waits for all walkers to arrive to *First* before it appends a new c-label to the c-sequence of *First*. Condition in Line 16 verifies when a walker reaches *First*. Condition in Line 17 verifies if that walker is not the last of the $C$ walkers (i.e., some are still on their way to *First*); if the considered walker is the $C(c)$th walker, then we can compute the new c-label on *First*, append its c-sequence, then add *First* to $Q$ in Line 26 (and thereby place new $C(c)$ walkers at the vertex *First*). Suppose then that we send $C(c)$ walkers from *First* when *First* has $p-1$ c-labels in its c-sequence and this yields a $p$th c-label in the c-sequence of *First* identical to the $(p-1)$th label. Since the $C(c)$ walkers traverse the graph in the exact same way to add the $(p-1)$th c-label as they do to add the $p$th c-label to the c-sequence of *First*, they will proceed in the same way when adding the $(p+1)$th label to the c-sequence of *First*. Because the ComputeLabel procedure will be applied on the same pairs of vertices and in the same sequence by the different walkers, i.e., the traversal will happen in the exact same manner, we conclude that if the procedure returns the same c-labels for all vertices when $(p-1)$th and $p$th labels in the c-sequence of *First* are identical, then when the algorithm traverses *First* for the $(p+1)$th time, it will return same c-labels for all vertices as it did when it traversed *First* for the $p$th time. The same conclusion can be drawn if we replace $(p+1)$ with $(p+2)$, $p$ with $(p+1)$, and $(p-1)$ with $p$. Finally, it is trivial to see that we will obtain the same conclusions when we replace $(p+i+1)$ with $(p+i+2)$, $(p+i)$ with $(p+i+1)$, and $(p+i-1)$ with $(p+i)$ for any positive integer $i > 0$.

It follows then that if $c$ has exactly two simple cycles and has stable c-labels, and they are discovered as soon as the the $p$th label is added to the c-sequence of *First*, then the c-labels returned by the procedure for all vertices, when the $p$th c-label is added to the c-sequence of

[7] The discussion applies to any variant of the shown structure, i.e., any $c$ that has exactly two simple cycles.

*First*, must be identical to all c-labels returned by the procedure when the $(p+i)$th c-label is added to the c-sequence of *First*, where $i > 0$ is a positive integer.

iii. $C(c) = n$ with $n > 2$, i.e., $c$ has more than two simple cycles. When there are $n$ simple cycles, $n$ walkers traverse the graph, and stop before a new c-label is appended to the c-sequence of *First*. Suppose that *First* has $p$ c-labels in its c-sequence, that the $(p-1)$th and $p$th c-labels are identical, and that the $C$ walkers are sent from *First* in order to obtain the $(p+1)$th label in the c-sequence of *First*. Suppose finally that the $(p+1)$th and $p$th label on *First* are identical, but that among the labels returned by the procedure when the $(p+1)$th is added to *First*, there is one c-label returned for some vertex other than *First* that is different from the c-label returned for that same vertex when the $p$th label was added to *First*. This can only be the case if at least one of the walkers did not take the same path at $p+1$ as it did at $p$, which in turn can only be the case if the path was modified. In order to modify the path, one needs to change the vertices and/or lines and/or ($\lambda_V$-)labels on vertices after the last traversal. It is therefore a contradiction to have the assumed situation when the $C(c)$ walkers are traversing the exact same graph and adding the $p$th and the $(p+1)$th c-label to the c-sequence of *First*. Observe that the given conclusion is independent of the value of $C(c)$. It follows that if $c$ has exactly $C(c)$ simple cycles and has stable c-labels, and they are discovered as soon as the the $p$th label is added to the c-sequence of *First* (hence, *First*'s $(p-1)$th and $p$th labels are identical), then the c-labels returned by the procedure for all vertices when the $p$th label is added to the c-sequence of *First* must be identical to all c-labels returned by the algorithm when the $(p+i)$th label is added to the c-sequence of *First*, where $i > 0$ is a positive integer.

3. *The last two c-labels on the c-sequence of First are different (Line 30 in Algorithm 5).* We must prove that *no stable c-labels can be found for vertices in c when condition in Line 30 verifies.* We prove this by induction on $p$.

   (a) $p = 2$*, i.e., the vertex First has two c-labels in its c-sequence.* We have shown in Point 2 above that two same c-labels in the c-sequence of *First* indicate that stable c-labels have been found for all vertices in $c$. We therefore suppose that the $(p-1)$th and $p$th c-labels in the c-sequence of *First* are different. Since we set $p = 2$ here, the first c-label is added at the initialization of the procedure and is by default **A**. As the second c-label must be different from **A**, it is either **AD** or **R**. The first column in Table 3 indicates these two possible combinations. That table then lists all possible c-labels that will be added after the procedure traverses *First* for the second and third time. The point with Table 3 is that patterns can be spotted with certainty as soon as the fourth c-label is added to the c-sequence of *First*. These patterns allow us to say whether there are stable labels in $g$. We see that one traversal is not conclusive: the first and second label differ and we cannot anticipate on this alone what the third label will be. We therefore perform an additional traversal of *First*, which leads to the following case.

**Table 3:** Possible and allowed c-sequence of the vertex *First* in a strongly connected component $c$. The procedure LabelComplexSCC ensures that the c-labels assigned at each traversal of *First* depend on the c-label added in the last previous traversal.

| Case: | Assumed c-sequence of *First* after LabelComplexSCC traverses for the first time the vertex *First*: | Possible and allowed c-sequences of *First* after LabelComplexSCC traverses for the second time the vertex *First* and given the c-sequence in the second column: | Possible and allowed c-sequence of *First* after LabelComplexSCC traverses for the third time the vertex *First* and given the c-sequence in the third column: |
|---|---|---|---|
| 1 | ⟨**A**, **R**⟩ | ⟨**A**, **R**, **A**⟩ | ⟨**A**, **R**, **A**, **R**⟩ |
| 2.1 | ⟨**A**, **R**⟩ | ⟨**A**, **R**, **AD**⟩ | ⟨**A**, **R**, **AD**, **A**⟩ |
| 2.2 | ⟨**A**, **R**⟩ | ⟨**A**, **R**, **AD**⟩ | ⟨**A**, **R**, **AD**, **AD**⟩ |
| 2.3 | ⟨**A**, **R**⟩ | ⟨**A**, **R**, **AD**⟩ | ⟨**A**, **R**, **AD**, **R**⟩ |
| 3 | ⟨**A**, **R**⟩ | ⟨**A**, **R**, **R**⟩ | ⟨**A**, **R**, **R**, **R**⟩ |
| 4 | ⟨**A**, **AD**⟩ | ⟨**A**, **AD**, **A**⟩ | ⟨**A**, **AD**, **A**, **AD**⟩ |
| 5 | ⟨**A**, **AD**⟩ | ⟨**A**, **AD**, **AD**⟩ | ⟨**A**, **AD**, **AD**, **AD**⟩ |
| 6.1 | ⟨**A**, **AD**⟩ | ⟨**A**, **AD**, **R**⟩ | ⟨**A**, **AD**, **R**, **A**⟩ |
| 6.2 | ⟨**A**, **AD**⟩ | ⟨**A**, **AD**, **R**⟩ | ⟨**A**, **AD**, **R**, **AD**⟩ |
| 6.3 | ⟨**A**, **AD**⟩ | ⟨**A**, **AD**, **R**⟩ | ⟨**A**, **AD**, **R**, **R**⟩ |

(b) $p = 3$*, i.e., the vertex First has three c-labels in its c-sequence.* The third column lists the possible combinations of c-labels when there are three c-labels in the c-sequence of *First*. These are the allowed cases when we assume that the first two c-labels are different. We see immediately the pattern in case 1: the traversal that started from **A** gave **R** and the second traversal gave **A**. The third traversal can only yield **R**. The only way for this not to happen is to have at least one different path to *First* different from any path traversed when the preceding c-label was added. A different path can only be present if the graph was modified after the first traversal. It is therefore a contradiction that a different path is present and that the graph is identical. This same rationale applies to the remaining cases 2.1–6.3. We see that first two different c-labels in the c-sequence of *First* can yield stable c-labels after the second traversal, if the second traversal yields c-labels shown in cases 3 and 5. In other cases, namely 2.1–2.3 and 6.1–6.3, all three different labels appear in all three first slots in the c-sequence of *First*. To spot the pattern there, we need to perform the third traversal of the vertes *First*.

(c) $p = 4$*, i.e., the vertex First has four c-labels in its c-sequence.* The fourth column lists the possible combinations of c-labels when there are four c-labels in the c-sequence of *First*. These are the allowed cases when we assume that the first two c-labels are different, and we have the third c-label as given in the third column. We immediately see the patterns for cases 2.1–2.3 and 6.1–6.3, where patterns could not be spotted after the second traversal of *First*. In 2.1 the pattern that will repeat is **A**, **R**, **AD**; in 2.2 the labels are stable; in 2.3 the pattern is **R**, **AD**. In 6.1 the pattern is **A**, **AD**, **R**; in 2.2 the patter is **AD**, **R**; in 6.3 we have stable labels in $g$. Table 3 provides *all* patterns when three distinct labels are available and the first two c-labels in the c-sequence of *First* are assumed different. If the first two c-labels are the same, we have immediately identified the pattern that gives us stable labels for vertices in $c$.

(d) $p = (4 + i)$ *for the positive integer* $i > 0$*, i.e., the vertex First has more than four c-labels in its c-sequence.* We have seen in the preceding case, when $p = 4$ that we can immediately anticipate the fifth c-label as soon as we have the first four. Patterns identified after the third traversal, when $p = 4$ will repeat because the $C$ walkers will traverse the same graph and produce the same results as in their prior traversals. We have indeed seen by contradiction in Point 3.b above that the next traversal will behave in the manner consistent with the previous ones. It follows that once we have reached the fourth c-label in the c-sequence of *First* and the last two c-labels in that c-sequence are not identical, no further traversals should be performed and the conclusion is the absence of stable c-labels in $c$.

We finally prove that the procedure LabelComplexSCC in Algorithm 5 *(iii) has the running time in* $O((|L(c)| + |V(c)|)(C(c) + 1) + C(c)|V(c)|)$. Let $c = (V(c), L(c))$. LabelComplexSCC first computes the number $C(c)$ of simple cycles via Johnson's algorithm [9] in $O((|L(c)| + |V(c)|)(C(c) + 1))$.

The worst case complexity of the **while** loop is when $c$ is a strongly connected tournament that has no stable c-labels. That $c$ is a *tournament* means that for any pair $v_i, v_j$ in $V(c)$, either $v_iv_j \in L(c)$ or $v_jv_i \in L(c)$. We take a *strongly connected* tournament for the worst case because it has the maximal number of simple cycles. A classical result is that a strongly connected tournament on $n$ vertices has a cycle of length $k$, $k$, for $k = 3, 4, \ldots, n$ [17]. When $c$ is a strongly connected tournament on $|V(c)|$ vertices, it has a cycle of length $k$, for $k = 3, 4, \ldots, |V(c)|$. The number of different lengths of simple cycles in $c$ is then $|V(c)|-2$, while the number $C(c)$ of distinct simple cycles in $c$ is above $|V(c)| - 2$, as there can be more than one distinct simple cycle of the same length. Let $C(c)_k$ be the number of distinct simple cycles in $c$ of length $k$, so that:

$$C = \sum_{k=3}^{|V(c)|} C(c)_k$$

is the number of distinct simple cycles in $c$. Another classical result is that a strongly connected tournament has a Hamiltonian cycle [2], so that $C_{|V(c)|} = 1$. Say now that we send $C(c)$ walkers from the vertex *First*, that is, we apply LabelComplexSCC and consider only the number of operations needed to add the second c-label to the c-sequence of *First*. Since $C(c)_{|V(c)|} = 1$, a single walker, call it H, will move along the Hamiltonian cycle. This being the longest simple cycle in $c$, H will be the last of the $C(c)$ walkers to arrive at the vertex *First*, and it is upon the arrival of H that the second c-label will be added to the c-sequence of *First*. H will traverse $|V(c)|$ vertices and that same number $|V(c)|$ of lines along the Hamiltonian cycle. It is important to observe that *First* may, but need not be on all simple cycles in $c$. (This is because a strogly connected tournament $c$ need not have a vertex that is on all simple cycles.) Consider another walker, say X, moving along a cycle of length, say $\frac{1}{3}|V(c)|$. We have the following two cases:

1. if *First* is on the cycle traversed by X, then X will traverse that cycle only once before H reaches *First*;
2. if *First* is not on the cycle traversed by X, then X will traverse that cycle three times before H reaches *First*.

The second case above is clearly worse in terms of time complexity than the first, as the same cycle will be visited three times intead of only once. For any given walker, the worst case is to traverse $\frac{|V(c)|}{k}$ times its cycle of length $k$. The more times the walkers traverse their respective cycles, the higher the number of operations that will be performed before the procedure LabelComplexSCC terminates. It follows that the upper bound on the number of operations that all $C(c)$ walkers will perform before the second c-label is added to the c-sequence of *First* is:

$$\sum_{k=3}^{|V(c)|} 2\frac{|V(c)|}{k}C(c)_k k = 2C(c)|V(c)|$$

It is in fact impossible to have the case where the above holds, i.e., where the procedure will traverse $\frac{|V(c)|}{k}$ times each simple cycle of length $k$. This makes the above a much too pessimistic estimate of the time complexity, but lets us avoid styding the relationship between the position of *First* on more cycles than the Hamiltonian cycle in $c$, which affects the number of times each walker will traverse its own cycle. If there are no stable c-labels in $c$, the procedure will perform three times the number $2C(c)|V(c)|$ of operations, as it must add three c-labels to the c-sequence of *First*. We therefore conclude that the procedure has the running time in $O((|L(c)| + |V(c)|)(C(c) + 1) + C(c)|V(c)|)$. □

# References

[1] B. Boehm, P. Grunbacher, and R. O. Briggs. Developing groupware for requirements negotiation: Lessons learned. *IEEE Software*, 2001.

[2] P. Camion. Chemins et circuits hamiltoniens des graphes complets. *C. R. Acad. Sc. Paris (A)*, 249:2151–2152, 1959.

[3] C. Chesñevar, J. McGinnis, S. Modgil, I. Rahwan, C. Reed, G. Simari, M. South, G. Vreeswijk, and S. Willmott. Towards an argument interchange format. *Knowl. Eng. Rev.*, 21(4):293–316, 2006.

[4] J. Conklin and M. L. Begeman. gibis: a hypertext tool for exploratory policy discussion. *ACM Trans. Inf. Syst.*, 6(4):303–331, 1988.

[5] R. Darimont and A. van Lamsweerde. Formal refinement patterns for goal-driven requirements elaboration. In *SIGSOFT FSE*, pages 179–190, 1996.

[6] Phan Minh Dung. On the acceptability of arguments and its fundamental role in nonmonotonic reasoning, logic programming and n-person games. *Artif. Intell.*, 77(2):321–358, 1995.

[7] V. Gervasi and B. Nuseibeh. Lightweight validation of natural language requirements. *Software—Practice & Exp.*, 32:113–133, 2002.

[8] J. A. Goguen and C. Linde. Techniques for requirements elicitation. In *Proc. Int. Symp. Req. Eng.*, pages 152–164, 1993.

[9] Donald B. Johnson. Finding all the elementary circuits of a directed graph. *SIAM Journal on Computing*, 4(1):77–84, 1975.

[10] I. Jureta and S. Faulkner. Tracing the rationale behind uml model change through argumentation. In *26th Int. Conf. on Conceptual Modeling (ER)*, pages 454–469, 2007.

[11] I. J. Jureta, S. Faulkner, and P-Y. Schobbens. Justifying goal models. In *IEEE Int. Req. Eng. Conf.*, pages 116–125, 2006.

[12] I. J. Jureta, S. Faulkner, and P-Y. Schobbens. Clear justification of modeling decisions for goal-oriented requirements engineering. *Requir. Eng.*, 13(2):87–115, 2008.

[13] Donald E. Knuth. *The Art Of Computer Programming*, volume 1. Boston: Addison-Wesley, 3rd edition, 1997.

[14] J. C. S. P. Leite and P. A. Freeman. Requirements validation through viewpoint resolution. *IEEE T. Softw. Eng.*, 17(12):1253–1269, 1991.

[15] P. Louridas and P. Loucopoulos. A generic model for reflective design. *ACM Trans. Softw. Eng. Methodol.*, 9(2):199–237, 2000.

[16] M. McGrath. Propositions. In Edward N. Zalta, editor, *The Stanford Encyclopedia of Philosophy*. Fall 2008.

[17] J. W. Moon. *Topics on Tournaments*. Holt, Reinhart & Winston, 1968.

[18] Esko Nuutila and Eljas Soisalon-Soininen. On finding the strongly connected components in a directed graph. *Inf. Process. Lett.*, 49(1):9–14, 1994.

[19] Robert Tarjan. Depth-first search and linear graph algorithms. *SIAM Journal on Computing*, 1(2):146–160, 1972.

[20] Robert E. Tarjan. Edge-disjoint spanning trees and depth-first search. *Acta Informatica*, 6(2):171–185, 1974.

[21] S. Uchitel, R. Chatley, J. Kramer, and J. Magee. Goal and scenario validation: a fluent combination. *Req. Eng.*, 11:123–137, 2006.

[22] P. Zave. Classification of research efforts in requirements engineering. *ACM Comput. Surv.*, 29(4):315–321, 1997.